\documentclass[useAMS,usenatbib]{mn2e}
\usepackage{graphicx}
\title[A {\it Spitzer}-IRS atlas of early-type galaxies in RSA]{A {\it Spitzer}-IRS Spectroscopic atlas of early-type galaxies \\
in the {\it Revised Shapley-Ames Catalog}}
\author[R. Rampazzo et al. ]{R. Rampazzo$^{1}$\thanks{E-mail:
roberto.rampazzol@oapd.inaf.it}, P.Panuzzo$^{2}$, O. Vega$^{3}$, A. Marino$^{4}$,
A. Bressan$^{5}$, M.S. Clemens$^{1}$
\\
$^{1}$INAF Osservatorio Astronomico di Padova, Vicolo
dell'Osservatorio 5, 35122 Padova, Italy\\
$^{2}$GEPI, Observatoire de Paris, CNRS, Univ. Paris Diderot,
Place Jules Janssen 92190 Meudon, France\\
$^{3}$Instituto Nacional de Astrof\'{\i}sica, Optica y Electr\'onica,
Apdos. Postales 51 y 216, C.P. 72000 Puebla, Pue., M\'exico\\
$^{4}$Universit\`a degli Studi di Padova, Dipartimento di Fisica e 
Astronomia ``G. Galilei" Vicolo dell'Osservatorio 3, 35122, Padova, Italy\\
$^{5}$ Scuola Internazionale Superiore di Studi Avanzati (SISSA), via Bonomea, 
265 - 34136 Trieste ITALY
}

\begin{document}
\topmargin=-1.5cm
\date{Accepted ; Received 2012; in original form 2012}

\pagerange{\pageref{firstpage}--\pageref{lastpage}} \pubyear{2013}

\maketitle

\label{firstpage}

\begin{abstract}
We produce an atlas of homogeneously reduced and calibrated 
low resolution IRS spectra of the nuclear regions of nearby early-type galaxies
(i.e. Es and S0s, ETGs), in order to build a reference 
sample in the mid-infrared window. 
From the {\it Spitzer} Heritage Archive we extract ETGs in the {\it Revised 
Shapley-Ames Catalog of Bright Galaxies}  having an IRS SL and/or LL spectrum. 
We recover 91 spectra out of 363 galaxies classified as ETGs  
in the catalog: 56 E (E0-E6), 8 mixed E/S0+S0/E, 27 S0 (both normal and
barred - SB0) plus mixed types SB0/Sa+SB0/SBa.
For each galaxy, we provide the fully reduced and calibrated spectrum, 
the intensity of nebular and molecular emission lines as well as of the Polycyclic 
Aromatic Hydrocarbons (PAHs) after a template spectrum of a passively evolving 
ETG has been subtracted. Spectra are  classified into five mid-infrared classes,
ranging from AGN (class-4) and star forming nuclei
(class-3), transition class-2 (with PAHs) and class-1 (no-PAHs)
 to passively evolving nuclei (class-0).\\ 
 A {\it demographic} study  of mid-infrared spectra shows that Es are 
 significantly more passive than S0s: 46$^{+11}_{-10}$\% of Es 
and 20$^{+11}_{-7}$\% of S0s have a  spectrum of class-0. 
Emission lines are revealed in 64$^{+12}_{-6}$\% of ETGs. The H$_2$S(1) line is found
with similar rate in Es (34$^{+10}_{-8}$\%) and in S0s (51$^{+15}_{-12}$\%). 
PAHs are detected in 47$^{+8}_{-7}$\% of ETGs, but only 9$^{+4}_{-3}$\%   have
PAHs ratios  typical  of  star forming galaxies. \\
Several indicators, such as peculiar morphologies and kinematics, dust--lane 
irregular shape, radio and X-ray properties, suggest that mid-infrared spectral classes 
are connected to phases of accretion/feedback phenomena occurring in the nuclei of ETGs.
\end{abstract}

\begin{keywords}
-- Infrared: galaxies
-- galaxies: elliptical and lenticular, cD
-- galaxies: fundamental parameters
--techniques: spectroscopic
\end{keywords}

\section{Introduction}

Early-type galaxies (Es and S0s, ETGs hereafter) are 
the most luminous and massive stellar aggregates in the local Universe and
posses a multiphase, sometimes conspicuous, interstellar medium (ISM). 
The {\it Spitzer} Space Telescope Heritage Archive (SHA) offers, in the mid-infrared (MIR) 
window, new tools in the endeavor of deciphering the evolutionary history of nearby ETGs.
Several {\it Spitzer}-IRS studies have been dedicated to unveil the MIR spectral 
characteristics of the nuclear region of ETGs, leading to the identification of
 Polycyclic Aromatic Hydrocarbons (PAHs) with both normal
and anomalous inter--band ratios, as well as emission lines from 
molecular hydrogen \citep[][]
{Bregman06,Bressan06,Kaneda05,Panuzzo07,Kaneda08,Panuzzo11}.

Using {\it Spitzer}--IRS spectra, \citet{Bressan06} studied the nuclear properties 
of 20 bright ETGs in the Virgo cluster, identifying a class of 
{\it passively evolving} ETGs. 
Their MIR spectra are characterized by the absence of ionic and molecular
emission as well as of PAHs. Only the silicate emission at 10$\mu$m 
from the circum--stellar dust of O-rich AGB stars  leave their imprint in
these spectra, superimposed on the photospheric stellar continuum 
generated by red giant stars \citep[][]{Knapp89,Athey02}. 
The study of \citet{Bressan06} also shows that ETGs with a passive
spectrum are very common in the Virgo cluster (16 out of 20). 
The remaining four galaxies include already known ``active'' 
objects like NGC 4486 (M~87) and NGC 4435, the early-type companion 
in {\it The Eyes} interacting pair. 
\cite{Buson2009} showed that the MIR continuum of M~87 
is the superposition of a passively evolving spectrum
and the synchrotron emission from the central AGN.
\citet[][]{Panuzzo07} showed that NGC 4435 has a MIR  spectrum
typical of star forming galaxies.
 
\begin{table*}
\caption{The sample of E galaxies}
\begin{tabular}{llccllcc}
\hline
 Ident. & Morpho.  & T   & D        & T88$^{(a)}$      & M$_{K}$  &  r$_e$ & $\sigma_{c}$ \\
           & RSA        &     & [Mpc]      &  Group &                  &  [kpc]  & [km~s$^{-1}$] \\
\hline
\large{\bf E} & & & & & \\
NGC 636  &  E1 & -4.8$\pm$0.5   &30&52 -7 7  & -23.94    & 2.8   & 165 \\
NGC 720  & E5 & -4.8$\pm$0.5  & 28 & 52 -9 7 &-24.94     & 4.9   &  241\\
NGC  821 & E6 & -4.8$\pm$0.4  &24* & 52 0 0 &-24.00      & 5.8   & 200\\
NGC 1209 & E6 & -4.5$\pm$1.0  &36  &51-14 14&-24.46   & 3.2   & 240 \\
 NGC 1275 & E pec      & -2.2$\pm$1.7  & 72*  &  &-26.16  & 5.9   & 259 \\
NGC 1297 & E2 &-2.5$\pm$0.9   &29 &51 -7 4  &-23.36     & 4.0   & 115 \\
 NGC 1339 & E4 & -4.3$\pm$0.5 &20   &51 2 1   &-22.78   & 3.9   & 346\\
NGC 1374 & E0 &-4.4$\pm$1.1  &20 &51 -1 1 & -23.30      & 2.4   & 186 \\
NGC 1379 & E0 &-4.8$\pm$0.5  &20 &51 -1 1 & -23.28      & 4.0   & 120 \\
NGC 1395 & E2 & -4.8$\pm$0.5  &20 &51 -4 4  &-24.64     & 4.7   & 245\\
NGC 1399 &  E1 & -4.6$\pm$0.5  &21&51 -1 1  & -25.30    & 4.1   & 345\\ 
NGC 1404 & E2 & -4.8$\pm$0.5  &20 &51 -1 1  &-24.71     & 2.3   & 234\\
NGC 1407 & E0 &-4.5$\pm$0.7  &29 &51 -8 4 & -25.62      & 9.9   & 286 \\
NGC 1426 & E4 & -4.8$\pm$0.5 &24   &51 -4 4  &-23.24    & 2.9   & 162 \\ 
 NGC 1427 & E5 & -4.0$\pm$0.9  & 20 & 51 -1 1 &-23.33   & 3.0   & 162\\
NGC 1453 & E2 & -4.7$\pm$0.7  &54 &               & -25.57  & 6.5   & 332\\
NGC 1549 & E2 & -4.3$\pm$0.9  &16*&53 -1 1  & -24.24*  & 3.5   & 202\\ 
NGC 1700 & E3 & -4.7$\pm$0.9  &44  &         &-25.16        & 3.9   & 239\\
 NGC 2300 & E3 & -3.4$\pm$1.4  &26  &42-16 16 &-24.51 & 4.0   & 261\\
 NGC 2974 & E4 & -4.2$\pm$1.2 &21   &31-18 18 &-25.42 & 2.5   & 220 \\
NGC 3193 & E2 & -4.8$\pm$0.5  &34 &21 -6 6  &-24.69     & 4.0   & 194\\
NGC 3258 &  E1 & -4.3$\pm$0.9  &32&31 -2    & -24.25     & 4.6   & 271\\  
NGC 3268 & E2 & -4.3$\pm$0.9  &35 &31 -2    &-24.60      & 6.1   & 227 \\
 NGC 3377 & E6 & -4.8$\pm$0.4  &11  &15 -1 1 &-22.71    & 1.8   & 139\\
NGC 3379 & E0 &-4.8$\pm$0.5  &11 &15 -1 1  & -24.04     & 1.9   & 209\\ 
NGC 3557 & E3 & -4.8$\pm$0.5  &46  & 31-10 10&-26.13  & 6.7   & 265\\ 
NGC 3608 &  E1 & -4.8$\pm$0.5  &23&21 -1 1  & -24.80    & 3.5   & 192\\ 
NGC 3818 & E5 & -4.6$\pm$0.8  & 36 & 22 -8 8 &-23.95    & 3.8   & 191\\
NGC 3904 & E2 & -4.7$\pm$0.6  &28 &22 -4 4  &-24.61     & 3.3   & 205\\
NGC 3962 &  E1 & -4.8$\pm$0.4  &35&22 1 1   & -25.09    & 6.0   & 225\\ 
NGC 4261 & E3 & -4.8$\pm$0.4  &32  &11-24 24 &-25.24  & 5.6   & 309 \\
NGC 4365 & E3 & -4.8$\pm$0.5  &23  & 11 -5 1 &-25.19    & 5.6   & 256\\
NGC 4374 & E1 & -4.3$\pm$1.2  &19 & 11 -1 1 &-25.13     & 4.7   & 282\\ 
NGC 4473 & E5 & -4.7$\pm$0.7  & 15 & 11 -1 1 &-23.77    & 1.9   & 179\\
NGC 4478 & E2 & -4.8$\pm$0.4  &20*&11 -1 1  & -23.15*  & 1.3   & 137\\
NGC 4486 & E0 &-4.3$\pm$0.6  &17 &11 -1 1 & -25.81      & 7.8   & 334\\ 
 NGC 4564 & E6 & -4.8$\pm$0.5  &16  & 11 -1 1&-23.09    & 1.5   & 157\\
NGC 4589 & E2 & -4.8$\pm$0.4  &22 &42-13 13 &-23.96   & 1.4   & 224\\
 NGC 4621 & E5 & -4.8$\pm$0.5  & 15 & 11 -1 1 &-24.13   & 2.9   & 225\\
 NGC 4660 & E5 & -4.7$\pm$0.5  & 15 & 11 -1 1 &-22.69   & 0.9   & 203\\
NGC 4696 & E3 & -3.8$\pm$0.6  &18* & 23 -1 1 &-24.13*  & 7.4   & 254\\
NGC 4697 & E6 & -4.4$\pm$0.8  &13  &11-11 10&-24.13   & 4.5   & 174 \\
NGC 5011 & E2 & -4.8$\pm$0.5  &42* &23 0 2  &-24.97*   & 4.8   & 249 \\
NGC 5018 & E4 & -4.4$\pm$1.0 &38   &11 0 0   &-25.23    & 4.2   & 209\\
NGC 5044 & E0 &-4.8$\pm$0.4  &31 &11-31 31& -24.79    & 12.4 & 239\\ 
NGC 5077 & E3 & -4.8$\pm$0.4  &39  & 11 0 0  &-24.73    & 4.3   & 260 \\
NGC 5090 & E2 & -4.9$\pm$0.3  &47* &23 0 2  &-25.79*  & 14.2  & 269\\
NGC 5638 & E1 & -4.8$\pm$0.4  &26 & 41 -3 1 &-23.86    & 3.5    & 165 \\
NGC 5812 & E0 &-4.8$\pm$0.4  &27 &41 0 1  & -24.22    & 3.3     & 200 \\ 
NGC 5813 & E1 & -4.8$\pm$0.4  &32 & 41 -1 1 &-25.15   & 8.9     & 239 \\ 
NGC 5831 & E4 & -4.8$\pm$0.5 &27   &41 -1 1  &-23.75  & 3.3     & 164\\
NGC 7619 & E3 & -4.7$\pm$0.6  &53  &         &-25.62      & 9.5     & 322\\
IC 1459  & E1 & -4.8$\pm$0.6  &29 & 61-17 16&-25.53    & 4.8     & 311\\ 
IC 2006  & E1 & -4.2$\pm$0.9  &20 & 51 -1 1 &-23.05      & 2.8     & 122\\
IC 3370&E2 pec &-4.8$\pm$0.5  &27  &23 0 5  &-24.32   & 5.1     & 202\\
IC  4296 &E0  &-4.8$\pm$0.4  &52 &	  & -26.09               & 10.4   & 340\\ 
\hline
\end{tabular}

\endcenter 
\footnotesize{
$^{(a)}$ The group/cluster identification \citep[][T88]{Tully88} is the following: 
11 Virgo cluster and Southern Extension, 12 Ursa Major Cloud, 
13 Ursa Major Southern Spur, 14 Coma--Sculptor
Cloud, 15 Leo Spur, 21 Leo Cloud, 22 Crater cloud, 23 Centaurus Cloud, 31 Antlia
-- Hydra Cloud, 41 Virgo -- Libra Cloud, 42 Canes Venatici, 43 Canes Venatici
Spur, 44 Draco Cloud, 51 Fornax and Eridanus Cloud, 52 Cetus -- Aries Cloud, 53
Dorado Cloud, 61 Telescopium -- Grus Cloud, 62 Pavo -- Indus Spur, 65 Pegasus
Spur.}
\label{tab1}
\end{table*}

\begin{table*}
\caption{The sample of E/S0, S0 and SB0 galaxies}
\begin{tabular}{llccllcc}
\hline
Ident. & Morpho.  & T   & D        & T88$^{(a)}$      & M$_{K}$ & r$_e$ & $\sigma_{c}$ \\
           & RSA         &     & [Mpc]  &  Group &                    &  [kpc]   & [km~s$^{-1}$] \\
\hline
 \large{\bf E/S0 and S0/E} & & & & &  & \\
 NGC 1052 & E3/S0 &-4.6$\pm$0.8          &19 &52 -1 1 &-24.00   & 3.1  & 215\\
 NGC 1351 & S0$_1$(6)/E6 &-3.1$\pm$0.6   &19 &51 -1 1 &-22.64  & 1.9 & 140 \\
 NGC 4472 & E1/S0$_1$(1)&-4.8$\pm$0.5    &17 &11 -1 1 &-25.73  & 8.2 & 294\\
 NGC 4550 & E7/S0$_1$(7) &-2.1$\pm$0.7   &15 &11 -1 1 &-22.25  & 1.1 & 96\\
 NGC 4570 & S0$_1$(7)/E7 & -2.0$\pm$0.7  &17 &11 -1 1 &-23.49  & 1.5 & 188 \\  
 NGC 4636 & E0/S0$_1$(6) & -4.8$\pm$0.5  &15 &11 2 1  &-24.42   & 6.4 & 209\\
 NGC 5353 & S0$_1$(7)/E7 & -2.1$\pm$0.6  &30 &42 -1 1 &-24.74  & 2.1 & 286\\
 NGC 6868 & E3/S0$_{2/3}$(3)&-4.8$\pm$0.6&38* &61 -1 1 & -25.58*& 6.2 & 277\\
\large{\bf S0,SB0,S0/Sa,SB0/Sa}  &                     &        &           & &  & \\
NGC 584  & S0$_1$(3,5)   &-4.6$\pm$0.9    &20 &52 -7 7 &-24.23   & 2.4     & 206 \\ 
NGC 1366 & S0$_1$(8) & -2.3$\pm$0.7       &17*&51 2 1  &-22.14*  & 0.9    & 120 \\
NGC 1389 & S0$_1$(5)/SB0$_1$ &-2.8$\pm$0.7&21 &51 -1 1 &-23.00 & 1.5& 139 \\ 
NGC 1533 & SB0$_2$(2)/SBa & -2.5$\pm$0.6  &11* &53 -1 1 &-22.58* & 1.6 & 174\\
NGC 1553 & S0$_{1/2}$ pec &-2.3$\pm$0.6   &18* &53 -1 1 &-25.00*   & 5.7  &180   \\
NGC 2685 & S0$_3$(7) pec & -1.0$\pm$0.8   &12 &13 -4 4 &-22.08    & 1.9    & 94\\
NGC 3245 & S0$_1$ & -2.1$\pm$0.5          &21 &21 -8 8 &-23.75   & 2.7         & 210\\
NGC 4036 & S0$_3$(8)/Sa &-2.6$\pm$0.7      &20&12 -5 1&-23.93  & 3.0      & 189\\
NGC 4339 & S0$_{1/2}(0)$ &-4.7$\pm$0.8     &16 &11 4 1&-22.55   & 2.5      & 114\\
NGC 4371 & SB0$_{2/3}$(r)(3) & -1.3$\pm$0.6&17 &11 -1 1&-23.44 & 1.9     & 135\\
NGC 4377 & S0$_1$(3) &-2.6$\pm$0.6         &18 &11 -1 1&-22.43    & 1.1     & 144\\
NGC 4382 & S0$_1$(3) pec &-1.3$\pm$0.6     &18 &11 -1 1&-25.13 & 4.8     & 178\\
NGC 4383 & S0  & -1.0$\pm$0.5             &22* &11 -1 1&-22.22*       & 1.2      & \dots\\
NGC 4435 & SB0$_1$(7) & -2.1$\pm$0.5       &12* &11 -1 1&-23.10* & 1.2    & 157\\
NGC 4442 & SB0$_1$(6) &-1.9$\pm$0.4       &15 &11 -1 1&-23.63   & 1.8      & 187\\
NGC 4474 & S0$_1$(8) &-2.0$\pm$0.5         &15 &11 -1 1&-22.27   & 1.5      & 88\\
NGC 4477 & SB0$_{1/2}$/SBa &-1.9$\pm$0.4   &19 &11 -1 1&-24.06 & 3.5   & 186\\
NGC 4552 & S0$_1$(0)  & -4.6$\pm$0.9 & 16 & 11 -1 1 & -24.31    & 2.3       & 264\\
NGC 4649 & S0$_1$(2) &-4.6$\pm$0.8&16 & 11 -1 1& -25.35         & 5.3       & 335\\
NGC 5128 & S0+S pec &-2.1$\pm$0.6 &7* &14-15 15 &-25.28*      &             & 120  \\
NGC 5273 & S0/a & -1.9$\pm$0.4 &17 &43 1 1 &-22.43                  & 2.5       & 66\\
NGC 5631 & S0$_3$(2)/Sa &-1.9$\pm$0.4 &28 &42 -7 3 &-23.76   & 2.6       & 171\\
NGC 5846 & S0$_1$(0) &-4.7$\pm$0.7 &25 &41 -1 1 & -25.07       & 7.6        & 250\\
NGC 5898 & S0$_{2/3}$(0) & -4.3$\pm$0.9 &29 &41-11 11 &-24.31& 3.1       & 220\\
NGC 7192 & S0$_2$(0) & -3.9$\pm$0.7 &38* & 62 -2 1 &-24.39*   & 5.3        & 257\\
NGC 7332 & S0$_{2/3}$ & -1.9$\pm$0.5&23 & 65 -2 1 &-23.81      & 1.6        & 136\\
IC 5063 & S0$_3$(3) pec/Sa &-1.1$\pm$0.5& 47* & &-24.61*        & 6.1         & 160\\
\hline
\end{tabular}

$^{(a)}$ As in Table~\ref{tab1}.
\label{tab2}
\end{table*}

In low density environments (LDEs), the MIR spectra of ETGs show a 
large variety of features. In a study of 40 ETGs, mainly located in LDEs, 
\citet{Panuzzo11} found that nearly 3/4 of the spectra do not show 
the passive  characteristics found in the Virgo sample of \citet{Bressan06}.
\citet{Annibali10}, using optical spectra, had already classified 
the activity of the ETGs in \citet{Panuzzo11}. Most of their nuclei
show generic LINER characteristics, while only few are either inert/passive or 
AGN systems. These latter,  identified in the optical using nebular line ratios through 
diagnostic diagrams \citep[see][and reference therein]{Baldwin81,Annibali10}, 
appear dominated in the MIR by hot dust emission and 
may show high ionization emission lines. 
In the MIR, several optical LINERs show star forming spectra, similar to NGC~4435, 
with PAH emission typical of late-type galaxies \citep[][]{Smith07}.  Therefore,
MIR spectra provide new clues for the understanding of the 
mechanisms that power LINERs. 

\citet{Panuzzo11} classified the MIR spectra of ETGs  into a five 
spectral classes. The majority of the optical LINERs can be classified into 
three MIR classes displaying  nebular and molecular emission 
(H$_2$) without PAHs or with 
either normal or anomalous PAHs \citep[][]{Bregman06,Kaneda05,Kaneda07,Vega10}.  
Among the spectra with PAHs, the most populated class of MIR spectra (50\%) 
shows anomalous PAH inter-band ratios (7.7$\mu$m/11.3$\mu$m $\leq$2.3).
The least populated class shows normal, star forming, PAH ratios.

Given the short life time of PAHs within the ISM of ETGs \citep[][]{Clemens10},
\citet{Panuzzo11} proposed that MIR classes may trace the evolutionary 
phases of a nucleus as the result of an accretion episode. 
Multi-wavelength observations of the  \citet{Panuzzo11} sample
 support  this hypothesis \citep[][]{Annibali07,Marino11,Rampazzo11}. 
 Accretion  in ETGs may be the result of 
 either secular evolution, driven by bar resonances,  interaction or 
 minor merger episodes which may induce  
nuclear star formation and/or AGN activity. Such activity, fading out with time, 
leaves traces in emission lines, PAH emission and the underlying continuum,
until the nucleus returns to a passive state. 
MIR spectral classes may also offer snapshots of so-called AGN feedback
i.e. the interaction between the energy and radiation generated 
by accretion onto the massive black hole with gas in the host galaxy 
\citep[][]{Fabian12}. AGN feedback may arise where the intense 
flux of photons and particles, produced by the AGN, strips the interstellar gas, 
halting both star formation and accretion onto the AGN itself.

Testing these hypotheses motivated us to query the {\it Spitzer} Archive 
for high S/N, low--resolution  IRS spectra in the quest for a larger sample 
of well studied, nearby ETGs. To this end
we used the   {\it Revised Shapley-Ames Catalogue of Bright Galaxies} (RSA
hereafter) \citep{Sandage87} as our starting point.  RSA classifies 363 nearby galaxies 
as ETGs, although it is not a complete catalogue
\citep[][]{Sandage87}, it is certainly representative of bright nearby
galaxies. We found 91 ETGs with {\it
Spitzer}-IRS spectra. 
This paper organizes this material, homogeneously reduced and calibrated.
The present atlas is intended to be a window on the MIR properties of the nuclear regions 
of nearby ETGs and to  contribute to the understanding of their evolution.
A local reference sample, made of well studied ETGs,
may be used to make comparisons with numerous distant sources
discovered by {\it Spitzer-IRS} \citep[e.g.][]{Sargsyan11}.  

The paper is organized as follows. In section~\ref{sample} we present the
characteristics of the {\it Spitzer}-IRS sample. Section~\ref{data-reduction}
 provides information about observations and the strategy for spectra 
extraction and calibration. 
ETG spectra are analyzed  in Section~\ref{analysis}, where we provide a
measure of atomic and molecular emission lines and PAH intensities. We
finally classify spectra into the MIR spectral classes devised by \citet{Panuzzo11}. 
In Section~\ref{demography} we perform a {\it demographic} 
study of the MIR spectral classes as a function of
morphological type, the galaxy environment, as well as  the
X-ray, CO and radio properties of the galaxies derived from the literature.
In Section~\ref{phenomenology} we investigate MIR classes in the
light of morphological and kinematical peculiarities widely used to infer the
recent evolutionary history of ETGs.

\begin{table*}
\centering
\caption {The {\it Spitzer}--IRS observations of E galaxies in RSA}
\begin{tabular}{llccccccc}
\hline
{ident} &  {PI}  &  {ID} & {SL1}   &   {SL2}      &  {LL2}                 &  {LL1}  & slit [3.6"$\times$18"] & Area${_{slit}}$/Area${_{r_e/8}}$\\
{}          & {}        &   {}     &  [s$\times$Cycle] &  [s$\times$Cycle] & [s$\times$Cycle]  & [s$\times$Cycle] &  [kpc$\times$kpc] &\\
\hline
\large{\bf E} & & & & & & &\\
 NGC~636  & Bregman   & 3535  &14$\times$8 & 14$\times$8     & 30$\times$6& ... & 0.5$\times$2.6 & 3.5\\
NGC~720 &   Bregman &  3535 & 14$\times$8  & 14$\times$8  & 30$\times$6& ... &    0.5$\times$2.4 & 1.0\\
NGC~821 &     Bregman &  3535 & 14$\times$8  & 14$\times$8  & 30$\times$6& ... &   0.4$\times$2.1& 0.5\\
 NGC~1209 & Rampazzo & 30256 &  60$\times$6  & 60$\times$6  & 120$\times$16& 120$\times$8 & 0.6$\times$3.1& 3.9\\
NGC~1275 &  Houck   &  14  &14$\times$2 & 14$\times$2     & 6$\times$4& 6$\times$4 & 1.3$\times$6.3 & 4.6\\
 NGC~1297 &  Rampazzo& 30256 & 60$\times$19  & 60$\times$19  & 120$\times$14& 120$\times$8 & 0.5$\times$2.5&1.6\\
 NGC~1339 & Bregman   & 3535  &14$\times$8 & 14$\times$8     & 30$\times$6& .. & 0.3$\times$1.7 & 0.8\\
 NGC~1374  &  Bregman   & 3535  &14$\times$8 & 14$\times$8     & 30$\times$6& ... & 0.3$\times$1.7 & 2.2\\
 NGC~1379  &   Bregman   & 3535  &14$\times$8 & 14$\times$8     & 30$\times$6& ...& 0.3$\times$1.7 & 0.8\\
  NGC~1395  &Kaneda &3619/30483 &60$\times$2 & 60$\times$2 & 30$\times$2& 30$\times$2& 0.3$\times$1.7 & 0.6\\
  NGC~1399  &  Bregman   & 3535  &14$\times$8 & 14$\times$8     & 30$\times$6& ... & 0.3$\times$1.7 & 0.8\\
  NGC~1404  & Bregman   & 3535  &14$\times$8 & 14$\times$8     & 30$\times$6& ...  & 0.4$\times$1.8 & 2.3\\
 NGC~1407 & Kaneda &3619/30483 &60$\times$2 & 60$\times$2 & 30$\times$2& 30$\times$2 & 0.5$\times$2.5 & 0.3\\ 
  NGC~1426 & Rampazzo& 30256 & 60$\times$12  & 60$\times$12 &120$\times$14& 120$\times$8 & 0.4$\times$2.1 & 2.1\\
NGC~1427 & Bregman &  3535 & 14$\times$8  & 14$\times$8  & 30$\times$6& ... & 0.3$\times$1.7 & 1.4\\
  NGC~1453 & Bregman   &  3535         & 14$\times$8 &  14$\times$8  &   30$\times$6 & ... & 0.9$\times$4.7 & 2.1\\
 NGC~1549  & Kaneda &30483  & 60$\times$4 & 60$\times$4 & 30$\times$4& 30$\times$4 & 0.3$\times$1.4 & 0.6\\
NGC~1700& Bregman   & 3535  &14$\times$8 & 14$\times$8     & 30$\times$6& ... & 0.8$\times$3.8 & 3.8\\
NGC~2300& Bregman   & 3535  &14$\times$8 & 14$\times$8     & 30$\times$6& ... & 0.4$\times$2.3 & 1.3\\
 NGC~2974 & Kaneda &3619/30483 &60$\times$2 & 60$\times$2 & 30$\times$3& 30$\times$3 & 0.4$\times$1.8 & 2.2\\
 NGC~3193 &  Appleton   & 50764  &14$\times$4 &14$\times$3      & 30$\times$2&30$\times$2 & 0.6$\times$3.0 & 2.2\\
   NGC~3258 & Rampazzo &30256 & 60$\times$8  & 60$\times$8  & 120$\times$14& 120$\times$8 & 0.6$\times$2.8 & 1.4\\
 NGC~3268 & Rampazzo &30256  & 60$\times$9  & 60$\times$9  & 120$\times$14& 120$\times$8& 0.6$\times$3.0 & 1.0\\
NGC~3377 &  Bregman &  3535 & 14$\times$8  & 14$\times$8  & 30$\times$6& ... & 0.2$\times$1.0 & 1.1\\
 NGC~3379  &   Bregman   & 3535  &14$\times$8 & 14$\times$8     & 30$\times$6& ...& 0.2$\times$1.0 & 1.1\\
  NGC~3557 & Kaneda & 30483 &   60$\times$3  & 60$\times$3  & 30$\times$3& 30$\times$3 & 0.8$\times$4.0 & 1.5\\ 
  NGC~3608  & Bregman   & 3535  &14$\times$8 & 14$\times$8     & 30$\times$6& ...   & 0.4$\times$2.0 & 1.3\\
 NGC~3818 & Rampazzo & 30256 &60$\times$19  & 60$\times$19  & 120$\times$14& 120$\times$8 & 0.6$\times$3.1& 2.7\\
 NGC~3904& Kaneda &30483  & 60$\times$3 & 60$\times$3 & 30$\times$4& 30$\times$4  & 0.5$\times$2.4 & 2.2 \\
 NGC~3962 & Kaneda &3619/30483 &60$\times$2 & 60$\times$2 & 30$\times$3& 30$\times$3 & 0.6$\times$3.0 & 1.1\\
  NGC~4261 &   Antonucci  &20525   & 240$\times$2 & 240$\times$3     &120$\times$3 &120$\times$2 & 0.6$\times$2.8 & 1.0\\
NGC~4365 & Bressan &3419 &  60$\times$3  & 60$\times$3  & 120$\times$3 & ... & 0.4$\times$2.0 & 0.5\\
  NGC~4374 & Rieke & 82 & 60$\times$4  & 60$\times$4  & 120$\times$4&120$\times$4 & 0.3$\times$1.7 & 0.5\\
NGC~4473 &  Bressan &3419 &  60$\times$3  & 60$\times$3  & 120$\times$3 & ... & 0.3$\times$1.3 & 1.9\\
 NGC~4478 & Bregman   & 3535  &14$\times$8 & 14$\times$8     & 30$\times$6& ... & 0.3$\times$1.7 & 7.3\\
 NGC~4486  & Bressan &3419 &  60$\times$3  & 60$\times$3  & 120$\times$3 & ... & 0.3$\times$1.5 & 0.1\\
 NGC~4564 &     Bressan &3419 &  60$\times$4  & 60$\times$4  & 120$\times$6 & ... & 0.3$\times$1.4 & 3.3\\
 NGC~4589 & Kaneda &3619/30483 &60$\times$2 & 60$\times$2 & 30$\times$2& 30$\times$2 & 0.4$\times$1.9 & 7.9\\
 NGC~4621 &  Bressan &3419 &  60$\times$3  & 60$\times$3  & 120$\times$3 & ... & 0.3$\times$1.3 & 0.8\\
NGC~4660 &   Bressan &3419 &  60$\times$3  & 60$\times$3  & 120$\times$5 & ... & 0.3$\times$1.3 & 8.8\\
  NGC~4696 & Kaneda &3619/30483 &60$\times$2 & 60$\times$2 & 30$\times$3& 30$\times$3 & 0.3$\times$1.6 & 0.2\\
 NGC~4697 & Bregman &  3535 &14$\times$8  & 14$\times$8  & 30$\times$6& ... & 0.2$\times$1.1 & 0.3\\
  NGC~5011 & Rampazzo&30256 &60$\times$6  & 60$\times$6  & 120$\times$12& 120$\times$8 & 0.7$\times$3.7 & 2.3\\
NGC~5018 &  Kaneda &30483  & 60$\times$4 & 60$\times$4 & 30$\times$3& 30$\times$3 & 0.7$\times$3.3 & 2.5\\
NGC~5044 & Rampazzo& 30256 & 19 & 19 & 14 & 8 & 0.5$\times$2.7 & 0.2\\
 NGC~5077 &Rampazzo & 30256 &60$\times$12  & 60$\times$12  & 120$\times$14& 120$\times$8 & 0.7$\times$3.4 & 2.5\\
 NGC~5090 & Kaneda &30483& 60$\times$4 & 60$\times$4 & 30$\times$3& 30$\times$3 & 0.8$\times$4.1 & 0.3\\
   NGC~5638 & Bregman &  3535 &14$\times$8  & 14$\times$8  & 30$\times$6& ... & 0.4$\times$2.3 & 1.7\\
 NGC~5812 & Bregman & 3535 &60$\times$6  & 60$\times$6  & 120$\times$12& 120$\times$8 & 0.5$\times$2.4 & 2.0\\ 
   NGC~5813 & Bregman & 3535 & 14$\times$8  & 14$\times$8  & 30$\times$6& ... & 0.6$\times$2.8 & 0.4\\
 NGC~5831 & Bregman &  3535 & 14$\times$8  & 14$\times$8  & 30$\times$6& ... & 0.5$\times$2.4 & 2.0\\
   NGC~7619 & Bregman   & 3535  &14$\times$8 & 14$\times$8     & 30$\times$6& ... & 0.9$\times$4.6 & 1.0\\
   IC~1459    & Kaneda &30483  & 60$\times$3 & 60$\times$3 & 30$\times$2& 30$\times$2 & 0.5$\times$2.5 & 1.1\\
IC~2006    & Bregman             &  3535         & 14$\times$8 &  14$\times$8  &   30$\times$6 & ...& 0.3$\times$1.7 & 1.6\\
   IC~3370    & Kaneda & 3619/30483 &60$\times$2 & 60$\times$2 & 30$\times$3& 30$\times$3 & 0.5$\times$2.4 & 0.9\\
 IC~4296    & Antonucci & 20525 & 240$\times$2  & 240$\times$2  & 120$\times$3& 120$\times$3 & 0.9$\times$4.5 &0.8 \\
\hline\hline
\end{tabular}
\label{tab3}
\end{table*}

\begin{table*}
\centering
\caption {The {\it Spitzer}--IRS observations of E/S0s and S0s in RSA}
\begin{tabular}{llccccccc}
\hline
{ident} &  {PI}  &  {ID} & {SL1}   &   {SL2}      &  {LL2}                    &  {LL1}  & slit [3.6"$\times$18"] & Area${_{slit}}$/Area${_{r_e/8}}$\\
{}          & {}        &   {}     &  [s$\times$Cycle] &  [s$\times$Cycle] & [s$\times$Cycle]  & [s$\times$Cycle] & [kpc$\times$kpc] &\\
\hline
 \large{\bf E/S0 and S0/E} & & & & & & & &\\
 NGC~1052 & Kaneda & 30483 &60$\times$2 & 60$\times$2 & 30$\times$2& 30$\times$2 & 0.3$\times$1.7 & 1.2\\
NGC~1351 & Bregman & 3535 &14$\times$8  & 14$\times$8  & 30$\times$6& ... & 0.3$\times$1.7 & 3.2 \\
NGC~4472 & Bregman & 3535 &14$\times$8  & 14$\times$8  & 30$\times$6& ... & 0.3$\times$1.5 & 0.1\\
NGC~4550 &  Bressan &3419 &  60$\times$20  & 60$\times$20  & 120$\times$14 & ...& 0.3$\times$1.3 & 5.8\\
NGC~4570 &  Bressan &3419 &  60$\times$3  & 60$\times$3  & 120$\times$5 & ...   & 0.3$\times$1.5 & 4.2\\
 NGC~4636 & Bressan &3419 &  60$\times$3  & 60$\times$3  & 120$\times$5 & ... & 0.2$\times$1.3 & 0.2\\
NGC~5353 &   Appleton  & 50764  & 60$\times$1& 60$\times$1     &30$\times$2 &30$\times$2 & 0.5$\times$2.6& 6.1\\
 NGC~6868 &  Rampazzo &30256 &60$\times$6  &  60$\times$6 &  120$\times$13 &120$\times$8 & 0.7$\times$3.3 & 1.2\\
\large{\bf S0,SB0,S0/Sa,SB0/Sa}  &                     &        &           & &   &\\
NGC~584 &  Bregman & 3535 &14$\times$8  & 14$\times$8  & 30$\times$6& ... & 0.3$\times$1.7 & 2.1\\
NGC~1366 & Rampazzo& 30256 & 60$\times$11& 60$\times$11& 120$\times$14& 120$\times$8 & 0.3$\times$1.5 & 11.7\\
 NGC~1389 & Rampazzo& 30256 &60$\times$9  & 60$\times$9 & 120$\times$14& 120$\times$8 & 0.4$\times$1.8 & 5.8\\
NGC~1533 &Rampazzo  & 30256 & 60$\times$3  & 60$\times$3& 120$\times$5& 120$\times$3 & 0.2$\times$1.0 & 1.5\\
NGC~1553 & Rampazzo& 30256 & 60$\times$3  & 60$\times$3 & 120$\times$3& 120$\times$3 & 0.3$\times$1.6 & 0.3\\
NGC~2685 & Rieke    &  40936 &14$\times$6 & 14$\times$6      &14$\times$6 &14$\times$6 & 0.2$\times$1.0 & 1.3\\
NGC~3245 & Sturm    & 3237  & 14$\times$2& 14$\times$2     & ... &... & 0.3$\times$1.8 & 1.8\\
NGC~4036 & Sturm    &  3237 & 14$\times$2& 14$\times$2     & ... &... & 0.3$\times$1.7 & 1.4\\
NGC~4339 &   Bressan &3419 &  60$\times$20  & 60$\times$20  & 120$\times$14 & ... & 0.3$\times$1.4 & 1.3\\
NGC~4371 &    Bressan &3419 &  60$\times$9  & 60$\times$9  & 120$\times$10 & ... & 0.3$\times$1.5 & 2.4\\
NGC~4377 &     Bressan &3419 &  60$\times$12  & 60$\times$12  & 120$\times$8 & ... & 0.3$\times$1.6 & 7.7\\
NGC~4382 &     Bressan &3419 &  60$\times$3  & 60$\times$3  & 120$\times$3 & ...  & 0.3$\times$1.6 & 0.4\\
NGC~4383 &     Weedman& 50834  &60$\times$2 & 60$\times$2   & 120$\times$1 &120$\times$1 & 0.4$\times$1.9 & 10.5\\
NGC~4435 &     Bressan &3419 &  60$\times$3  & 60$\times$3  & 120$\times$5 & ... & 0.2$\times$1.0 & 3.1\\
NGC~4442 &      Bressan &3419 &  60$\times$3  & 60$\times$3  & 120$\times$3 & ... & 0.3$\times$1.3 & 2.2\\
NGC~4474 &      Bressan &3419 &  60$\times$20  & 60$\times$20  & 120$\times$24 & ...& 0.3$\times$1.3 & 3.1\\
NGC~4477 &  Riecke   & 40936  &14$\times$6 & 14$\times$6   &14$\times$6 & 14$\times$6 & 0.3$\times$1.7 & 0.9\\
NGC~4552 & Bregman & 3535 &14$\times$8  & 14$\times$8  & 30$\times$6& ...& 0.3$\times$1.4 & 1.5\\
NGC~4649 & Bregman & 3535 &14$\times$8  & 14$\times$8  & 30$\times$6& ...& 0.3$\times$1.4 & 0.3\\
NGC~5128 & Houck/Lacy&  14 &6$\times$4/6$\times$4 & 6$\times$4/6$\times$4     & 6$\times$4& 6$\times$4 & 0.1$\times$0.6 & \\
NGC~5273 &  Riecke   & 40936  &6$\times$2 &      &6$\times$2 & 6$\times$2 & 0.3$\times$1.5 & 1.4 \\
NGC~5631 &   Riecke   & 40936  &14$\times$6 & 14$\times$6   &14$\times$6 & 14$\times$6 & 0.5$\times$2.4  & 3.7 \\
 NGC~5846 &Bregman &3535 & 14$\times$8  & 14$\times$8  & 30$\times$6& ... & 0.4$\times$2.2 & 0.3\\
NGC~5898 & Rampazzo & 30256 &60$\times$11  & 60$\times$11  & 120$\times$14& 120$\times$8 & 0.5$\times$2.5 & 2.7\\
NGC~7192 & Rampazzo&  30256 & 60$\times$12 &60$\times$12  & 120$\times$14  & 120$\times$8 & 0.7$\times$3.3 & 1.6\\
NGC~7332 & Rampazzo & 30256 &60$\times$7 &60$\times$7  & 120$\times$14  & 120$\times$8 & 0.4$\times$2.0 & 6.1\\
IC~5063    & Gorjan & 30572 & 14$\times$2  & 14$\times$2  & 30$\times$1& 30$\times$1 & 0.8$\times$4.1 &1.8\\
\hline\hline
\end{tabular}
\label{tab4}
\end{table*}

\begin{figure*}
\centering
\includegraphics[height=15cm]{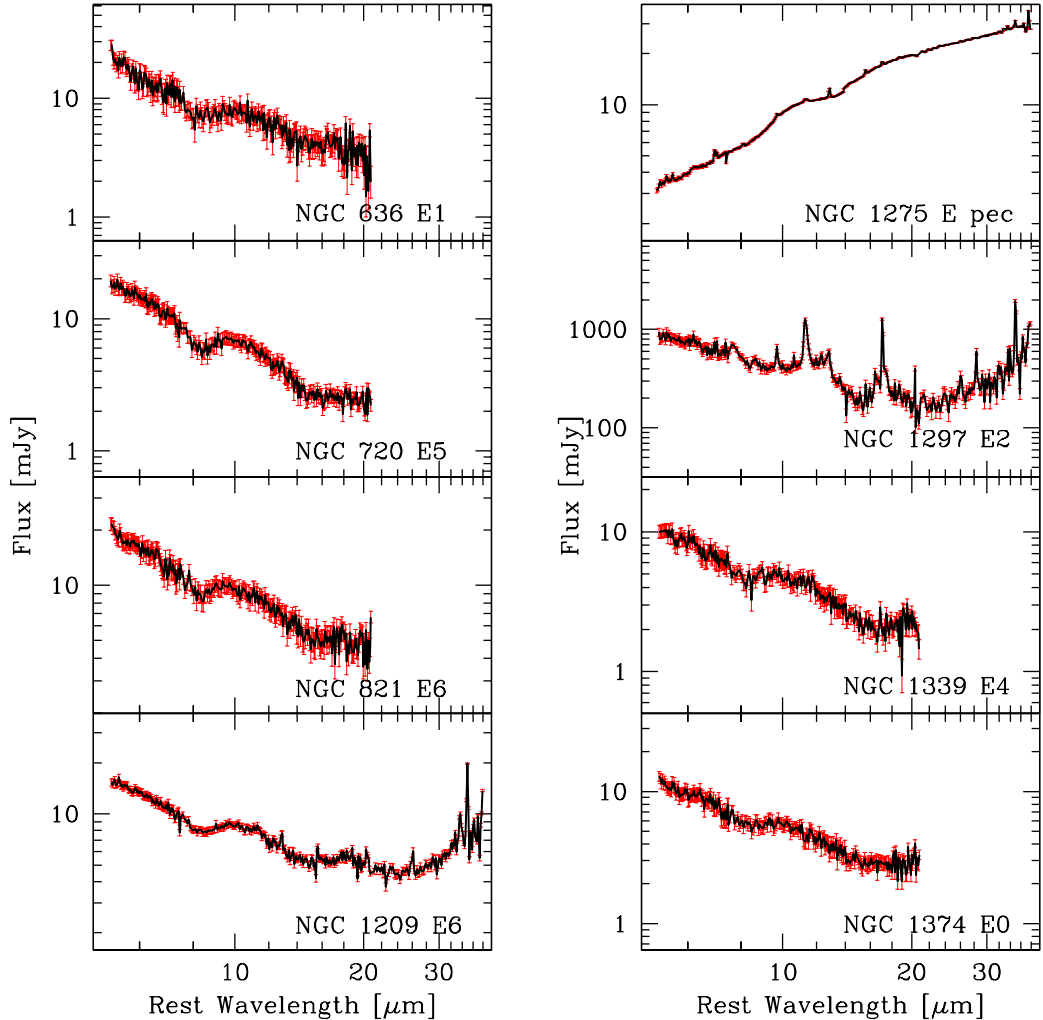}
\caption{Spectra of Es. Flux vs. rest wavelength as obtained from 
{Spitzer-IRS} low resolution modules. Bars represent 1$\sigma$ error.
The LL modules have been scaled to match the SL fluxes.}
\label{fig1}
\end{figure*}
%

\begin{figure*}
\centering
\includegraphics[height=15cm]{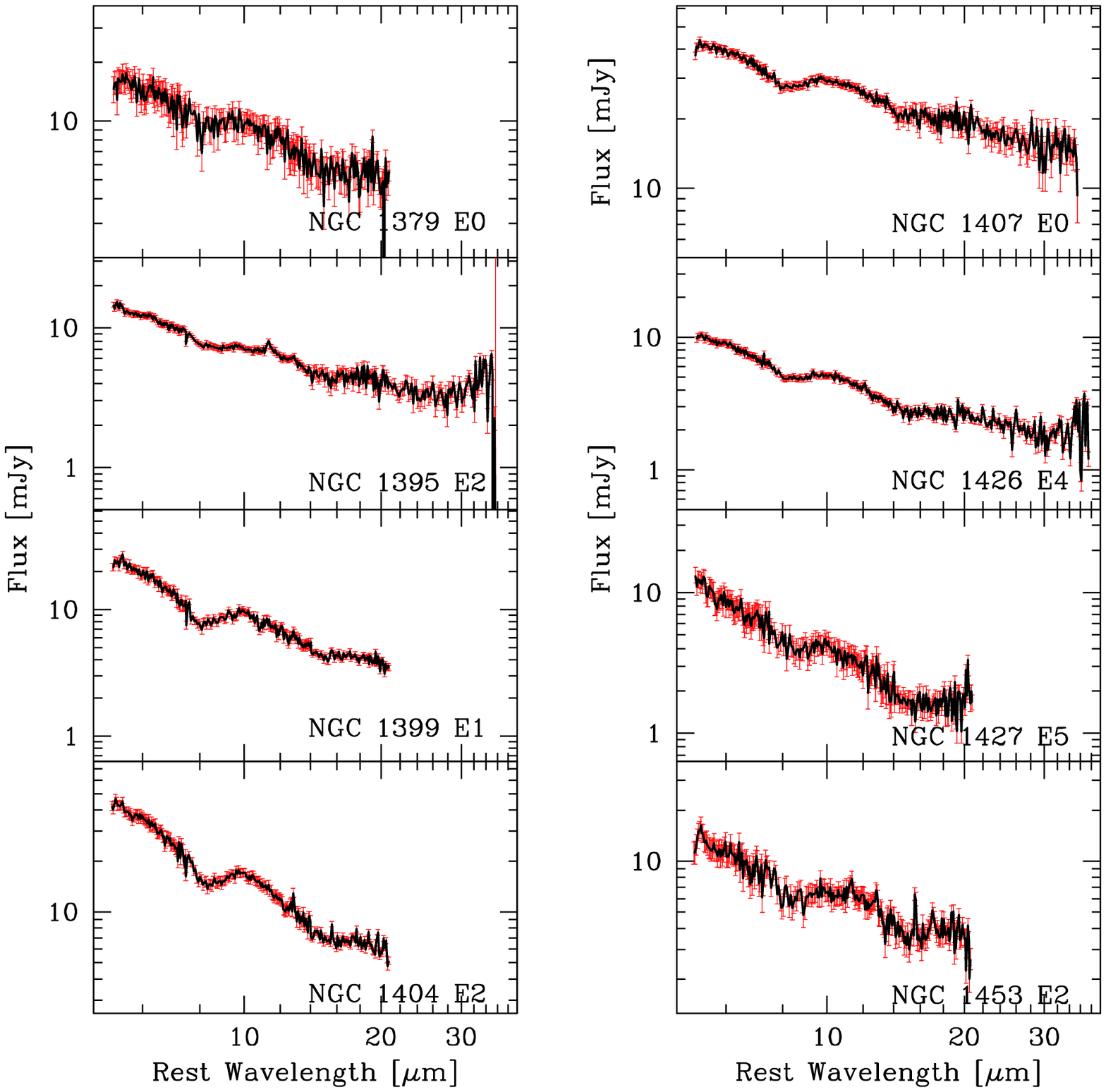}
\addtocounter{figure}{-1}
\caption{{\it (cont.)} MIR spectra of Es.}
\end{figure*}
%

\begin{figure*}
\centering
\includegraphics[height=15cm]{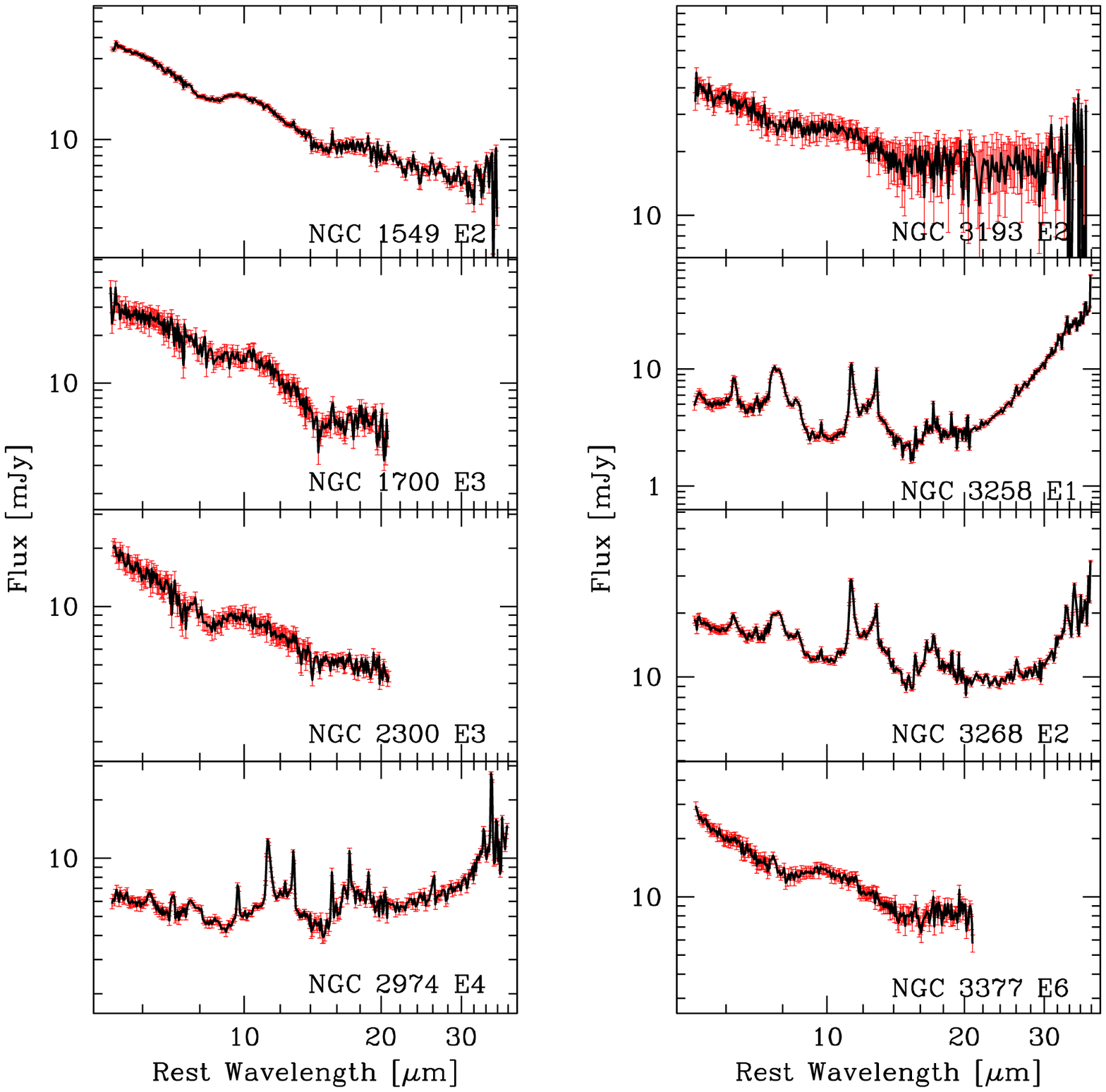}
\addtocounter{figure}{-1}
\caption{{\it (cont.)} MIR spectra of Es.}
\end{figure*}

\begin{figure*}
\centering
\includegraphics[height=15cm]{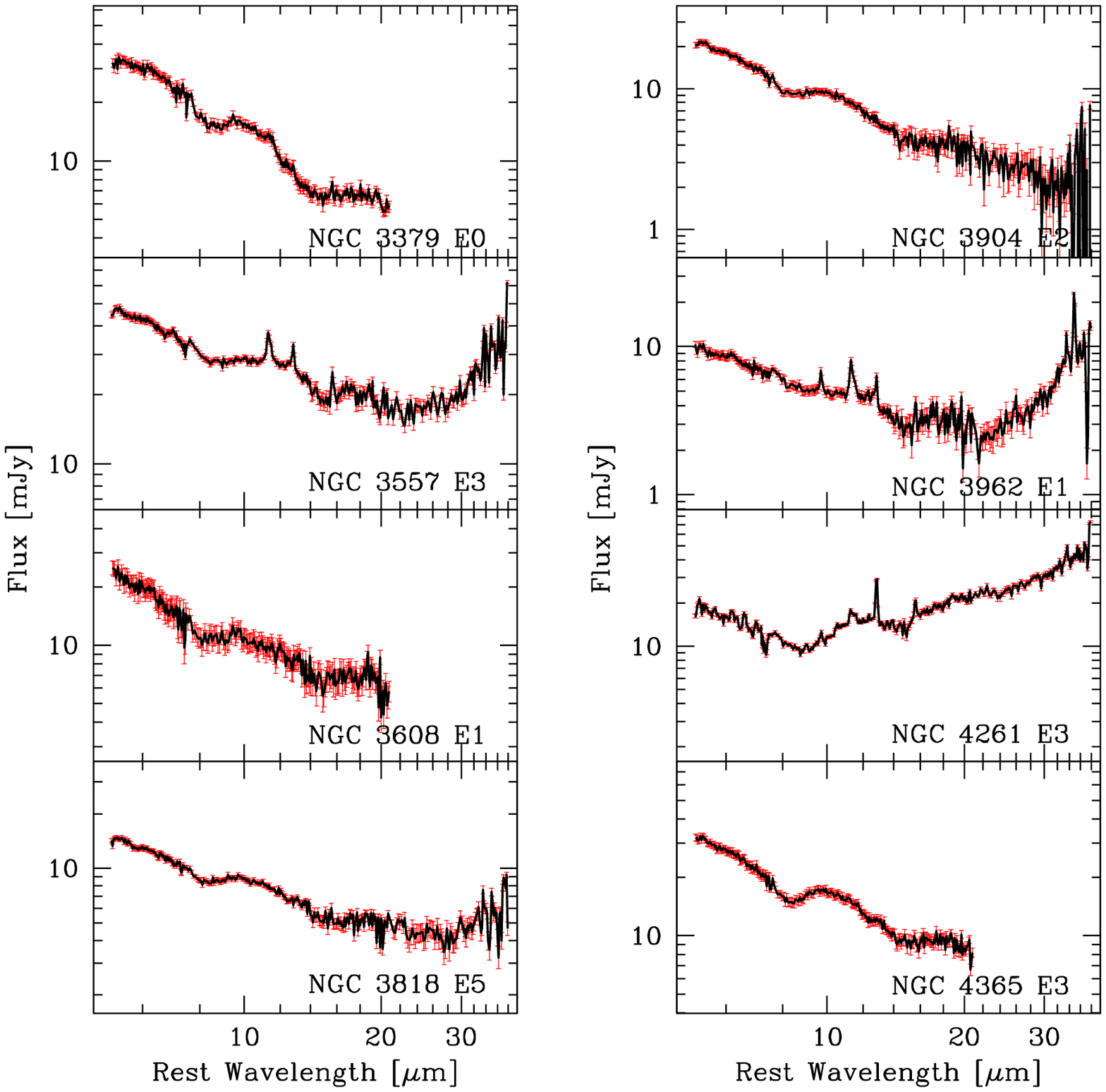}
\addtocounter{figure}{-1}
\caption{{\it (cont.)} MIR spectra of Es.}
\end{figure*}

\begin{figure*}
\centering
\includegraphics[height=15cm]{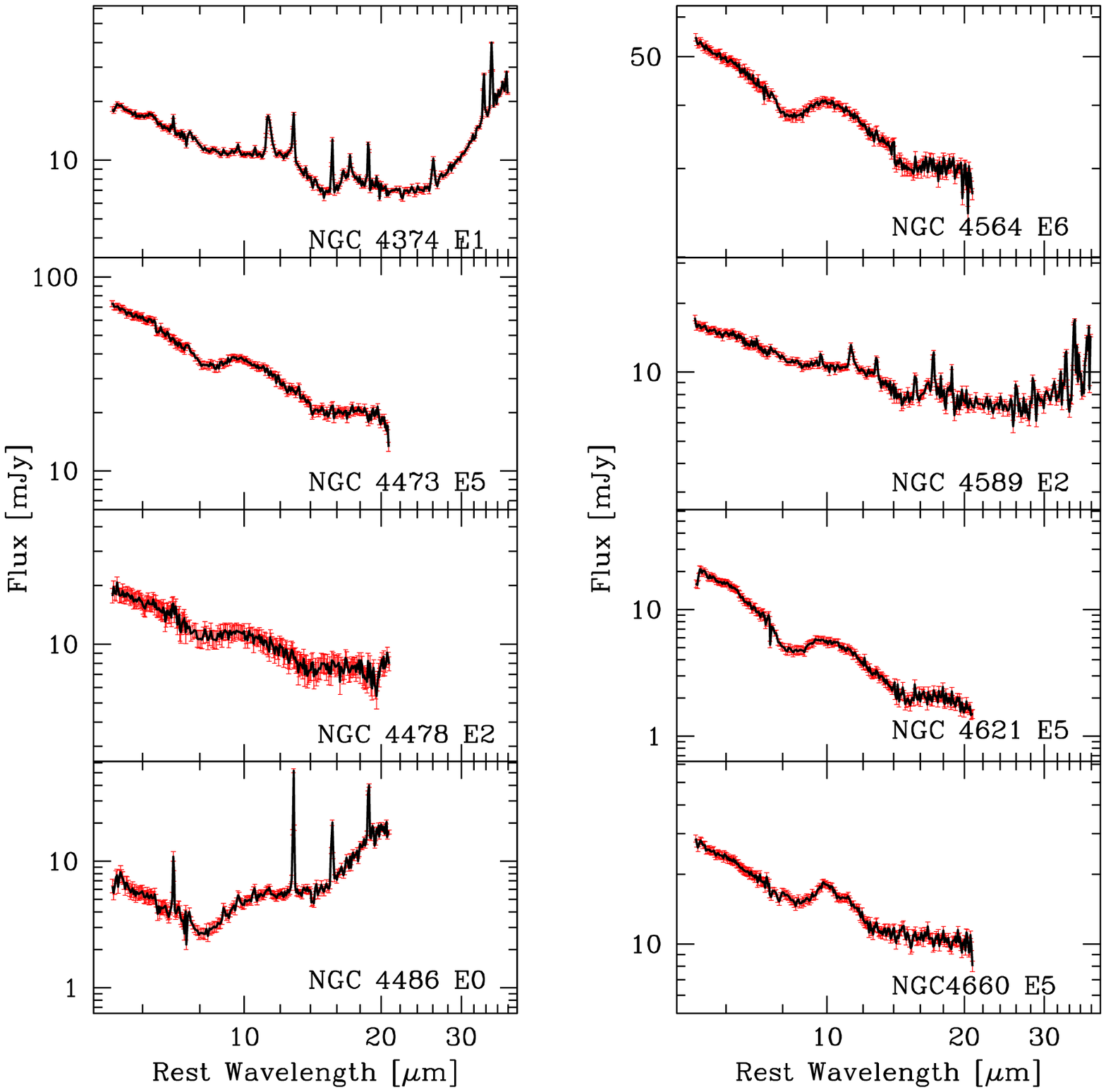}
\addtocounter{figure}{-1}
\caption{{\it (cont.)} MIR spectra of Es.}
\end{figure*}

\begin{figure*}
\centering
\includegraphics[height=15cm]{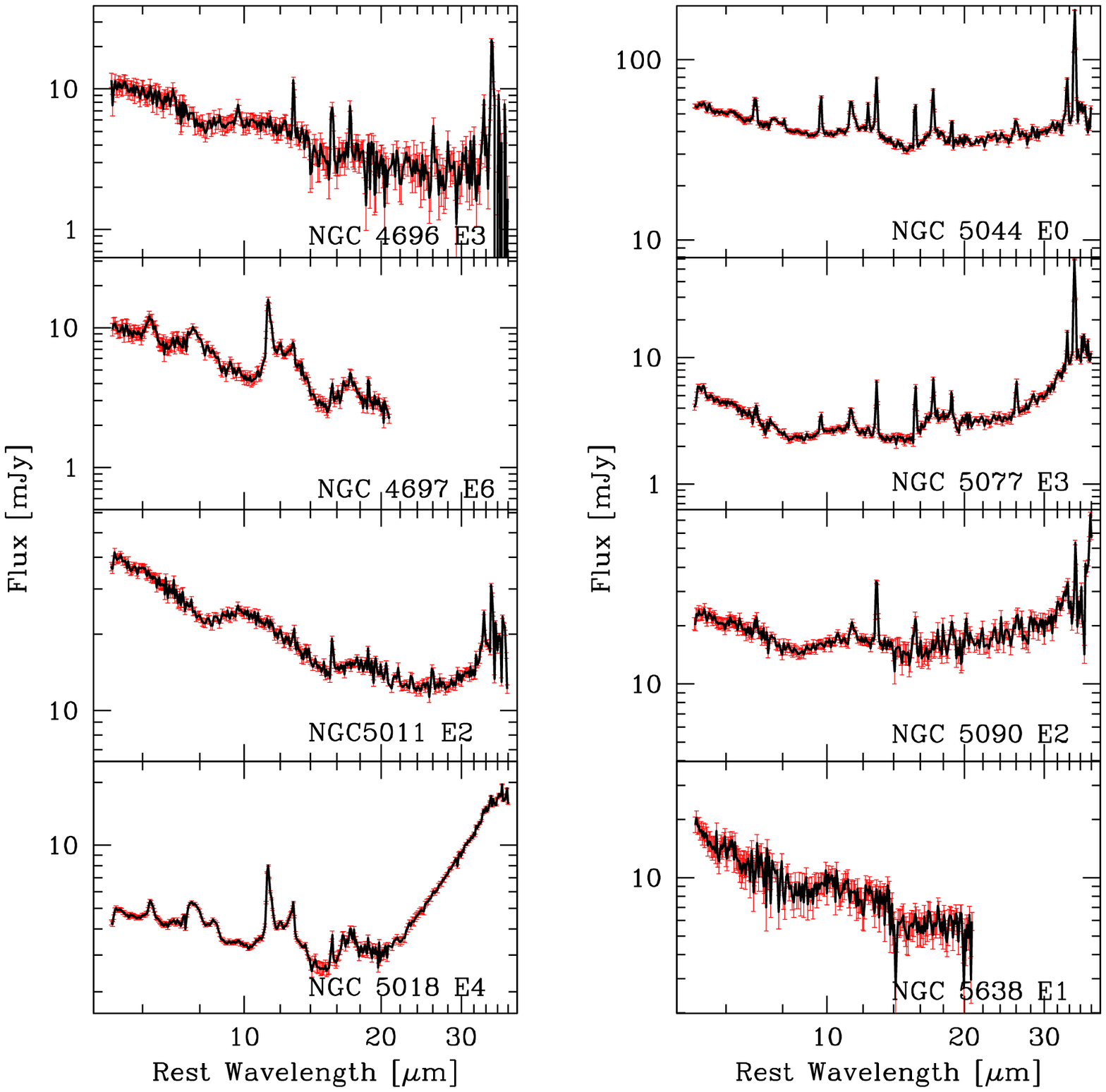}
\addtocounter{figure}{-1}
\caption{{\it (cont.)} MIR spectra of Es.}
\end{figure*}

\begin{figure*}
\centering
\includegraphics[height=15cm]{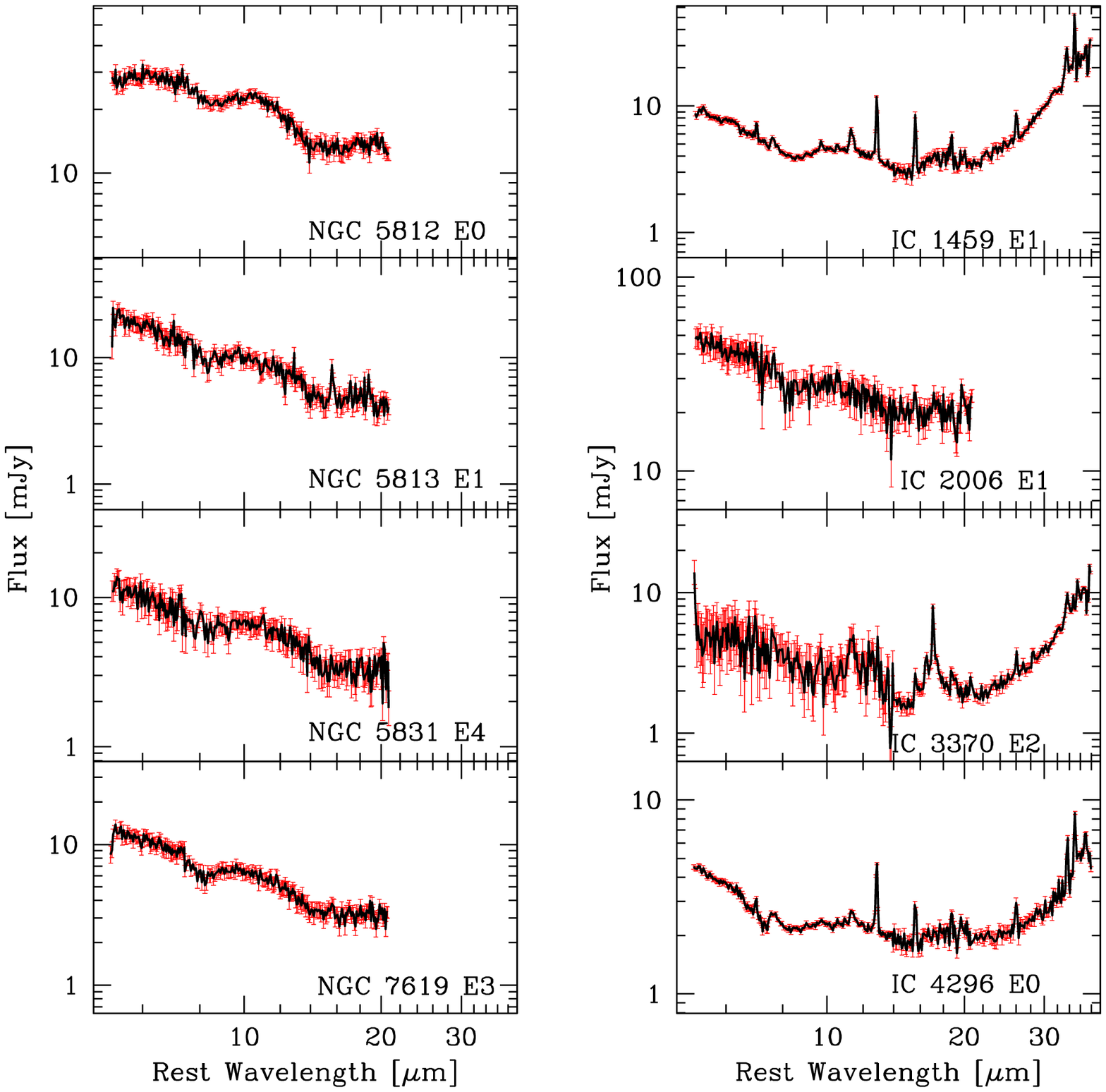}
\addtocounter{figure}{-1}
\caption{{\it (cont.)} MIR spectra of Es. For IC~3370 \citep[see][]{Kaneda08},
the SL observations miss the center of the galaxy by 10\arcsec, which caused 
significant reduction in the S/N ratio of the SL spectrum.}
\end{figure*}

\begin{figure*}
\centering
\includegraphics[height=15cm]{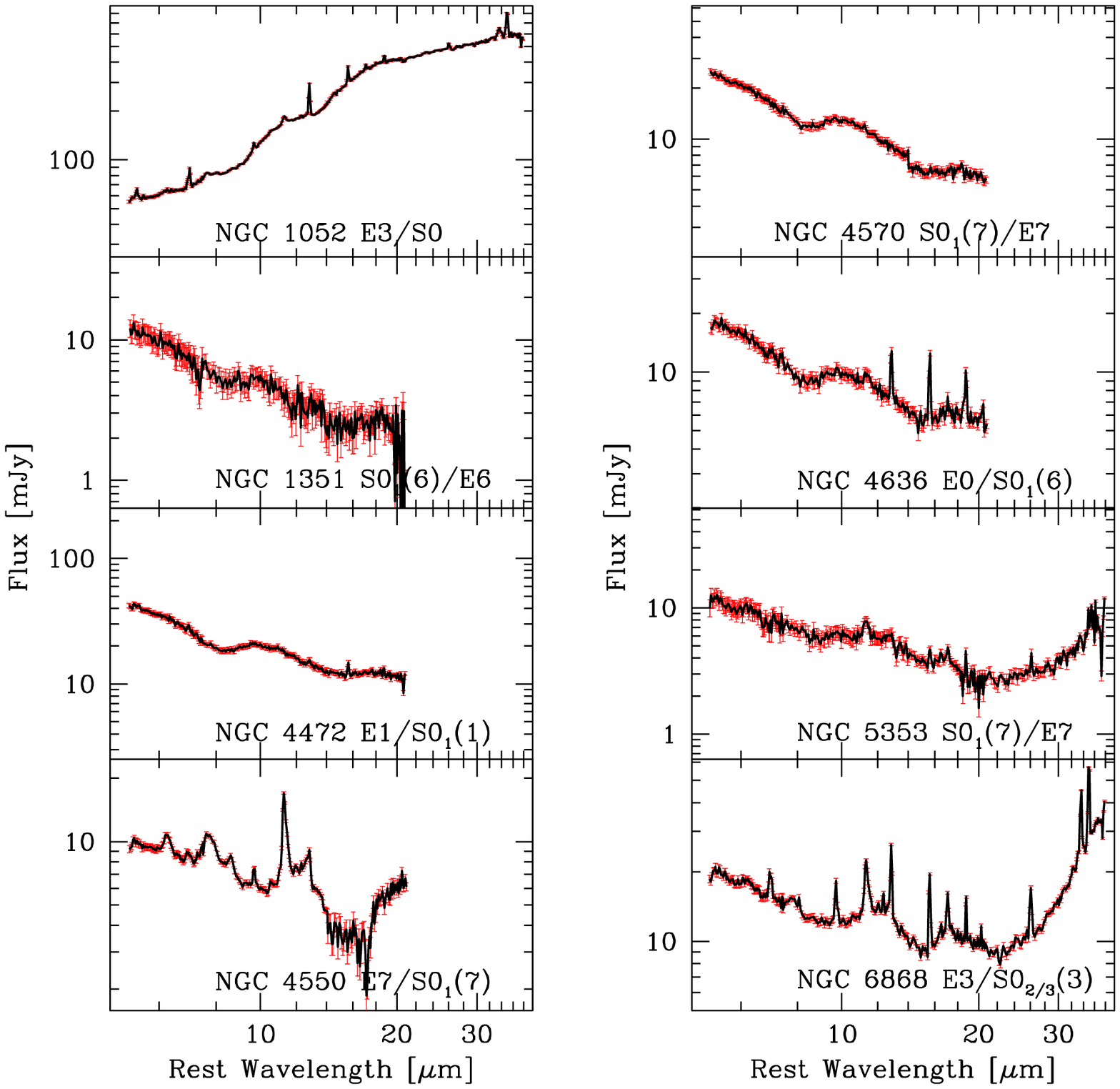}
\addtocounter{figure}{-1}
\caption{{\it (cont.)} MIR spectra of E/S0s. 
In the case of NGC~4472 the LL2 module has been 
arbitrarily scaled.}
\end{figure*}

\begin{figure*}
\centering
\includegraphics[height=17cm]{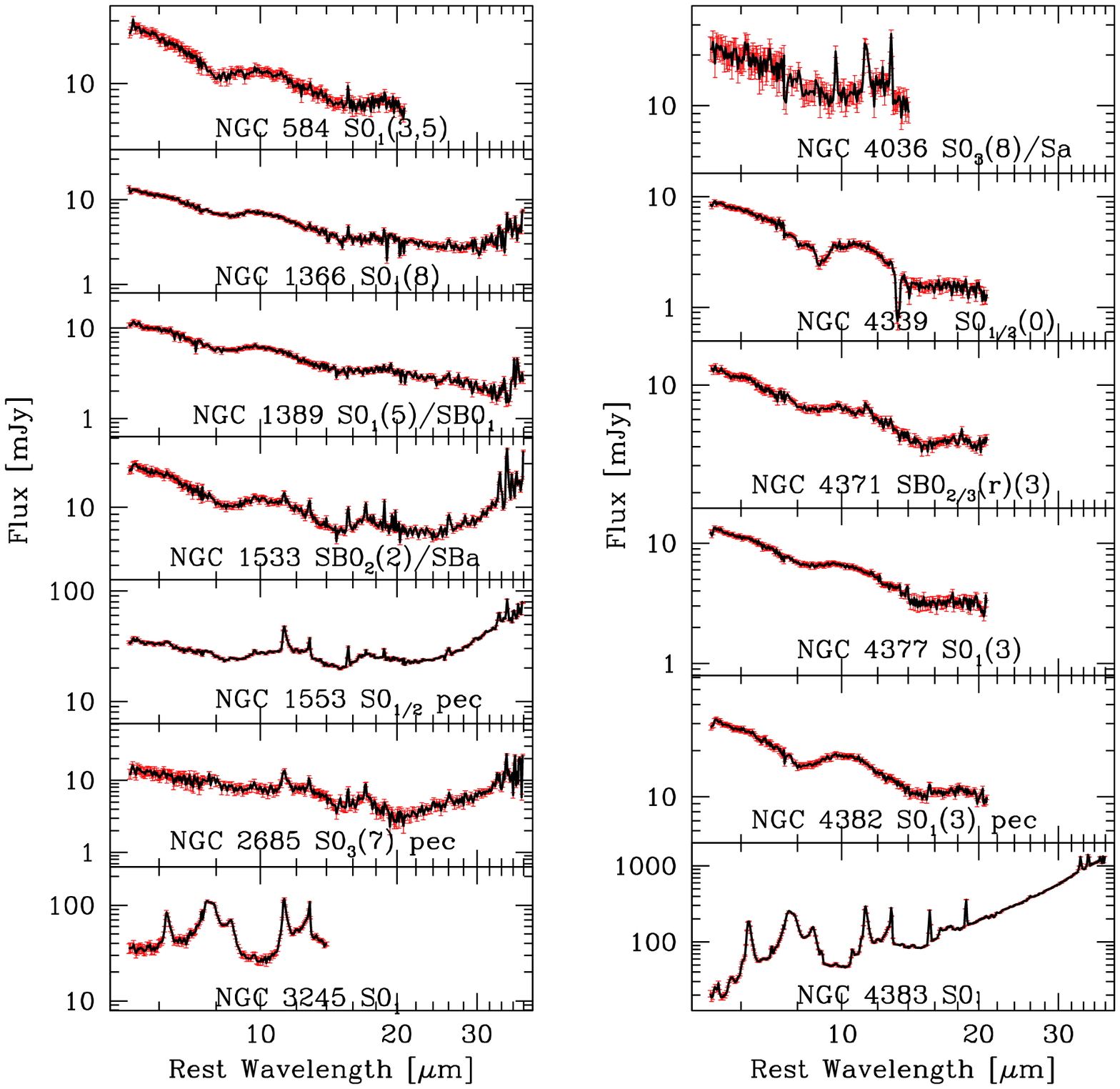}
\addtocounter{figure}{-1}
\caption{{\it (cont.)}  MIR spectra of S0s and SB0s. The spectrum of NGC4339 shows 
absorption features at 9 and 13 $\mu$m which are
data reduction artifacts due to a bright source falling in the peak-up
blue FOV during the observation. }
\end{figure*}

\begin{figure*}
\centering
\includegraphics[height=17cm]{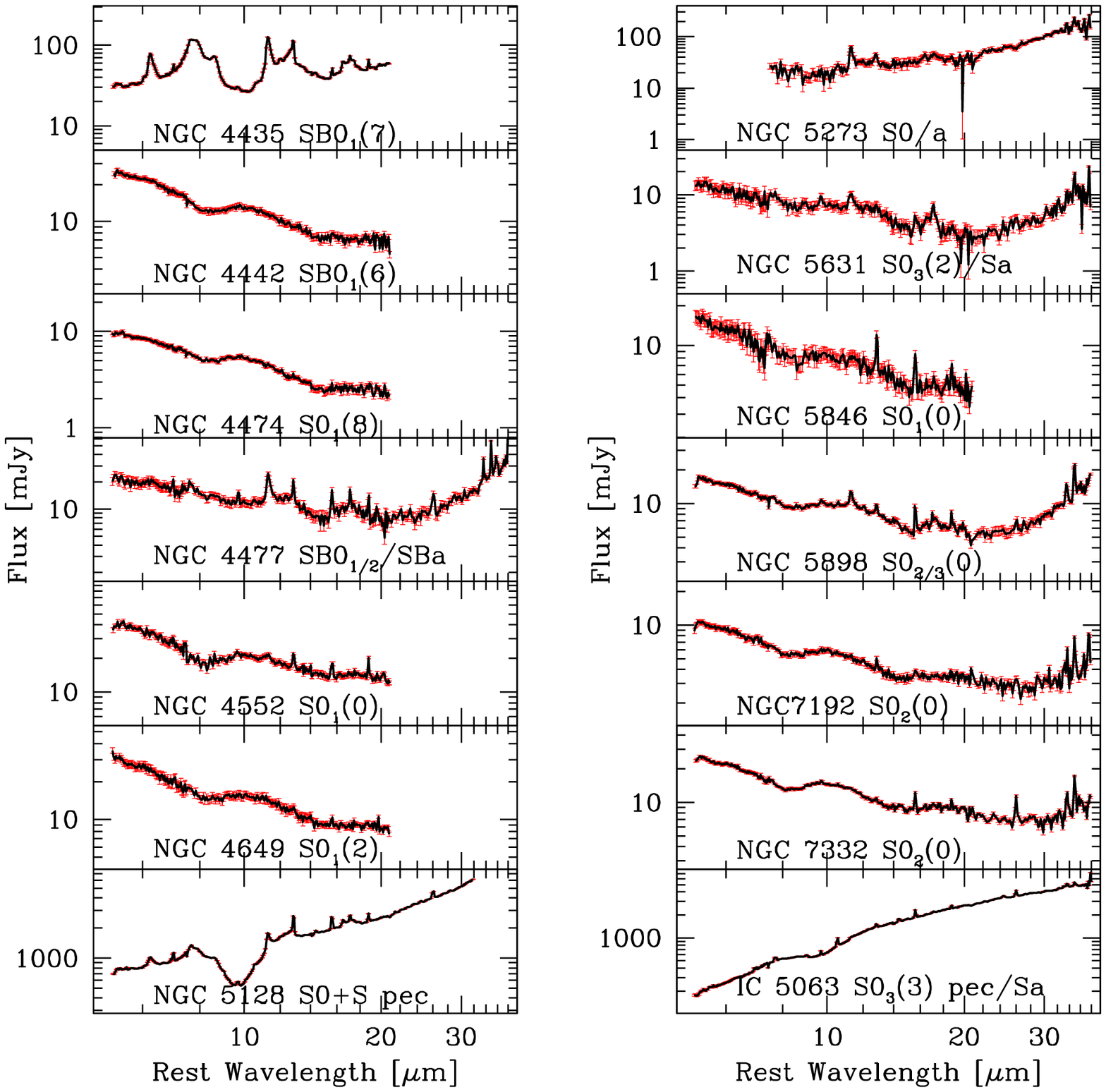}
\addtocounter{figure}{-1}
\caption{{\it (cont.)} MIR spectra of S0s and SB0s.}
\end{figure*}

\begin{figure*}
\includegraphics[width=8cm]{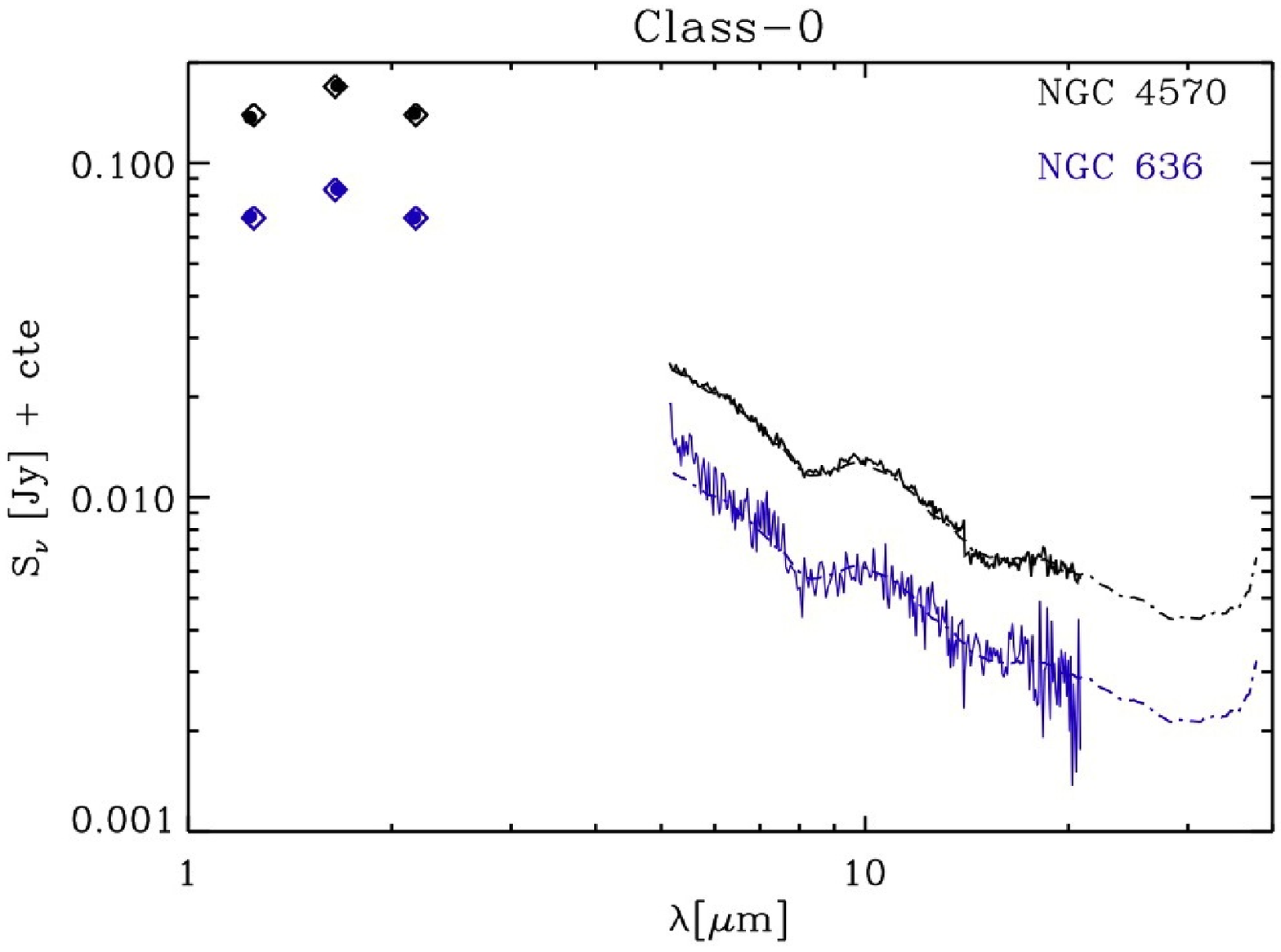}
\includegraphics[width=8cm]{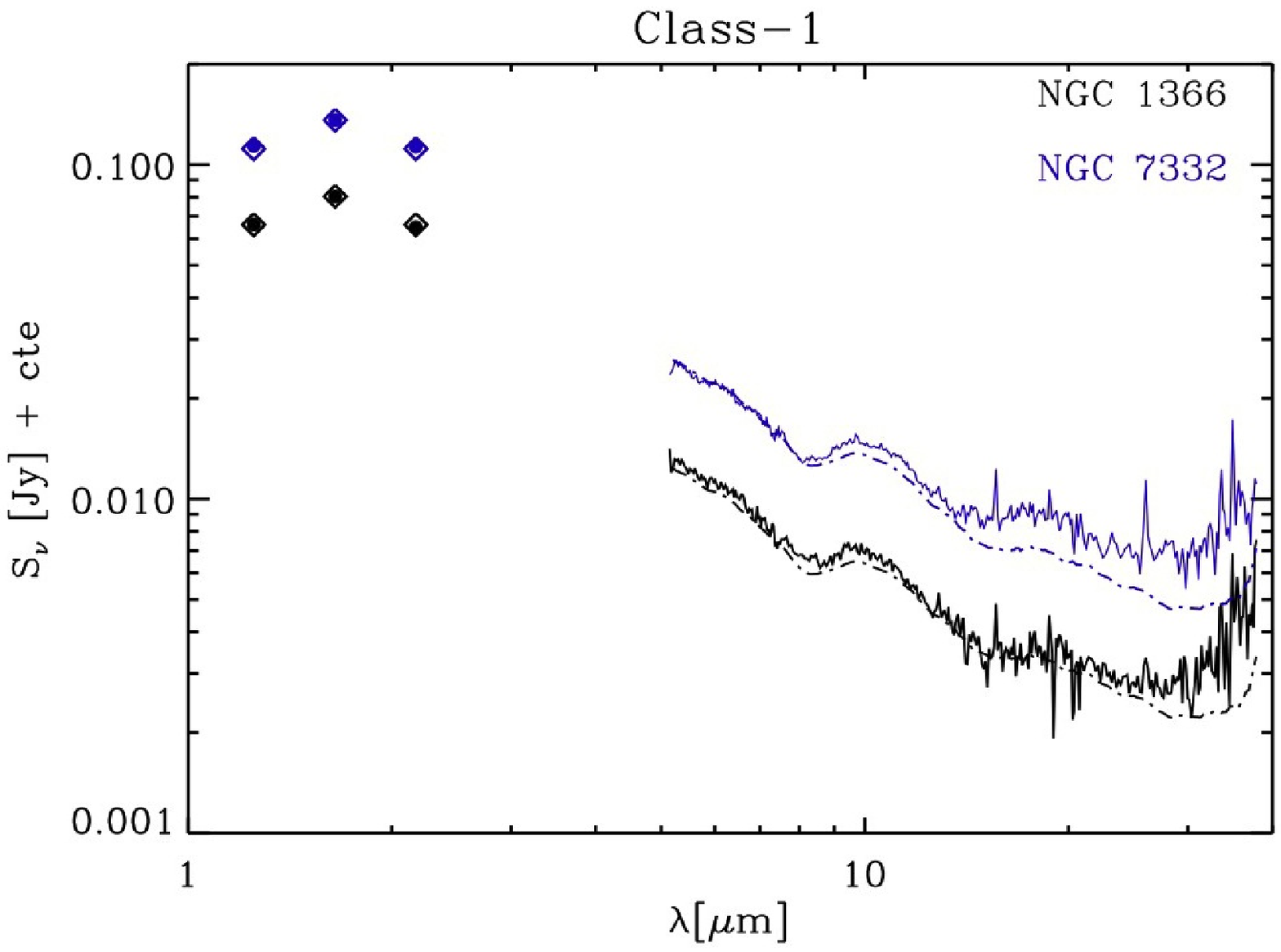}
\includegraphics[width=8cm]{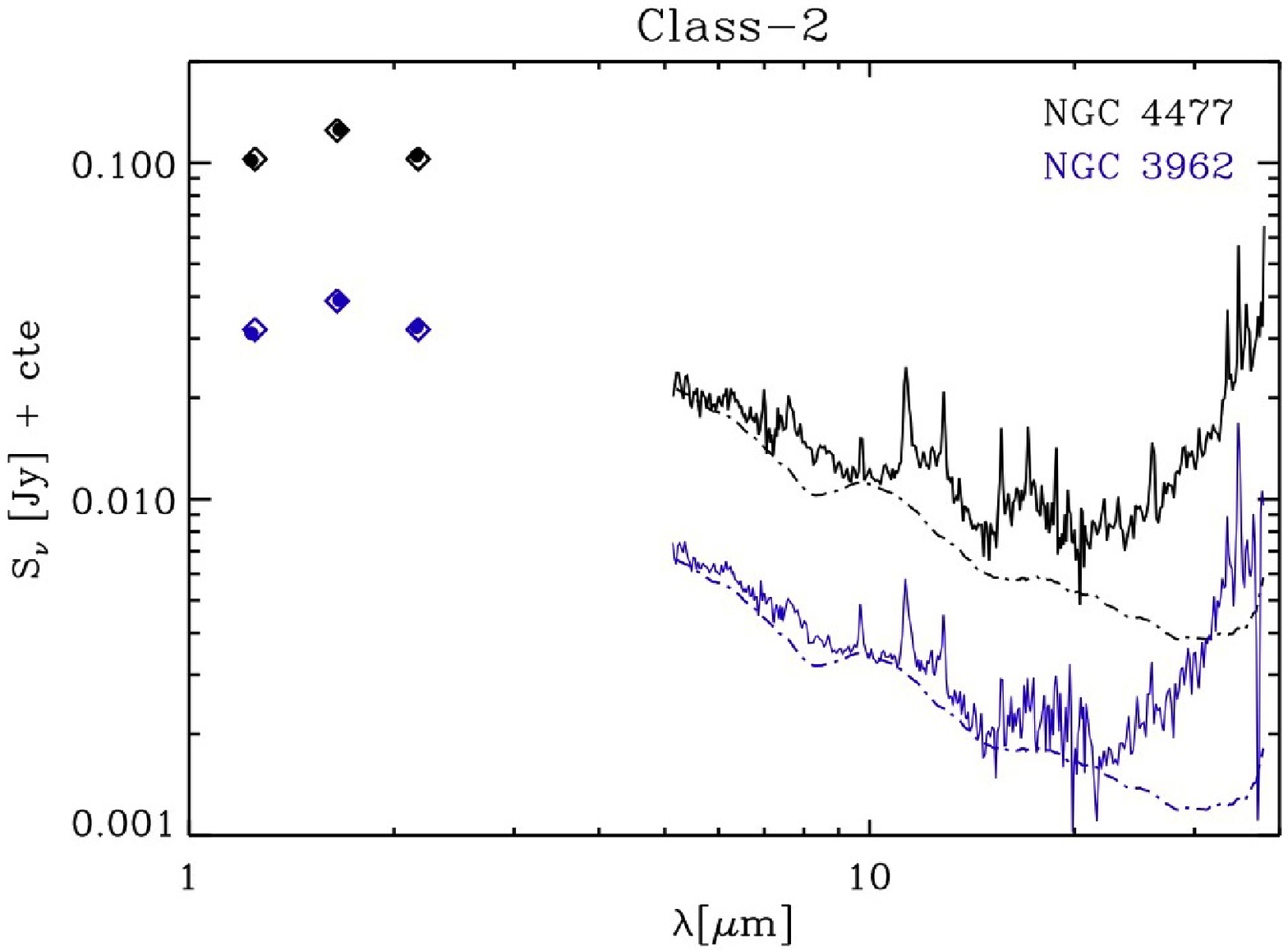}
\includegraphics[width=8cm]{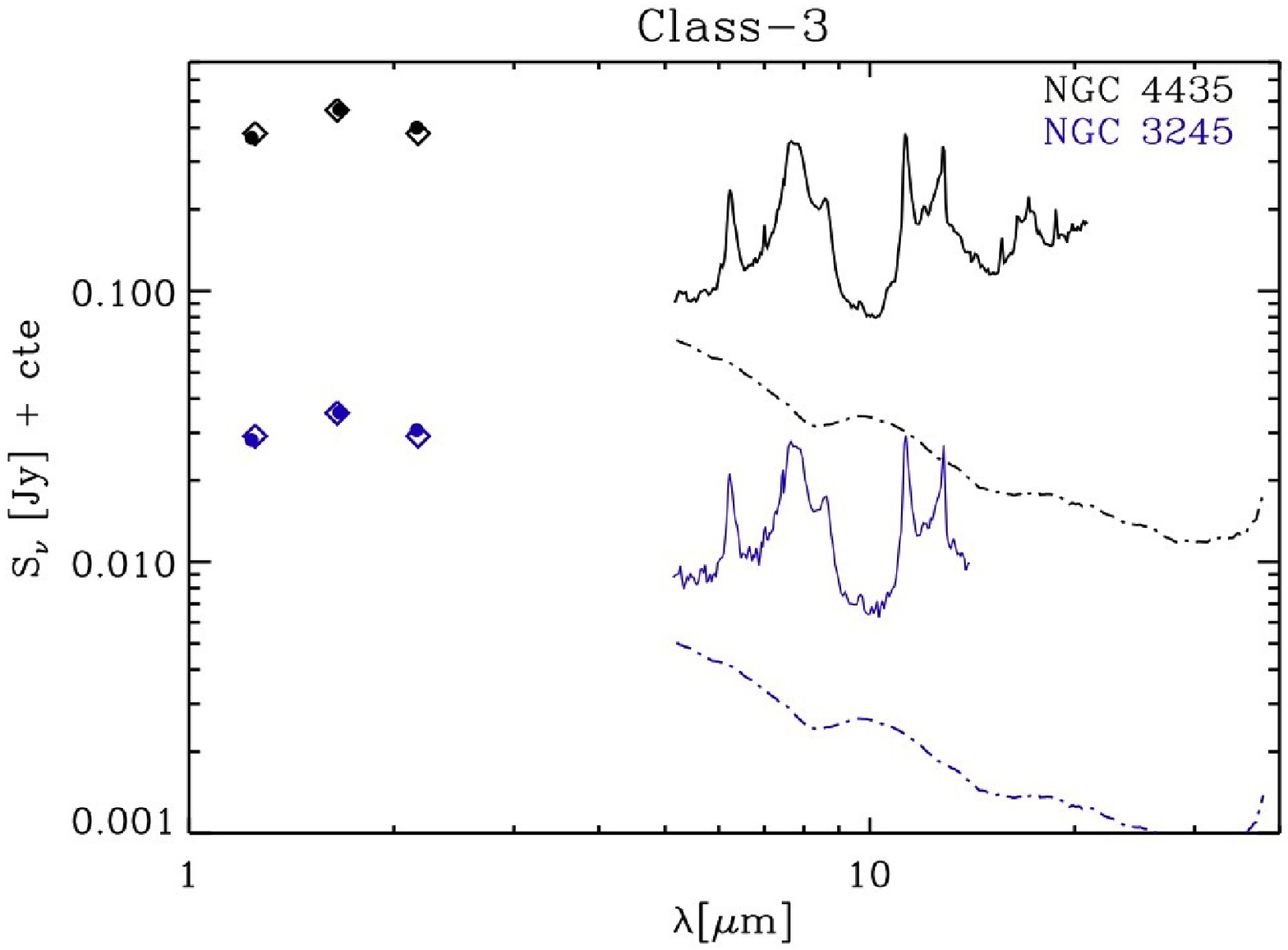}
\includegraphics[width=8cm]{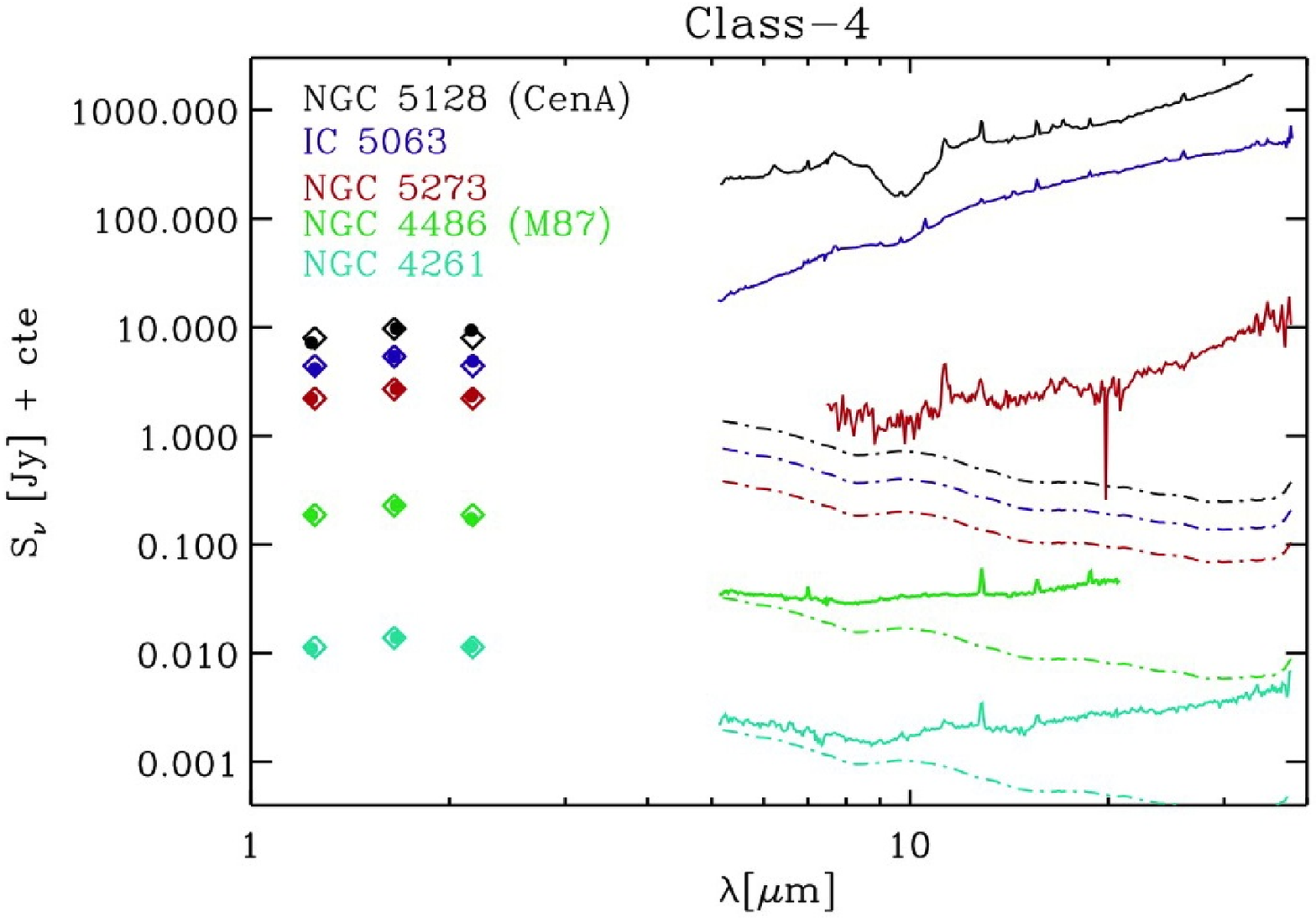}
\caption{Comparison of the passive template used in this work (dotted-dashed line) 
and the MIR spectra (solid line) of different galaxies with different degrees of MIR 
activity. The template has been normalized to the H-band flux of each galaxy. The
2MASS J, H, K-band fluxes of the galaxies, within the
central 5\arcsec\ radius, are indicated with filled circles, while the corresponding normalized 
values of the template are plotted as open diamonds.
} 
\label{fig2}
\end{figure*}

\begin{figure}
\centering
\includegraphics[width=8.6cm]{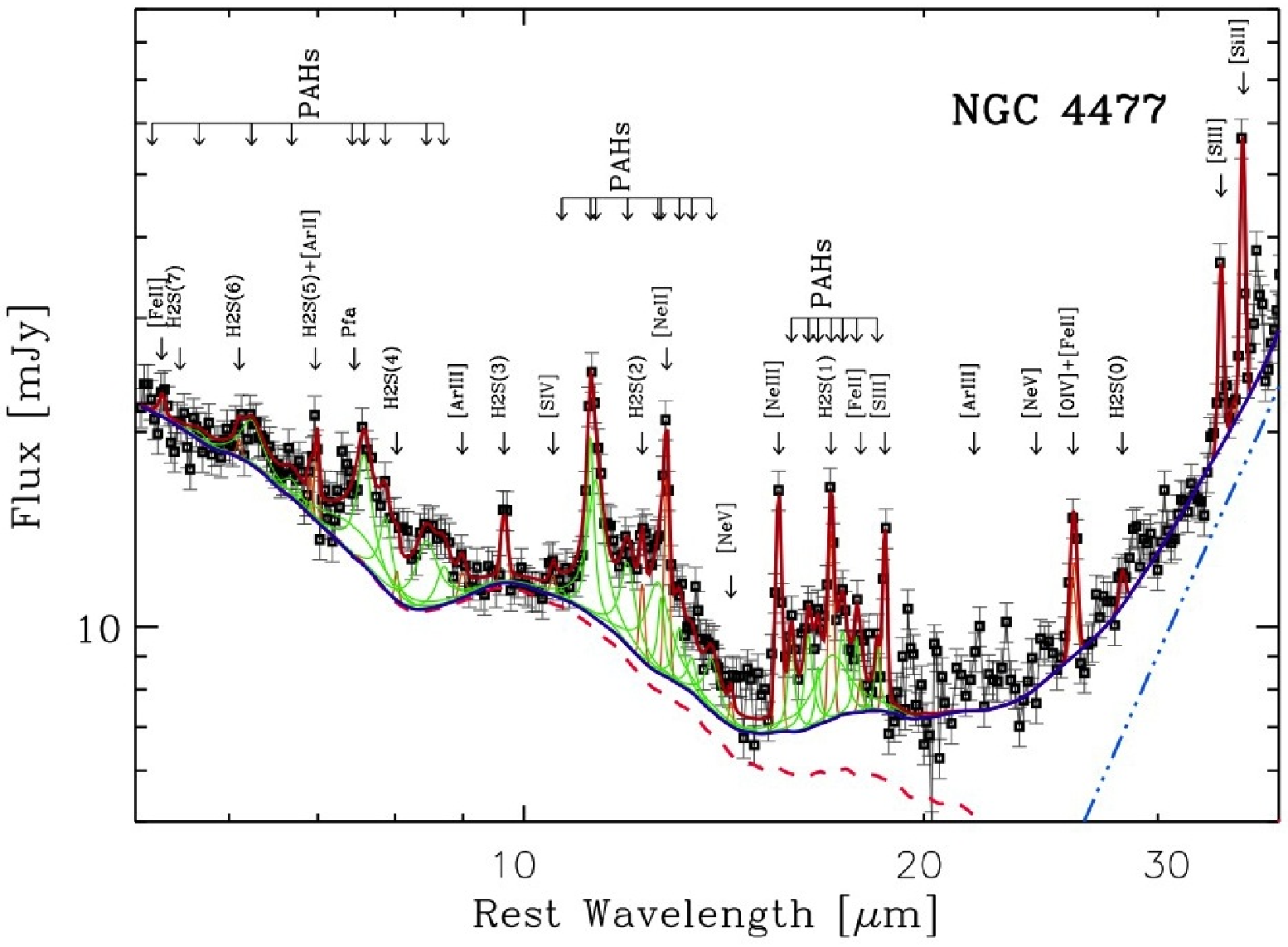}
\includegraphics[width=8.6cm]{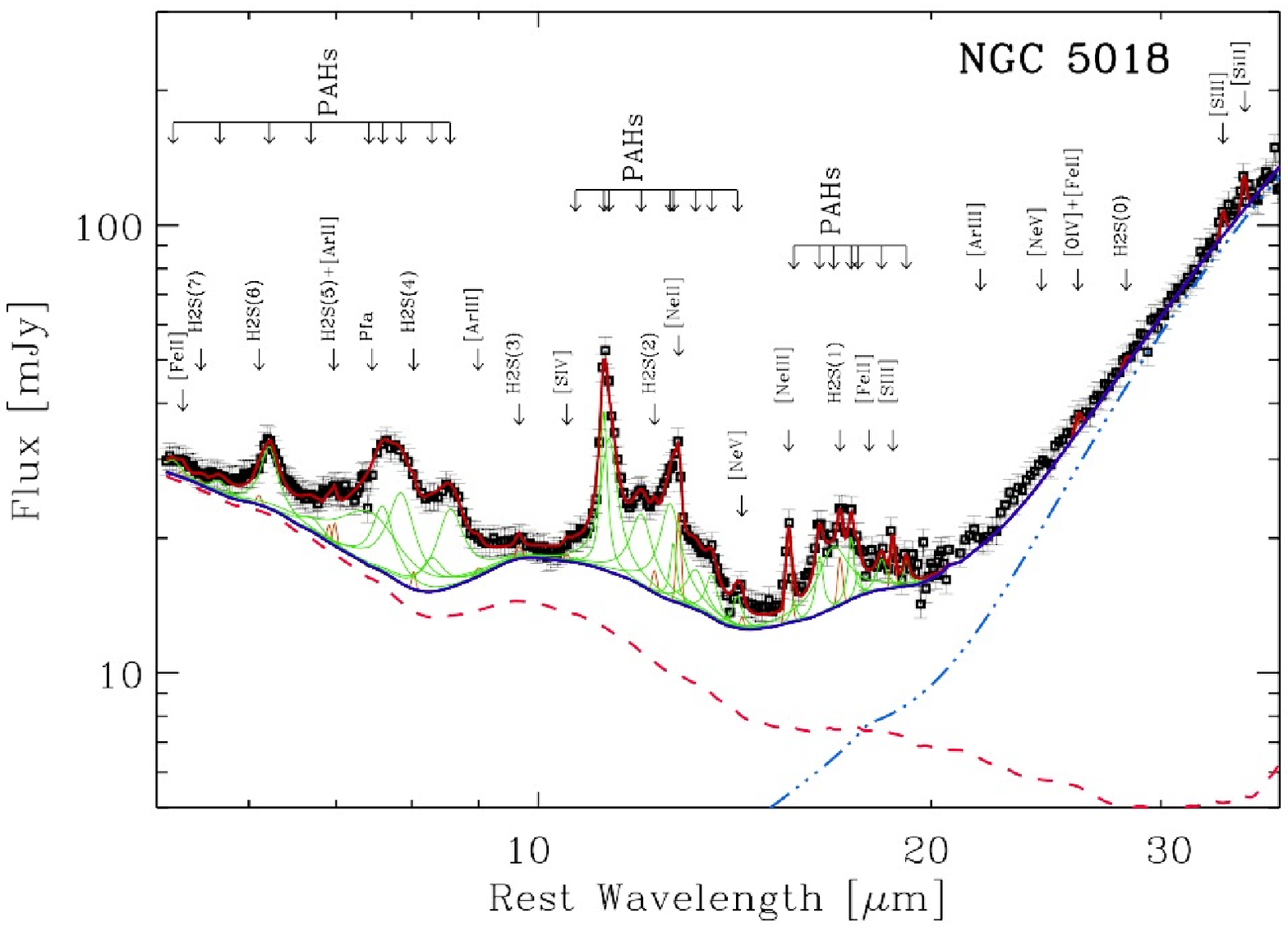}
\includegraphics[width=8.6cm]{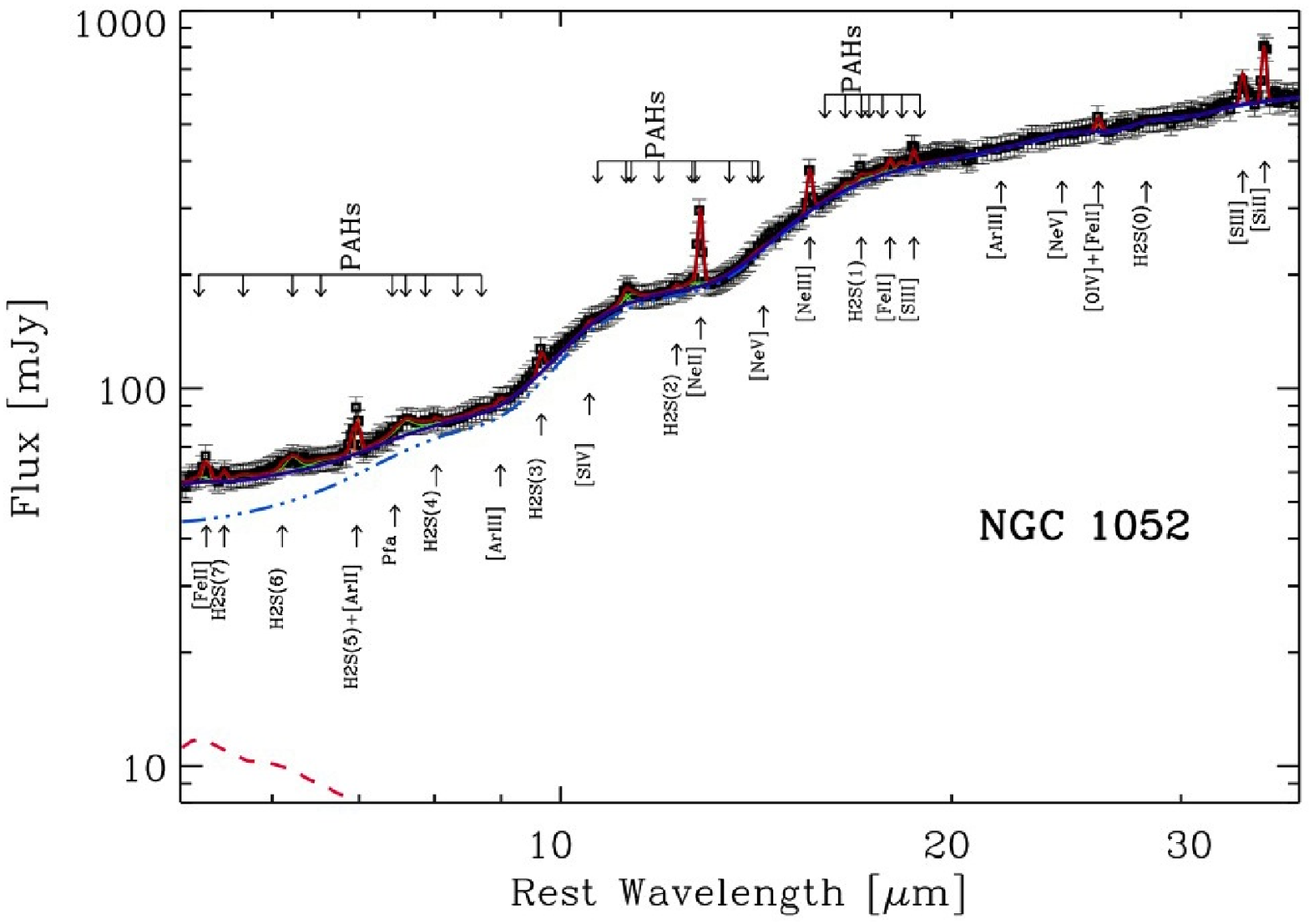}
\caption{Best fit of the spectrum of three ETGs with 
anomalous PAH emissions (upper panel), normal PAH emissions (middle panel), 
and steep MIR continuum (bottom panel). Open squares and the solid
thick red line are the observed MIR spectra and our final best
fit, respectively. The fit is calculated as the sum of an underlying
continuum (solid thick blue line), the PAH features (solid thin
green line) and the emission lines (solid thin orange line). The
two components of the continuum, old stellar population (dashed
red line) and diffuse dust emission (dot-dashed blue line), are also plotted.} 
\label{fig3}
\end{figure}

\section{The sample}
\label{sample}

We cross match the RSA catalogue with the {\it Spitzer} Heritage Archive, 
including IRS spectra either with just SL segments 
or both SL and LL segments, as our priority is in characterizing spectral features 
and obtaining the MIR classification of the spectra as in \citet{Panuzzo11}. 
In Tables~\ref{tab1} and ~\ref{tab2} we list the resulting sample of 91 ETGs and
in Tables~\ref{tab3} and \ref{tab4} the basic observational parameters. Although by no 
means complete, this atlas doubles the \citet{Panuzzo11} sample, the largest 
study to date. 

ETGs are divided according to their RSA morphological classification (column 2)
into Es and S0s. In the S0 family we include both {\it bona fide} S0s,
in their normal and barred forms, and the ``mixed'' sub-classes E/S0 and 
S0/E. Tables~\ref{tab1} and ~\ref{tab2} also report the morphological type code, $T$ 
(column 3),  provided in {\tt HYPERCAT}. 
The sample of Es is composed of 56 galaxies: different sub-classes, from
 E0 to E6, are all populated. The sample of S0s consists of 35 galaxies of several
sub-types, including mixed cases i.e. E/S0, S0/Sa or SB0/Sa. In all cases
the morphological type code, $T$, is $\leq 0$, as expected for ETGs. 
Most of the E galaxies have $T\leq -3$. Remarkable exceptions are NGC~1297 (T=$-2.5\pm$0.9)
 and NGC~1275 ($T=-2.2\pm$1.7) classified as E-S0 and S0, respectively,
 in {\tt HYPERCAT}. S0s in RSA span all the $T<0$ range, consistently
  with their mixed nature. About 1/3 of E/S0 + S0 (see Table\ref{tab2}) 
  are considered  truly Es ($T\leq -3$) in {\tt HYPERCAT}. 

The redshift independent distances, D, (column 4) and the absolute 
K-mag, M$_K$, (column 6) are from the Extragalactic Distance Database 
 \citep{Tully2009} and the effective radius, $r_e$ (column 7), from RC3 \citep[][]{RC3}. 
For galaxies, labeled with an asterisk,  we use a Hubble  constant of 
73 km~s$^{-1}$~Mpc$^{-1}$ and the heliocentric velocity, V$_{hel}$,  from {\tt NED}.
Most of the galaxies in the sample have a heliocentric systemic velocity lower
than 3500 km~s$^{-1}$. Only NGC 1275, NGC 1453, NGC 1700, NGC 7619 
and IC~4296 exceed this limit. In the global sample, distances of the 
galaxies are less than 72 Mpc. For each galaxy we report the
cluster/group association to which the galaxy should be gravitationally bound (see 
column 5 and the note of Table 1).
These data are provided in the T3000 catalogue \citep[][]{Tully2009}.
Es and S0s are located in different environments. 
Our sample includes 31 cluster members, 21 located in the denser regions
of Virgo (T88 group 11 -1 1, 11 +2 +1) and  8 of Fornax (T88 group 51 -1 1) clusters.  
 According to \citet{Tully88}, the environmental density associated with
 cluster members  is $\rho \geq 1.33$ (gal~Mpc$^{-3}$). 
 Two additional galaxies, NGC~1275 (Perseus cluster),  NGC~7619 
 (Pegasus~I cluster), too distant to be found in Tully's Catalogue, 
 have been included in the present cluster sample.
 The remaining 60 ETGs are located in LDEs and have  
 $\rho \leq 0.97$  (gal~Mpc$^{-3}$). 

Our sample of Es and S0s presents fundamental characteristics
noticed in different samples. In Tables 1 and 2 we report the 
M$_{K_T}$ (column 6), a proxy of the stellar mass, and the
central velocity dispersion (column 8), $\sigma_{c}$, (from {\tt HYPERCAT}) 
 a proxy of the total galaxy mass \citep[][]{Clemens06,Clemens09}. 
We applied a Mann-Whitney U-Test \citep[][]{Wall77} to  both the
 M$_{K_T}$ and $\sigma_c$  distributions of Es and of S0s 
 to compare the properties of the two samples. We verified  the null hypothesis 
that the two  samples result from the same parent population
can be rejected at the 99\% confidence level.
The median of M$_{K_T}$ and of $\sigma_c$ distributions for  Es is
larger than those of S0s, suggesting that Es are less dominated by 
Dark Matter than S0s of similar total mass. \citet[][their Figure~1]{Shankar04}
noticed that the different dominance of Dark Matter in Es with respect to S0s 
suggest that the two morphological families had different evolution 
mechanisms. 

\section{Observations and data reduction}
\label{data-reduction}

The details of the {\it Spitzer}-IRS observations for each galaxy are provided 
in Tables~\ref{tab3} and \ref{tab4} for Es and S0s, respectively.  
In column~8 we provide the slit aperture
in kpc. In column~9, the ratio between the area covered by the slit
and the circular aperture of radius $r_{e}$/8 is given. 
Although we used a fixed aperture for the extraction of the spectra,
the slit covers central portions of the ETGs, with a
size, on the average, of about 2-3 $\times$ $\pi (r_{e}/8)^2$ 
(1.9$\pm$1.8 for Es and 2.9$\pm$2.8 for S0s). 
We refer  to this portion as the ``nuclear'' part of the galaxy throughout 
the paper.

Observations were performed in Standard Staring mode with low resolution
($R\sim$ 64--128) modules SL1 (7.4--14.5$\mu$m), SL2 (5--8.7$\mu$m), LL2
(14.1--21.3$\mu$m) and LL1 (19.5--38$\mu$m). Observations do not
include, in general, all IRS modules.

The data reduction procedure is fully described in \citet{Panuzzo11}.
Briefly  in the following we recall the main steps of the reduction.
After the removal of bad pixels from the co-added images,
the sky background was removed by subtracting co-added images
taken with the source placed in different orders at the same nod
position. Offset exposures, if available,
were used to measure the sky background without contamination from the
target galaxy itself. For those galaxies for which only the LL2 module was
used, the background subtraction was done by subtracting co--added
images taken with the source at different nod positions. 

In order to derive calibrated spectral energy distributions, we 
taken into account that the galaxies are extended, compared
to the IRS point spread function (PSF). Since the IRS spectra are
calibrated on point-sources, we have devised an {\it ad hoc} procedure
to correct for the effects of the variation with wavelength of the
IRS PSF.  We simulated the effect of convolving an intrinsic surface
brightness profile (a modified King law, \citet{Elson87}) with the
instrumental PSF, and of the extraction in a fixed width of  3.6\arcsec. 
This provides us a correction factor to be applied to the extracted
spectrum as a function of the wavelength and the profile parameters. 
By fitting the observed profiles with the simulated ones, we can
reconstruct the intrinsic profiles and correct the extracted spectra to
correspond to the intrinsic SED. This
procedure also has the advantage of determining whether a particular
feature is spatially extended or not.

This procedure was applied to SL modules but not to LL modules where the
PSF is too large to determine the intrinsic surface brightness profile.
For this reason, the spectra in the LL range are extracted as if the
sources were point-like and then manually scaled to match the SL segment.
This is equivalent in assuming that the light distribution at LL
wavelengths is the same as that at the end of the SL segment.
Note that for those sources that are fully dominated by the nucleus (i.e.
AGN-dominated like NGC 1275) there is no need to rescale the LL spectra.

The rest frame, flux calibrated, IRS spectra of the sample galaxies are 
shown in Figure~\ref{fig1}. 

\section{Spectral properties of the atlas}
\label{analysis}

  The present atlas aims to build, from homogeneously reduced
  Spitzer-IRS spectra, a data-set of emission lines and
  PAH intensities as well as to divide  nearby ETG nuclei 
  into MIR physical classes as in \citet{Panuzzo11}.

\subsection{Analysis of MIR spectra}

To analyze spectra we adopted a model of the MIR emission that takes into account 
the following components: 1) the underlying stellar continuum due to the
old population characteristic of a passively evolving ETG
includes the contribution of the dusty AGB stars which dominate the
emission at wavelengths $\leq$ 6 $\mu$m \citep{Bressan06}; 2) a
featureless thermal continuum representing a putative dust
contribution at longer wavelengths; 3) the emission from molecular and
atomic lines and 4) the PAH features. The best fit of each spectrum is obtained  
using the Levenberg-Marquardt algorithm. 

The model is fully described in \citet{Vega10} and \citet{Panuzzo11} and 
 is similar to that used by \citet{Smith07}. The main difference
 consists in the selection of the underlying stellar continuum component.
The stellar component is described by \citet{Smith07} as the emission of a blackbody at 
5000~K. This approximation, suitable for the analysis of star-forming galaxies, 
where the hot dust and PAH features dominate the MIR emission,  is not adequate
for ETGs. In these galaxies, the underlying stellar component usually 
dominates the MIR continuum at shorter wavelengths, and the MIR 
emission is characterized by a dip at 8 $\mu$m, likely due to photospheric SiO absorption
bands \citep{Verhoelst09}, and a bump at $\sim$10 $\mu$m from the silicate emission
 from the  dusty circumstellar envelopes of O-rich AGB stars (e.g. Bressan et al. 1998, 
 Bressan et al. 2006). We therefore adopt  a semi-empirical, high S/N template derived
from the analysis of passive ETGs in \citet{Panuzzo11} to describe the stellar
continuum. The template is built by averaging the NIR (J-H-K
2MASS) data, within the central $5"$ radius, and the 5--40 $\mu$m 
{\it Spitzer}-IRS spectra of three passively evolving ETGs,
namely NGC~1389, NGC~1426, and NGC~3818.
The subtraction of the underlying stellar continuum is performed by 
assuming that the NIR fluxes are completely due to the stellar component.
Thus, we normalize our stellar continuum template to the
observed flux in the H-band, and calculate the contribution
of the stellar continuum to the MIR spectra. 
Figure~\ref{fig2}  provides some examples of the MIR continuum 
in different type of spectra and Figure~\ref{fig3} shows some 
examples of the fitting procedure ( see also next section).

Measurements of PAH features and of the nebular and molecular emission lines 
are collected in Tables~\ref{tabA1} and \ref{tabA2}, respectively. 

\subsection{MIR spectral classes}

\citet{Panuzzo11} subdivided low resolution {\it Spitzer}-IRS spectra into 
five classes, from class-0 to class-4.  Class-0 defines passively evolving
nuclei. Classes from 1 to 4 aim to categorize spectra 
related to different powering mechanism active in the nuclear regions. 
The classification scheme is based on  
(1) the detection of emission (atomic and/or molecular) 
lines and PAH features, (2) the value of the PAH inter-band ratios, (3) the presence 
of an excess in the  MIR continuum over the underlying old stellar population. 

We briefly resume below the  classification criteria of \citet{Panuzzo11}.
Class-0 spectra show neither ionic, molecular nor PAH emission lines
superimposed to the photospheric stellar continuum generated by red giant stars.
Only the silicate features at $\approx$10 $\mu$m \citep[][]{Bressan06} 
and at 18 $\mu$m \citep{Panuzzo11},  arising from the circum-stellar 
dust of O-rich AGB stars  \citep{Bressan98}, are present in these spectra.

Class-1 spectra show either nebular or molecular emission lines but no
PAHs. Ionic, in particular [Ne II]12.8$\mu$m, [Ne III] 15.5$\mu$m, 
[SIII] 18.7$\mu$m, 33.5$\mu$m and molecular H$_2$ 0-0 rotational 
emission lines (S(n)) are detected. The continuum of this class is
similar to that of passive class-0 ETGs, at least up to 25 $\mu$m, showing, in
some cases, dust emission at longer wavelengths. 

All other classes show PAH complexes plus gas, ionic
and molecular, emission lines. These latter include forbidden nebular emission
lines of several elements like Ar, Fe, N, O, S and Si.
PAH emission features are typically detected at 6.2, 7.7, 8.6, 11.3, 12.7 and 17
$\mu$m \citep[][]{Bregman06,Bressan06,Kaneda05,Panuzzo07,Kaneda08,Panuzzo11}.

Classes 2 and 3 are distinguished via their different PAH inter-band ratios. 
Class-2 spectra have an anomalous PAH inter-band ratio, 
7.7$\mu$m/11.3$\mu$m $\leq$2.3. We set this value as the lower limit
for the class-3  since, in their figure 14, \citet[][]{Smith07} show that some
HII dominated sources and star-forming galaxies may reach this low PAH inter-band ratio. 
Class-3 spectra are dominated by the 7.7$\mu$m, 11.3$\mu$m,
12.7$\mu$m and 17$\mu$m PAH complexes and have normal PAH inter-band ratios
typical of star forming galaxies  \citep[][]{Smith07}.

Finally, spectra dominated by a hot dust continuum are collected in the 
MIR class-4 (see discussion below). Emission lines with high ionization like 
[S IV] and [Ne V] and PAH features may sometimes be visible.

Figure \ref{fig2}  illustrates the variation of the continuum in the different MIR 
classes. The presence of emission lines, starts from class-1, and the 
importance of PAH features increases from class 2 to 3. Detailed fits of line and PAH
emission features are shown in Figure~\ref{fig3} for NGC-4477 (class-2, top panel), 
NGC~5018 (class-3, mid panel) and  NGC~1052 (class-4, bottom panel).

\begin{figure*}
\centering
\includegraphics[width=14cm]{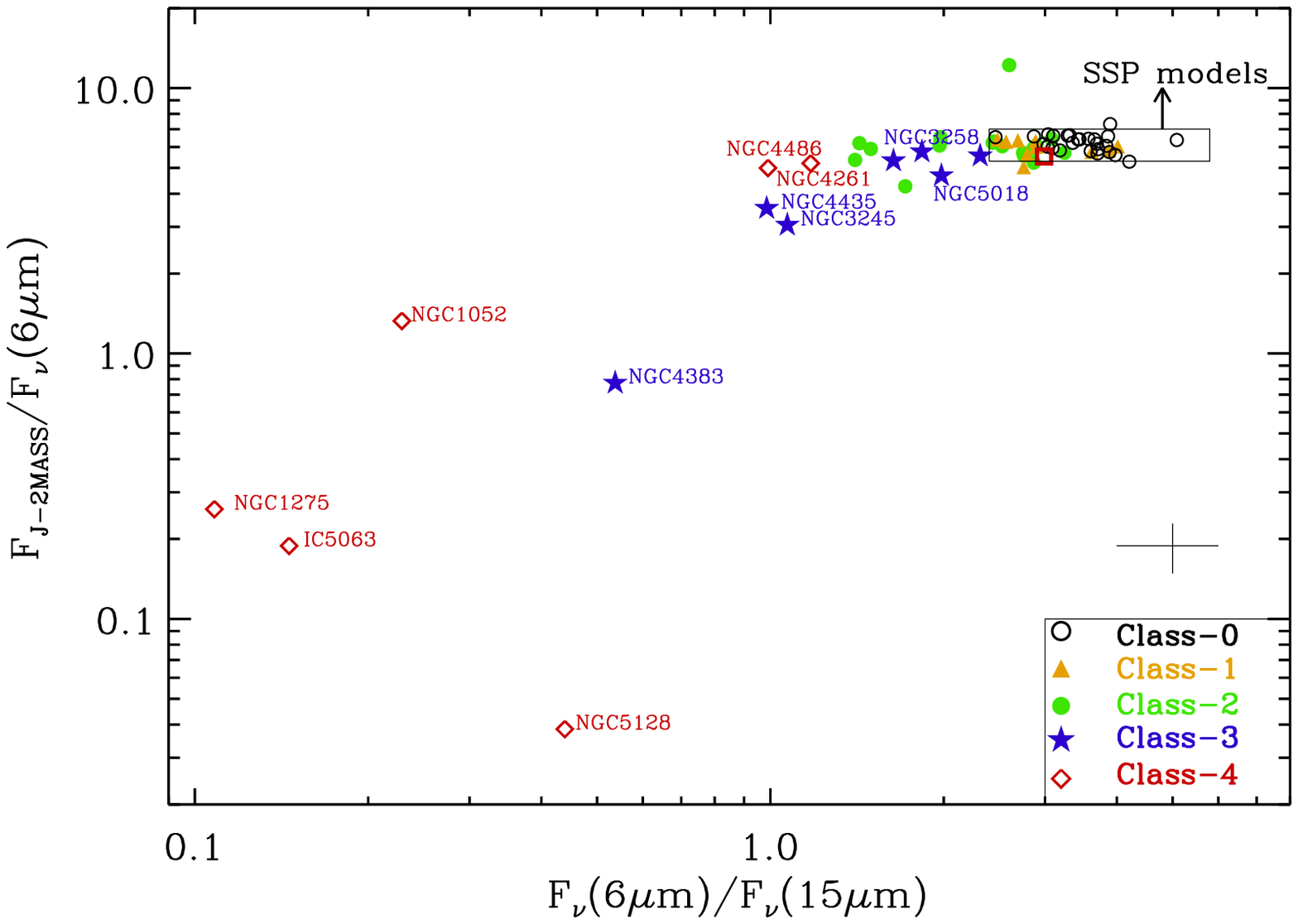}
\includegraphics[width=14cm]{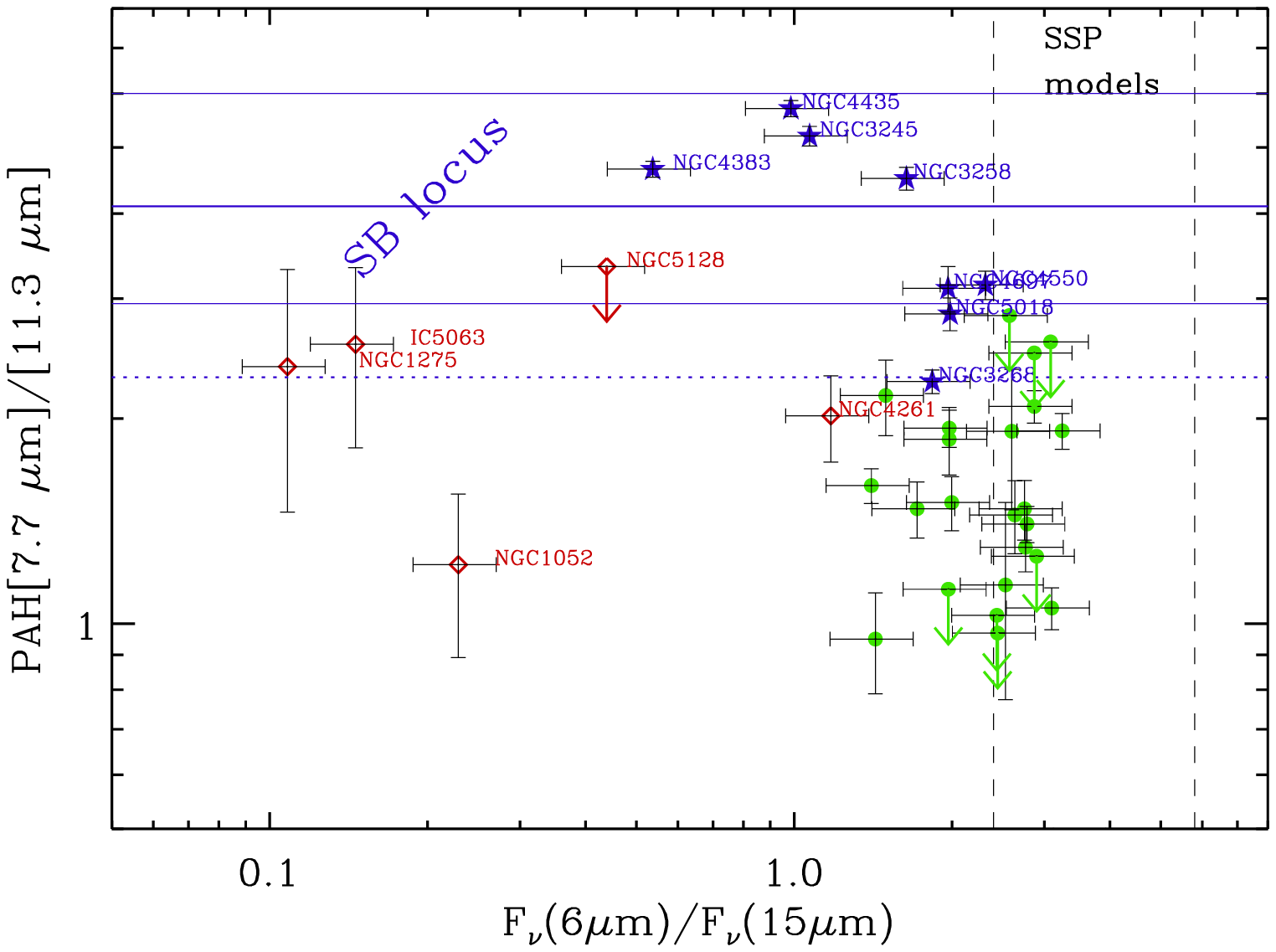}
\caption{({\it Top panel}) Color-color plot of the 2MASS J-band/6$\mu$m  vs. 
the 6$\mu$m/15$\mu$m flux ratios. The rectangle encloses Single Stellar Population 
(SSP) models of ETGs  with ages in the interval 10--14.5 Gyr and metallicty, $Z$, 
in the range 0.004--0.05. The red open square within the rectangle 
represents a 12 Gyr old SSP model with solar metallicity. The typical error is also shown.
({\it Bottom panel}) The 7.7$\mu$m/11.3$\mu$m PAH ratio vs.  the 6$\mu$m/15$\mu$m 
flux ratio. The dotted horizontal line correspond to cirrus emission
\citep{Lu03}. Solid horizontal lines mark the region of the HII dominated sources or  
starburst galaxies (SB), the central line is the median 4.2 value \citep{Smith07}. 
The dashed vertical lines mark the area of old SSP models described above.}
\label{fig4}
\end{figure*}

The contribution of the old stellar population to the MIR continuum 
decreases from class 0 to class 3, becoming almost negligible in 
class-4. Figure~\ref{fig4} shows ETGs in the color-color 
2MASS J-band/6$\mu$m vs. 6$\mu$m/15$\mu$m plane,
which considers the MIR excess over the underlying old
stellar population due to the hot dust. ETG nuclei of classes 3 and 4 are clearly
separated  from more  quiescent ones (classes 0 and 1) and partly 
from  class-2 nuclei. More quiescent ETGs are 
all located within a region (the rectangle in the top panel of Figure~\ref{fig4}) 
delimited by old Single Stellar Population (SSP) models 
(age: 10 -- 14.5 Gyr and metallicity 0.004$\leq$ Z $\leq$0.05  
i.e. $\approx$1/4 - 3 Z$_\odot$ \citep[][]{Bressan98}).
The 6$\mu$m/15$\mu$m value of NGC~4261 sets our empirical upper limit
of class-4 ETGs. This value of the MIR excess is comparable,
 within errors, to NGC 4486 for which \citet{Buson2009} show that 
the MIR continuum is just the superposition of a passively
evolving spectrum and a synchrotron emission from the central AGN. 

The bottom panel of Figure~\ref{fig4} shows the discrimination
offered by 7.7$\mu$m/11.3$\mu$m PAH ratio in disentangling 
active MIR classes 2, 3 and 4.  All class-2 nuclei have anomalous
7.7$\mu$m/11.3$\mu$m PAH ratios and  6$\mu$m/15$\mu$m values 
 higher than NGC~4261. All, but one, class-3 ETGs are located in the typical 
 star burst (SB) region (solid horizontal lines) \citep{Smith07}. NGC~3268 is borderline 
between class-3 and class-2 since its 7.7$\mu$m/11.3$\mu$m PAH ratio = 2.3. 
NGC~5128 has both a high 6$\mu$m/15$\mu$m excess
and a 7.7$\mu$m/11.3$\mu$m PAH ratio typical of star forming galaxies
(see next section). 

The result of our MIR spectral classification is collected in Table~5. 
 In Figure~\ref{fig5} we provide a synoptic view  of  all ETG nuclei
 in classes 3 and 4.

\subsection{Individual notes about MIR classification}

\begin{figure*}
\centering
\includegraphics[width=12cm]{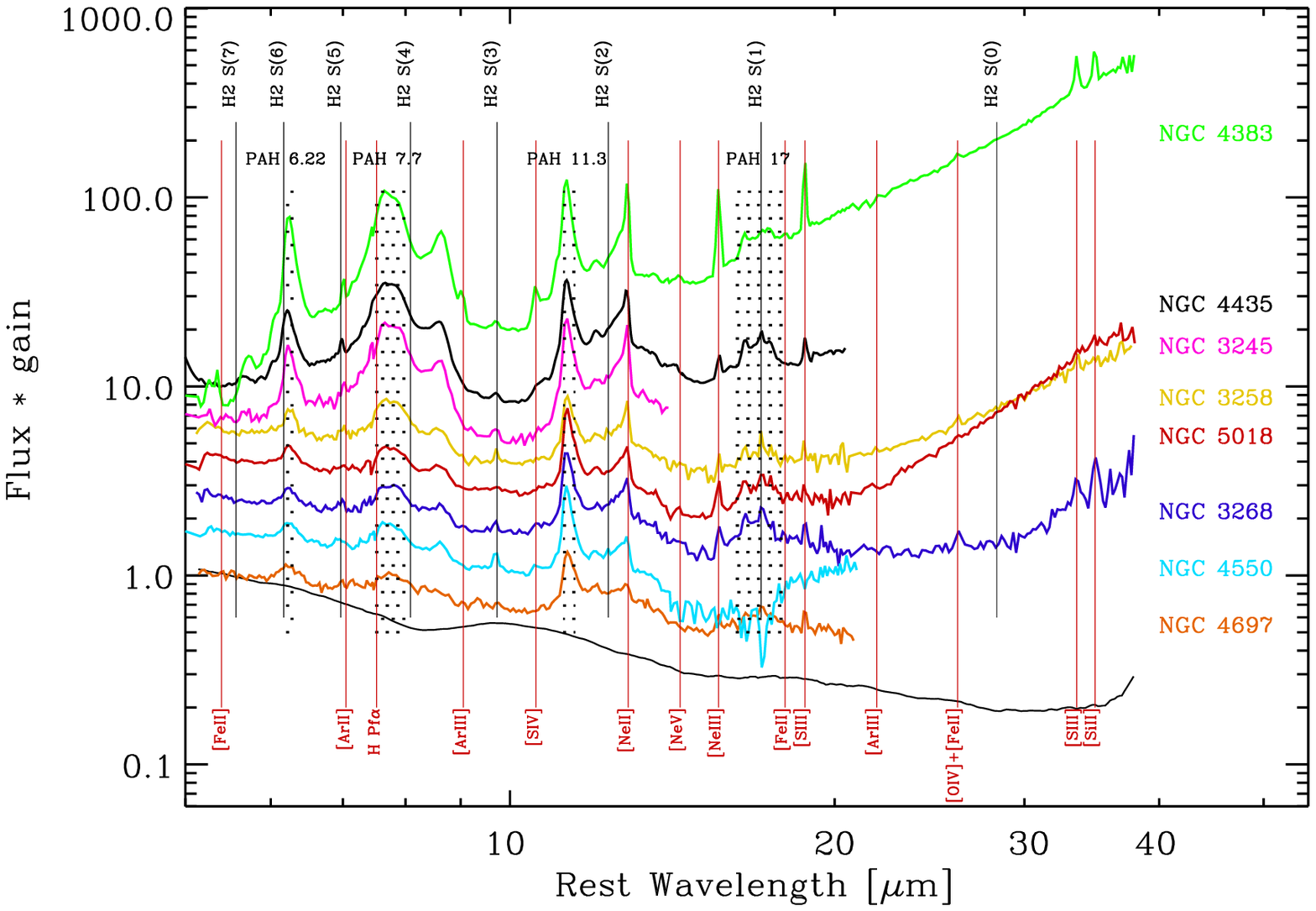}
\includegraphics[width=12cm]{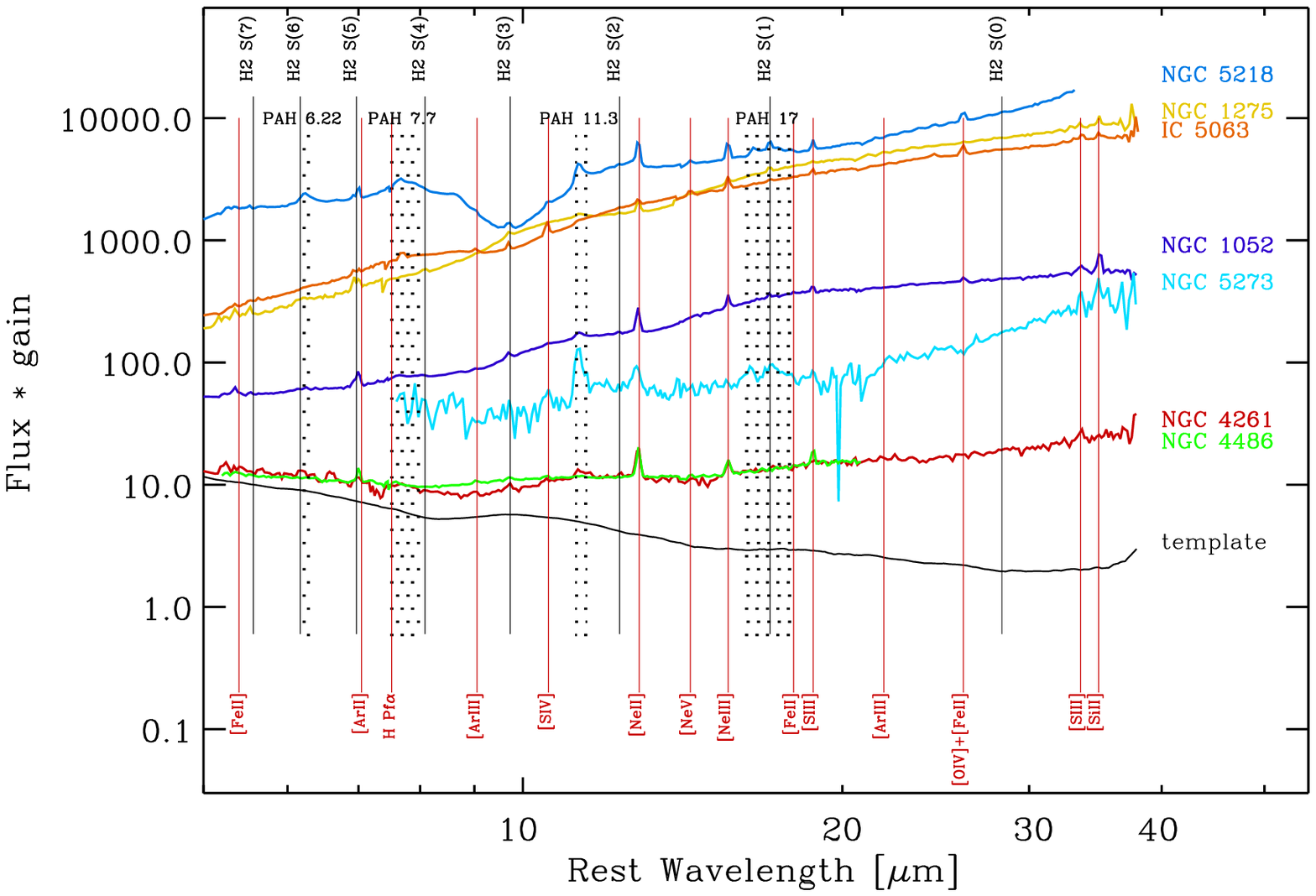}
\caption{Examples of ETGs with MIR class-3 (top panel) and class-4 (bottom panel)
spectra. Fluxes have been normalized to the template and then arbitrarily scaled. 
PAH features (dotted area) typical of star forming galaxies together with ionic 
and/or molecular H$_2$ emission lines (solid lines), are detected in these spectra.
The template passive spectrum is also shown.} 
\label{fig5}
\end{figure*}

\medskip
\underbar{NGC~4261} (E3) ~~~ The galaxy (see Figure~\ref{fig4}) is borderline 
between class-4 and class-2. The galaxy displays a MIR spectrum  similar to that of
 NGC 4486 (see Figure~\ref{fig5}). 
 At odds with NGC~4486, the spectrum of this galaxy shows the high 
 ionization [NeV] 14.32 $\mu$m emission line and likely a dusty torus is needed 
 to fully account for the observed continuum, in addition to a synchrotron component.  

\medskip
\underbar{NGC~4383} (S0:) ~~~ This galaxy was included in MIR class-3 because of 
its PAH inter-band ratio (PAH$_{7.7 \mu m}/$PAH$_{11.3 \mu m}= 4.6$). 
However, it presents the steepest and strongest MIR continuum of any ETG
in class-3. The analysis of the spectrum shows that the slope of
 the MIR continuum, calculated as F$_\nu (30\mu m)/$F$_\nu (15\mu m)= 6.7$, 
 is  consistent with the values found in starburst galaxies \citep[see e.g.][]{Brandl06}. 
Furthermore, the [NeIII] 15.5 $\mu$m /[NeII] 12.8 $\mu$m vs. [SIII]
 33.5 $\mu$m/ [SiII] 34.8 $\mu$m diagnostic diagram \citep[see][]{Dale06} 
 locates the galaxy in the  starburst region. All of these indicate that the class-3
classification points to the presence of on-going star formation in the nucleus.
 
 The lower values of the MIR slopes shown by the remaining class-3 objects 
 in Figure~\ref{fig5} (2.3 for NGC 3258, 1.2 for NGC 3268, 4.2 for NGC 5018 
 and 4.5  for NGC 4435 \citep{Panuzzo07}) may indicate that the starburst phase 
 in these objects is fading and there are not as many ionizing stars to heat the dust 
 as in the peak of the starburst phase. This has been discussed in some detail by 
 \citet{Panuzzo07} for the case of NGC 4435.
 
\medskip
 \underbar{NGC~5128 Cen A} (S0+S pec) ~~~ The most striking feature in 
 its IRS spectrum is the strong silicate  absorption feature at 9.7 $\mu$m. 
 This feature is weak or seen in emission in the other class-4 
 galaxies of   the sample. \citet{Sargsyan11} found this feature typical of 
 the so called ``absorption AGN'' while ``emission AGN'' have spectra similar 
 to our MIR class-4 spectra. Such a strong absorption is an indication of an 
 enormous amount of extinction   towards the nucleus, which is probably 
 embedded in a very dusty and compact torus  \citep[see e.g.][]{Armus07}. 

\medskip
\underbar{NGC~5273}(S0/a) ~~~ The  galaxy  was included in class-4 because of its 
large  MIR excess over the underlying stellar population continuum. Unfortunately 
the spectrum does not include the SL2 module (see Table~4).

\section{Demography of MIR spectral properties of ETGs}
\label{demography}

In the following sub-sections we investigate the MIR spectra and their classes 
as a function of the properties of the host galaxy in a multi-wavelength context. 
Adopting a Poisson statistics, for each percentage we obtained single-sided upper and lower limits, 
corresponding to 1$\sigma$  Gaussian  errors \citep{Gehrels86}.

\subsection{MIR classes vs. morphological classification and galaxy environment}

We summarize the data concerning MIR classes versus E and S0 morphological 
classes and the galaxy environment in Table~\ref{tab6}.

Es are significantly more passive than S0s: 46$^{+11}_{-10}$\% of Es 
and 20$^{+11}_{-7}$\% of S0s have a class-0 spectrum. 

S0s show the tendency  to be more gas rich than Es: 
80$^{+18}_{-15}$\% of S0s and  54$^{+11}_{-10}$\% of Es 
belong to classes 1, 2, 3 and 4 showing emission 
lines. Considering the whole sample of ETGs, 64$^{+12}_{-6}$\% 
of the nuclei  show emission lines, although with different intensity.
This is in agreement with optical studies, in which, 
depending on the sample (sometimes strongly biased against 
passively evolving objects) the ionized gas is detected in  $\approx$50-90\% of ETGs \citep{ph86,mac96,Sarzi06,Sarzi10,yan06,serra08,Annibali10}. 

41$^{+10}_{-9}$\% of Es and 57$^{+16}_{-13}$\% of S0s show 
PAH emission, indicating that in about half (47$^{+8}_{-7}$\%) of ETGs 
a star formation episode has occurred about 10$^8$ years ago \citep[][]{Kaneda08,Panuzzo11}.
Galaxies with normal PAH ratios, class-3 spectra, are a minority both
amongst Es and S0s: only 9$^{+4}_{-3}$\% show star forming spectra.

\citet{Panuzzo11} considered the ratio between the class-2 and class-3 spectra
in the hypothesis that anomalous PAHs in class-2 are produced by carbon stars
\citep[][]{Vega10}. For solar metallicity, carbon stars are present in stellar populations
with ages between $\sim 250$ Myr and $\sim 1.3$ Gyr \citep{Marigo07}.
Assuming a characteristic life-time of 200 Myr for a star formation episode 
\citep[][]{Panuzzo07} the expected ratio between ETGs with
PAH$_{anomalous}$/PAH$_{normal}$ is between 1 and 7. Here we obtain a
 ratio of 3.5, half of that reported in \citet{Panuzzo11}.

All MIR class-4 spectra show PAHs, suggesting that the AGN
phenomenon is associated with star formation. 
Class-4 MIR spectra (Figure~\ref{fig5} bottom panel) include 3 E and 4 S0 galaxies, 
corresponding to 8$^{+4}_{-3}$\% of our ETG sample.  

The fraction of Es with PAH features (classes 2, 3 and 4)   
in clusters tends to be lower than that in LDEs 
(20$^{+19}_{-11}$\% vs. 49$^{+13}_{-11}$\%),
suggesting that  star formation episodes are triggered in LDEs.
The PAH features of S0s show a  similar percentage in clusters and 
in LDEs (44$^{+23}_{-16}$\% vs. 68$^{+25}_{-19}$\%).
These results are consistent with other indicators 
that find ETGs located in high density environments 
older \citep[][]{Clemens06,Clemens09} and, in general, less ``active'' 
\citep[][to mention theoretical approaches]{Moore96,Feldmann2011} 
than their counterparts in groups and in the field.  

\subsection{MIR classes and the hot nuclear gas}

In the top panel of Figure~\ref{fig6}, we plot the nuclear X-ray luminosity, $L_{X, nuc}$
versus the MIR classes. The $L_{X, nuc}$ (Table B1, columns 4 and 9, Pellegrini 2010)  
refers to a central aperture ($\sim$2\arcsec\ radius)  in 
the 2-10 keV bands of {\it Chandra} observations (Pellegrini, private communication).
 The $L_{X, nuc}$ aperture lies within the {\it Spitzer}-IRS slit width and allows us  
 to investigate the presence of the AGN engine 
 and the role of the X-ray radiation field on the MIR spectra.
 
ETGs in class-4  have the higher values of  $L_{X, nuc}$,
as expected from the presence of an AGN.
The analysis of X-ray spectra in  a sub-sample of
our ETGs, obtained with 
 {\it Chandra} and XMM-{\it Newton} data, show that a power law 
component is needed to account for the 
observed emission of all ETGs in class-4
 \citep[][]{Gonzalez09,Boroson11,Machacek04,Capetti06,Gru07,Ghosh05}. 

ETGs in class-0  show low values of  $L_{X, nuc}$ 
($\sim 10^{38} - 10^{39}$ erg~s$^{-1}$, see Figure~\ref{fig6}), with some
exceptions.  The X-ray spectral analysis of the class-0 NGC 720, NGC 821, 
NGC 1399 need an AGN contribution at odds with NGC 1427, NGC 3379,  NGC 3377, 
NGC 3608, NGC 4473. \citet{Pellegrini10} remarks indeed the large spread of 
$L_{X, nuc}$ and suggests that it could be produced by the nuclear activity 
cycle, where $L_{X, nuc}$ is regulated by the joint actions of the feedback and the
fuel availability.

High values of $L_{X, nuc}$ are also found in ETGs of classes 1 and 2  and 
partly of class-3, likely reflecting the contribution of an AGN. In particular,
an AGN contribution has been detected in the X-ray spectra of  NGC 1404,
NGC 4649   and NGC 5846 of class-1;  NGC 1553,  NGC 2685, NGC 4036, 
NGC 4374, NGC 4552, IC 1459, NGC 5090 of class-2;  NGC 3245,  NGC 4435, 
NGC 4697, NGC 5018 of class-3.  No AGN contribution is required 
neither to model the X-ray spectra of  NGC 584 (class-1) nor  
NGC 4589 and NGC 4696 (class-2).

X-ray cavities are considered signatures of the AGN feedback 
\citep[][and references therein]{Cavagnolo10}. {\it Chandra} observations detected
X-ray cavities in many of our ETGs  (NGC 3608  in class-0;  NGC 4472, 
NGC 4649, NGC 5813, NGC 5846 in class-1; NGC 1553, NGC 4552, NGC 4636, 
NGC 5044, IC 4296 in class-2;  NGC 4261 in class-4) suggesting an
evolutionary link between MIR classes.

\subsubsection{11.3$\mu$m/7.7$\mu$m PAH ratio and the hot nuclear gas }

\begin{figure}
\centering
\includegraphics[width=8.7cm]{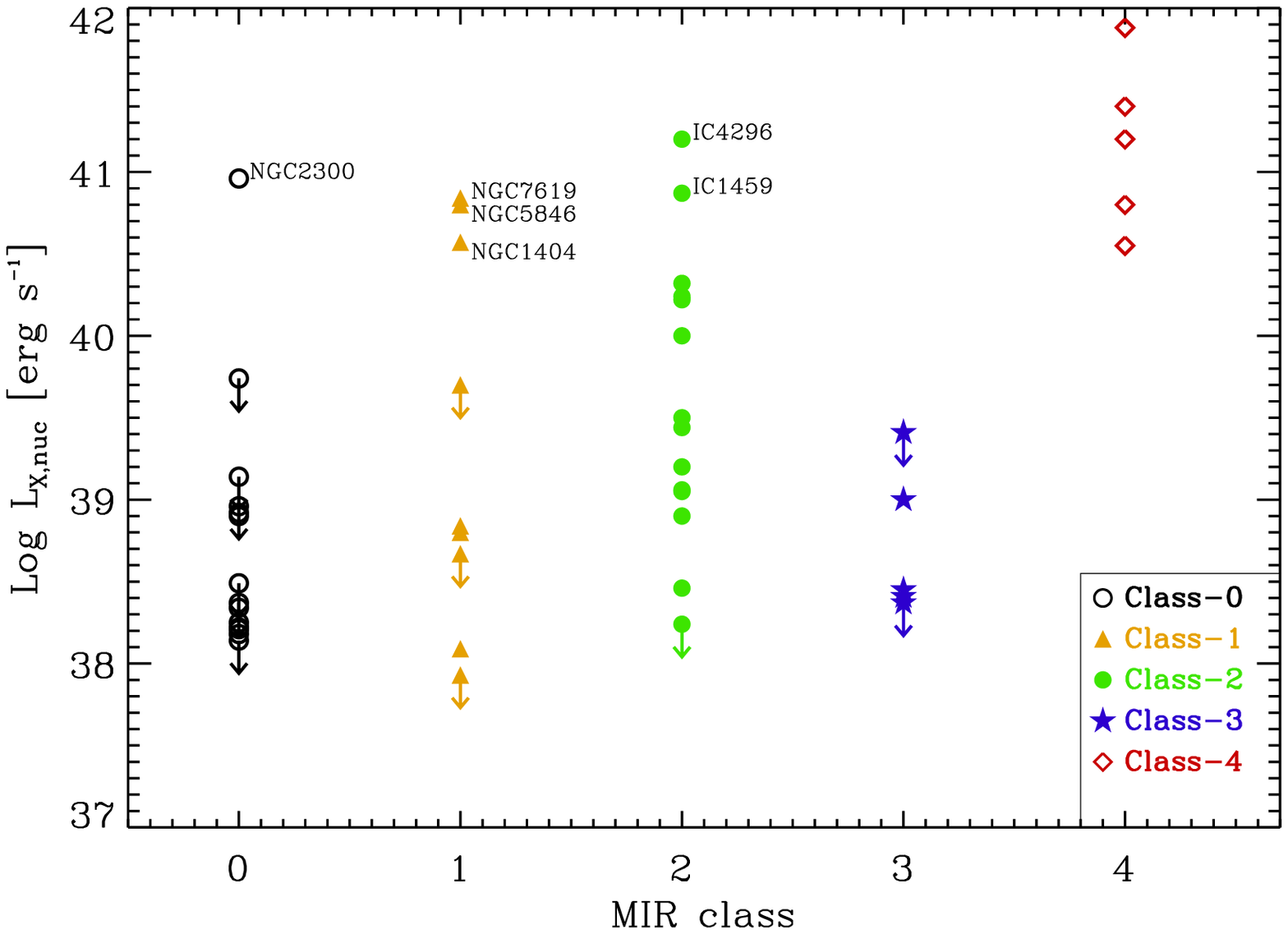}
\includegraphics[width=8.7cm]{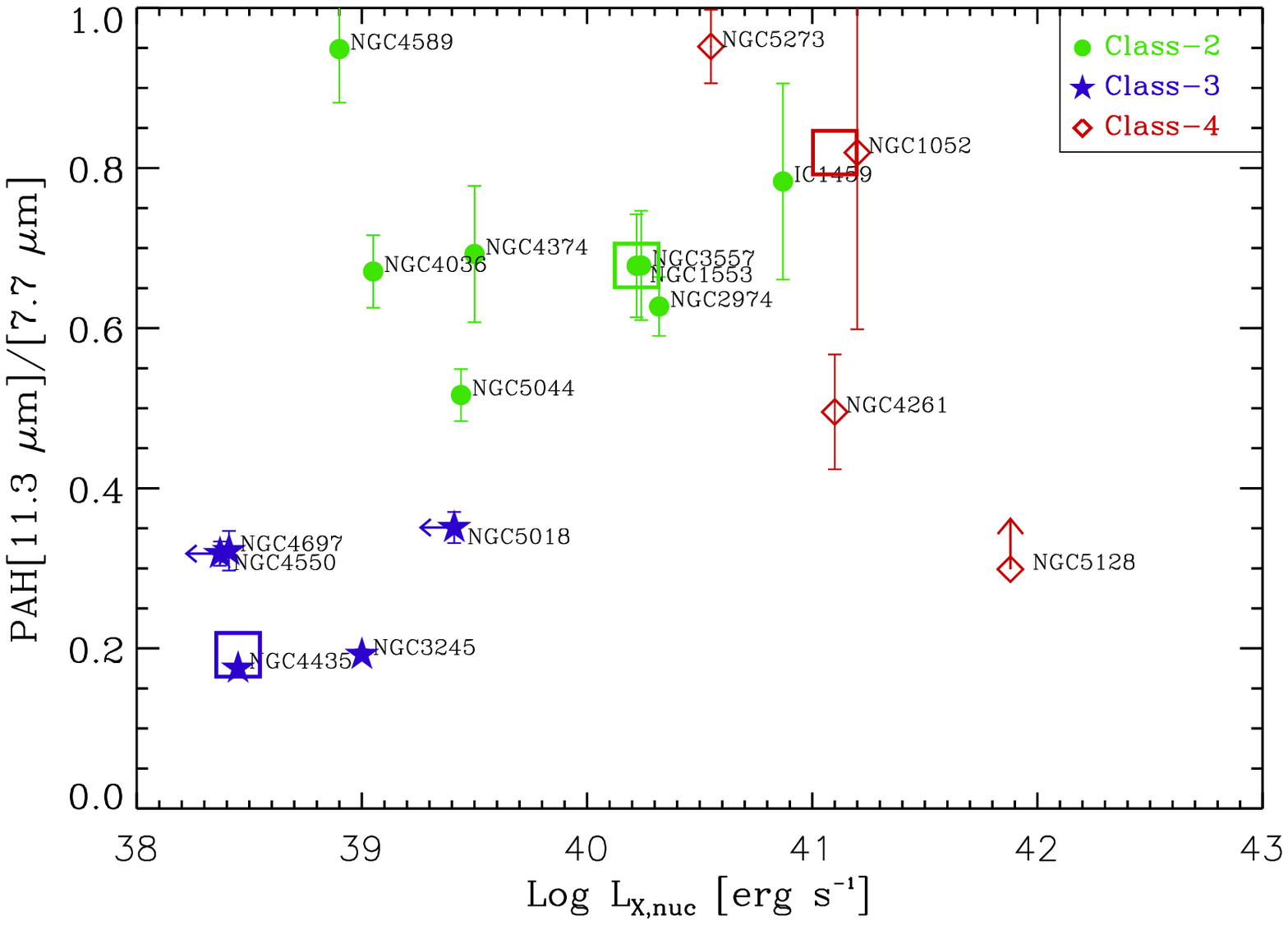}
\caption{
 ({\it Top panel}) Galaxy nuclear X-ray luminosity, Log~$L_{X, nuc}$,  
\citep{Pellegrini10} vs. MIR classes. X-ray bright galaxies of classes 0, 1, 2
are labeled. ({\it Bottom panel}) The ratio of PAH features at 11.3 and 7.7 $\mu$m
vs. Log~$L_{X, nuc}$.
Arrows represent upper limits and are colour coded as the detections. 
The large, open squares represent the median value of detected objects in
each class. The bottom right diamond refers to the PAH ratio measured for NGC~5128.
The  strong absorption centred at 10$\mu$m (see Figure~\ref{fig1}) probably perturbs 
the measure of the PAH line strength so that the PAH ratio is likely a lower limit. 
Therefore, in addition to the diamond, we label it with an arrow.} 
\label{fig6}
\end{figure}

In  the bottom panel of Figure~\ref{fig6}, we plot the ratio between the PAH 
features at  11.3$\mu$m and 7.7$\mu$m versus $L_{X, nuc}$. 
Although the dispersion is quite large, the figure shows 
a weak relation (the Pearson correlation coefficient is $r=0.58$) between 
the PAH 11.3/7.7 $\mu$m ratio and $L_{X, nuc}$. 
\citet{Draine07} suggested that variations in PAH emission ratios can be 
accounted for by the combined effects of variations in exciting radiation, 
PAH size and ionization state. 
The nuclear X-ray radiation field seems indeed to play a role in the 
determination of the PAH 11.3/7.7 $\mu$m ratio, taking into account the 
large variation of $L_{X, nuc}$ reported by \citet{Pellegrini10}. 
ETG nuclei in class-3 (normal PAH ratios),  have a  
lower  $L_{X, nuc}$ emission compared to ETGs with 
anomalous PAH ratios, as shown by the median values 
(large open squares in Figure~\ref{fig6} bottom panel) of classes 2 and 4.

\subsection{MIR classes, MIR atomic line diagnostics and H$_2$ emission}

We report in Table~B1 (columns 3 and  8) the optical 
activity class  of the ETG nuclei \citep{Ho97,Annibali10}. 
A large fraction of them  are LINERs in the optical window since they show low 
ionization emission lines. Among optical LINERs, MIR classes well 
identify both passively evolving ETGs and star forming nuclei 
\citep[][their Figure~2]{Rampazzo11}.
However, what powers the remaining objects of the vast class 
of LINERs is still debated.

\begin{figure*}
\centering
\includegraphics[width=13cm]{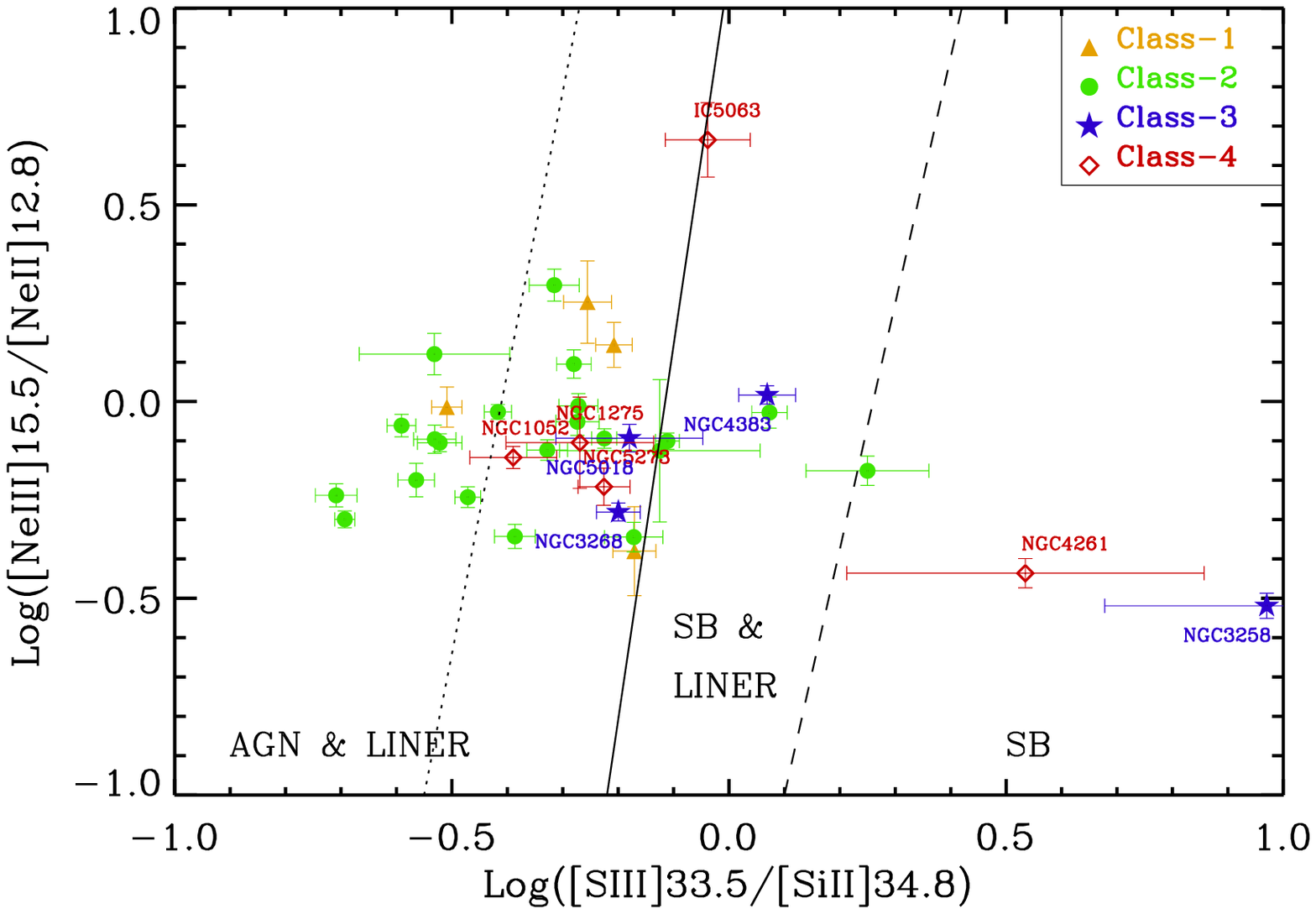}
\caption{Emission line diagnostic diagram \citep[from][]{Dale06} separating 
AGN from LINERs and star burst (SB) ETGs. We indicate MIR classes
and have labeled some relevant galaxies.}
\label{fig7}
\end{figure*}

In Figure~\ref{fig7}, MIR classes 1 to 4 are plotted in the 
Ne[III]15.5$\mu$m/[NeII]12.8$\mu$m vs. [SIII]33.5$\mu$m/[SiII]34.8$\mu$m 
plane \citep{Dale06}. The plot indicates the areas in which AGN, LINERs and
 starburst (SB) galaxies are typically located. Similarly to optical diagnostic 
 diagrams \citep[e.g.][]{Annibali10}
the MIR line diagnostic is unable to fully separate powering mechanisms. 
ETGs in class-3, but NGC~3258, inhabit the SB sector, mix together
with class-2 galaxies, although the two classes have totally different PAH ratios
i.e. different star forming properties. ETGs in class-2 mix also with class-4, 
suggesting the presence of an AGN contribution.  
The class-4 NGC~4261,  a LINER in the optical, is located in the 
SB region of Figure~\ref{fig7}. In the MIR, its AGN nature is suggested by
the presence of an excess over the passive stellar continuum, 
similarly to NGC~4486, and by the high ionization [NeV] 14.32 $\mu$m 
emission line, as well as, in X-rays, by the detection  of a central AGN and cavities. 

In the current literature, low accretion rate AGNs are the favoured mechanism 
for ionizing LINERs. Considering that massive black holes are thought to be 
present in galaxies with a spheroidal component \citep[e.g.][]{Kormendy04},
the above mechanism is supported by the presence of compact X-ray and/or 
nuclear radio sources \citep[][]{Gonzalez09}, UV and X-ray variability 
\citep[][]{Maoz05,Pian10} and broad emission lines in the 
optical spectra.  However, there is evidence for a deficit of ionizing photons 
from weak AGN suggesting that more than one excitation mechanism
may operate in LINERs \citep[see][]{Eracleous10}. 
Also the photoionization by old post-asymptotic giant
branch (post-AGB) stars \citep{Trinchieri91,Binette94,Stasinska08} can 
account for the ionizing photon budget only in the weakest LINERs, or 
in off-nuclear regions \citep{Annibali10}. 

For a long time, shocks \citep{Koski76,Heckman80,Dopita95,Allen08}
have been  proposed as a possible gas ionization mechanism in ETG nuclei.
The shock scenario may be related to the presence of the 
H$_2$ molecular emission that is thought to be an important coolant 
in post-shock regions \citep{Roussel07,Ogle07}.   We found H$_2$S(1) line 
with similar  rates in Es (34$^{+10}_{-8}$\%)  and S0s (51$^{+15}_{-12}$\%).
Several galaxies with X-ray cavities indeed show H$_2$ emission, 
like NGC 1553, NGC 4374,  NGC 4552, NGC 4636, NGC 5044, NGC 5813 and
 IC 4296, as well as  MIR class-4 galaxies 
with X-ray detected AGN, like NGC 1275, NGC 1052, NGC 4036 and NGC 5128.
However, the largest fraction of H$_2$ emitters is found among MIR class-2
galaxies. 

\subsection{MIR classes and CO}

In this section we report about CO observations, with the aim 
of investigating MIR classes vs. ETGs cold gas content.    

41 galaxies of our sample  are included in the survey of 
260 ETGs in  ATLAS$^{3D}$ \citep{Young11}. 
 Only 6 out of 41 (15$^{+9}_{-6}$\%) were detected: 
4 (NGC 4036, NGC 4435, NGC 4477, NGC 5273) in both
in CO ($J=1-0$) and ($J=2-1$) and 
2 (NGC 2685 and NGC~3245) in the ($J=1-0$) line.
Detected galaxies are S0s and  none  belong to MIR classes 0 or 1. 
NGC~2685, NGC~4036 and NGC~4477 are of MIR class-2;  NGC~3245, 
 NGC~4435 of class-3 and NGC~5273 of class-4. 
 Their  molecular gas masses are in the range 
7.27 $<$ log~M(H$_2$) $< $ 8.13 M$_\odot$ \citep{Young11}. 

NGC 4550 and NGC 4697 of class-3 are undetected  
in \citet{Young11}. \cite{Wiklind01} and \citet{Sofue93} detected  
NGC~4550  and  NGC 4697 (together with class-2 NGC~4589), respectively,  
in the CO ($J=1-0$) line. In this line, it is also detected the class-3 S0
NGC 4383  \citep{Thronson89}.

NGC~4261 and NGC 4486, in class-4, were undetected  
by \citet{Young11} (see also \citet{Ocana10} for NGC~4261). 
\citet{Smolcic11} detected NGC 4486  in the ($J=1-0$) line.  Several 
ETGs of class-4 have been detected in CO, namely    
NGC~1052 \citep[][($J=2-1$)]{Welch10}, NGC~1275 \citep{Salome06} and
NGC~5128 \citep[][]{Morganti10} and IC~5063 \citep[][in CO ($J=2-1$)]{Morganti13}.
Their H$_2$ masses differ by orders of magnitude scaling down from about 10$^{10}$ 
M$_\odot$ in NGC~1275, to 10$^{8}$ M$_\odot$ in NGC~5128 and in IC 5063 and 
to about 10$^{7}$ M$_\odot$ in  NGC 1052.  

Summarizing, all ETGs detected in CO lines show PAHs, either with normal
or anomalous ratios,  in their MIR nuclear spectra.

\subsection{MIR classes and the radio continuum at 1.4 GHz}

A large fraction of ETGs in the present sample are radio emitters,
although with different radio intensities, morphologies and classes.
In Table~B1 (columns 5 and 10) we report the 1.4~GHz radio power 
catalogued by \citet{Brown11}. These authors found that
the radio powers of ETGs, with similar M$_K$ magnitude, may vary
by roughly than two orders of magnitude  over long periods of time. 
The radio power may keep trace of the
past activity (AGN and/or star formation) useful to investigate
the possible link between MIR classes. 

\begin{figure*}
\centering
\includegraphics[width=8.8cm]{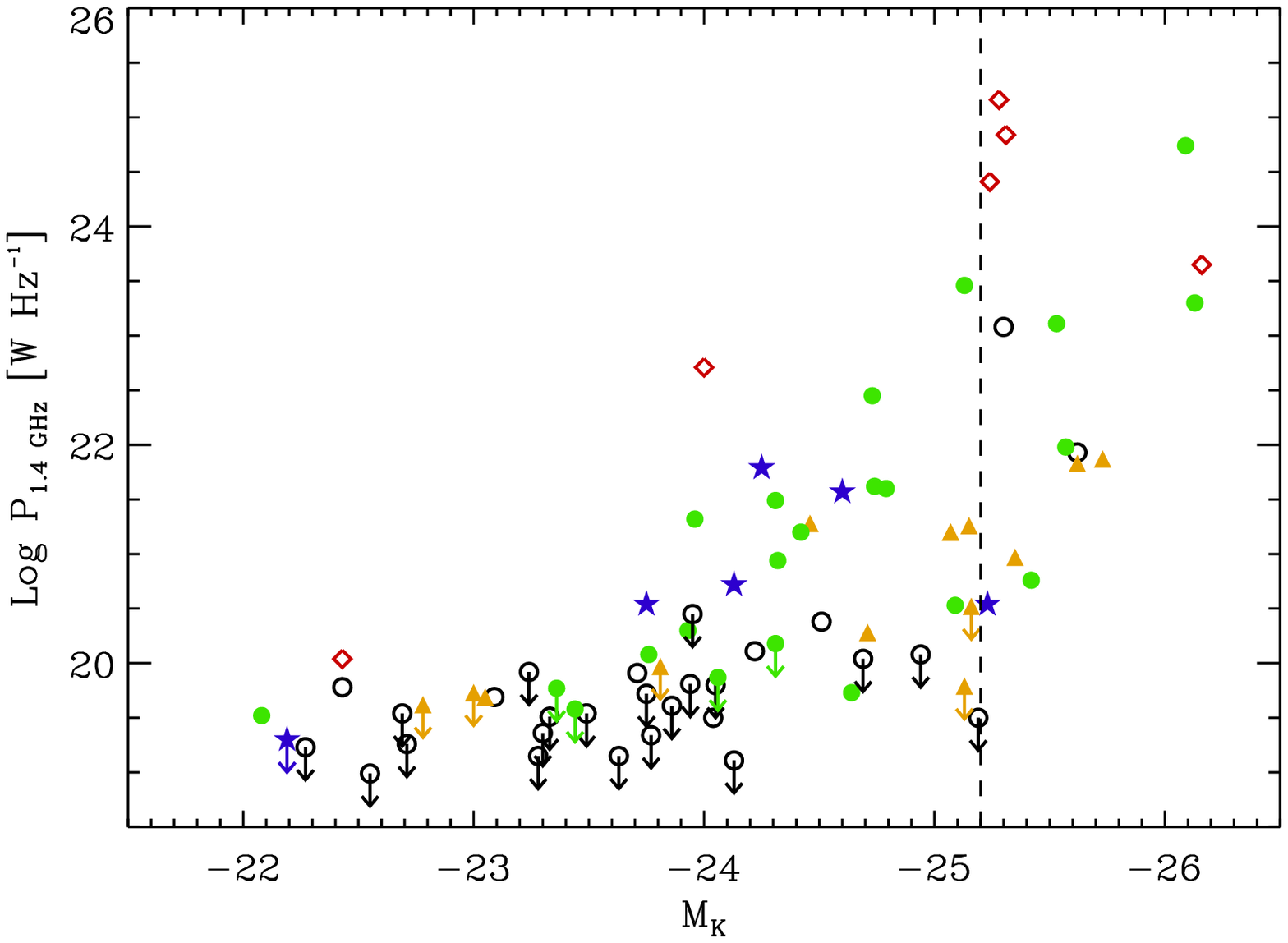}
\includegraphics[width=8.8cm]{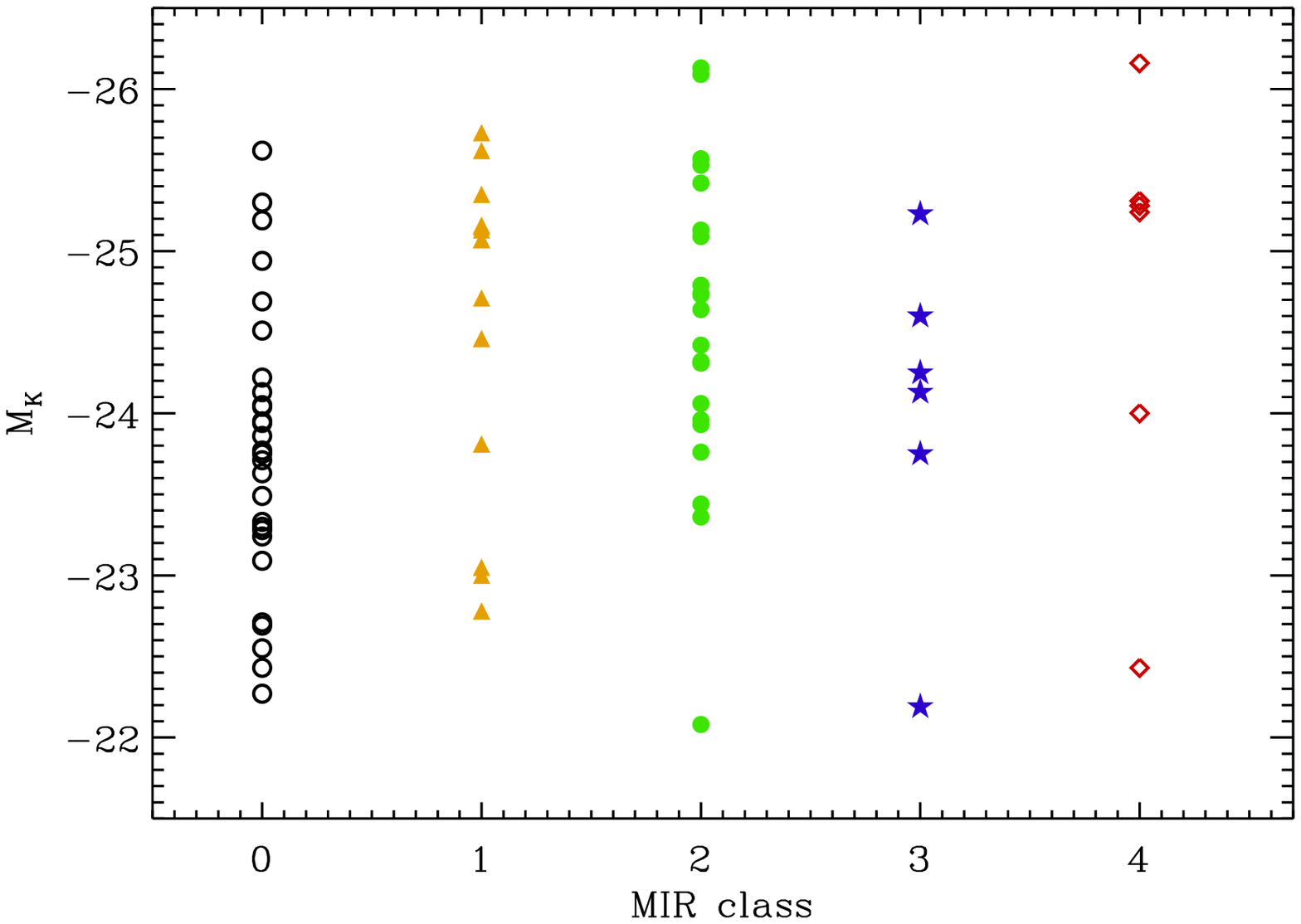}
\includegraphics[width=14cm]{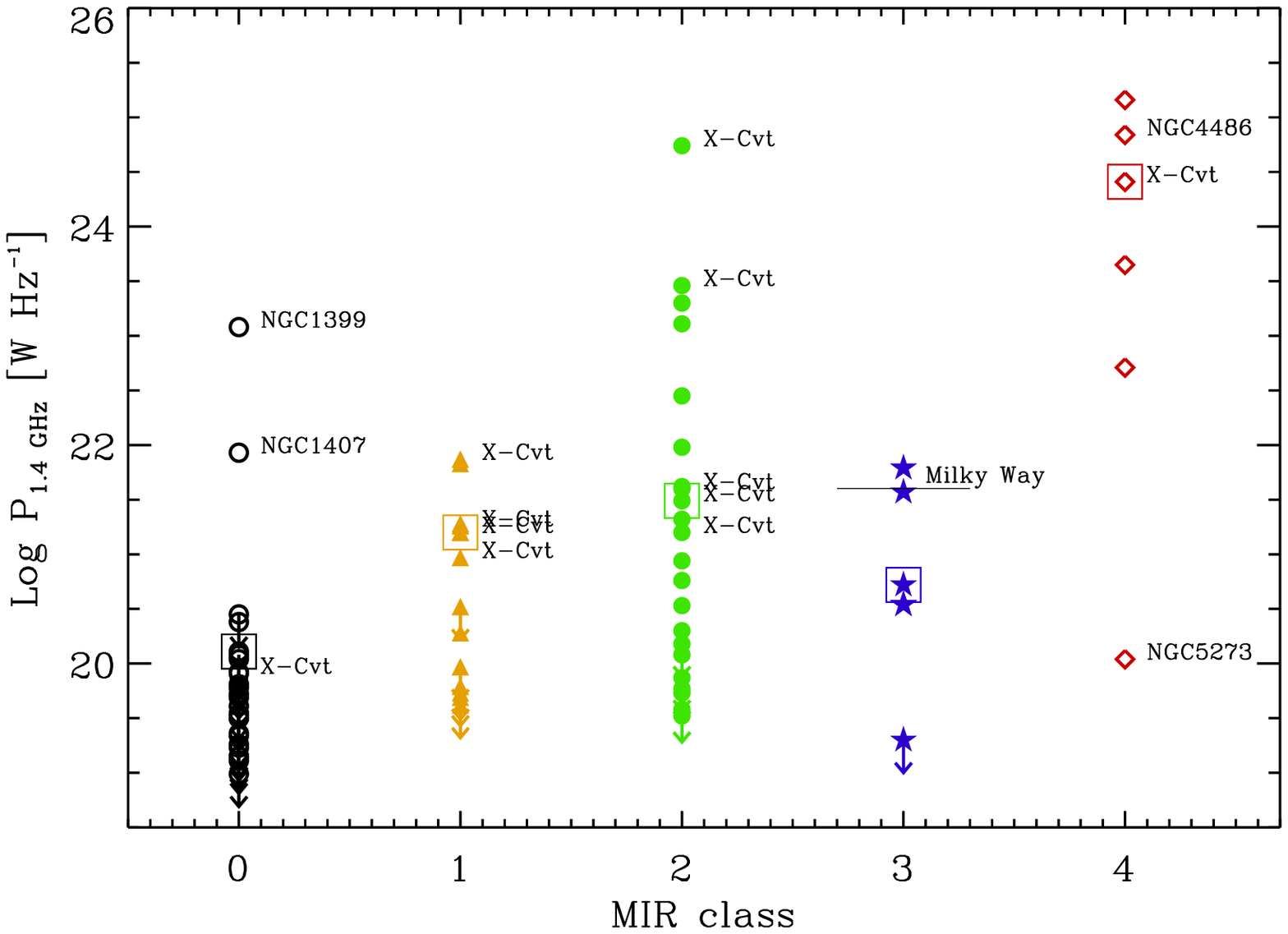}
\caption{ ({\it Top left})  1.4 GHz radio power vs. M$_K$ (open circle = class-0,
triangle = class-1, full dot = class-2, star = class-3 and open diamond =
 class-4). Arrows represent upper limits. ETGs with M$_K \leq -25.2$, the vertical  dashed line,
 are always detected \citep[see][]{Brown11}. ({\it Top right}) M$_K$ vs. MIR class  for the 
 radio sample (see Table~B1). ({\it Bottom}) The radio power vs. the MIR class.
 The radio power of the Milky Way is indicated with a horizontal line (see text). 
 Large open squares represent the median value of the detected objects in each MIR class.
 ETGs with X-ray  detected cavities \citep{Cavagnolo10} are labeled with X-Cvt.} 
\label{fig8}
\end{figure*}

Figure~\ref{fig8} (top left panel) shows 
the distribution of the 1.4~GHz radio power vs. M$_K$ magnitude. 
All ETGs brighter than -25.2 mag. are detected.
According to \citet{Brown11}, massive galaxies always host an AGN or have 
recently undergone star formation. 
The top right panel of Figure~\ref{fig8} shows M$_K$ magnitude vs.  
MIR classes. Each MIR class covers roughly the full  range of M$_K$. 

\begin{table}
\label{tab5}
\centering
\caption{MIR spectral classes}
\begin{tabular}{llcllc}
\hline
 Ident. & Morpho. & MIR   & Ident.  & Morpho. & MIR  \\
           & RSA       &  class  &            &RSA       &  class  \\
\hline
NGC 636   &  E1 & 0 &NGC 5090 & E2 & 2    \\
 NGC 720  & E5 &  0  &NGC 5638 & E1 &  0    \\
NGC  821 & E6 & 0 &NGC 5812 &  E0 & 0 \\
NGC 1209 & E6 &  1 & NGC 5813 & E1 &  1 \\
NGC 1275 & E pec  &4 &   NGC 5831 & E4 &  0\\
NGC 1297 & E2 &  2 &  NGC 7619 & E3 &  1   \\
NGC 1339 & E4 &  0   &   IC 1459    & E1 & 2\\
NGC 1374 & E0 & 0  & IC 2006   &  E1 &  1 \\
NGC  1379 & E0 &0  &  IC 3370 & E2 pec & 2  \\
NGC 1395 & E2 &  2  &IC 4296   &  E0 &  2 \\
NGC 1399 &  E1 & 0  &                &       &     \\
NGC 1404 & E2 & 1    & NGC 1052 & E3/S0 &  4 \\
NGC  1407 & E0 & 0  & NGC 1351 & S0$_1$/E6  & 0  \\
NGC 1426 & E4 &  0 & NGC 4472 & E1/S0$_1$ &1  \\
NGC 1427 & E5 & 0   & NGC 4550    & E/S0  &3  \\
NGC 1453 & E2 &  2   & NGC 4570 & E7/S0$_1$  &0   \\
NGC 1549 & E2 &  2   & NGC 4636 & E0/S0$_1$ & 2  \\
NGC 1700 & E3 & 1  & NGC 5353 & S0$_1$/E7 & 2  \\
NGC 2300 & E3 & 0  &NGC 6868 & E3/S0$_{2/3}$  & 2  \\
NGC 2974 & E4 &  2   & NGC 584 & S0$_1$  & 1  \\
NGC 3193 & E2 & 0     & NGC 1366 & S0$_1$ & 1  \\
NGC 3258 &  E1 &  3  & NGC 1389 & S0$_1$/SB0$_1$ &0  \\
NGC 3268 & E2 &  3  & NGC 1533 & SB0$_2$/SBa & 2  \\
NGC 3377 & E6 &  0   & NGC 1553 & S0$_{1/2}$ pec &2   \\
NGC  3379 & E0 & 0    & NGC 2685 & S0$_3$ pec& 2  \\
NGC 3557 & E3 &   2   & NGC 3245 & S0$_1$ &  3 \\
NGC 3608 &  E1 & 0  & NGC 4036 & S0$_3$/Sa  & 2  \\
NGC 3818 & E5 &  0   & NGC 4339 & S0$_{1/2}$  & 0  \\
NGC 3904 & E2 &  0   &NGC 4371 & SB0$_{2/3}$(r) & 2  \\
NGC 3962 &  E1 &  2  &NGC 4377 & S0$_1$  &0  \\
NGC 4261 & E3 &  4   &NGC 4382 & S0$_1$ pec & 1    \\
NGC 4365 & E3 &  0  &NGC 4383 & S0:  & 3   \\
NGC 4374 & E1 &  2       & NGC 4435 & SB0$_1$  & 3  \\
NGC 4473 & E5&   0  &NGC 4442 &  SB0$_1$ & 0  \\
NGC 4478 & E2  & 0 &NGC 4474 & S0$_1$   &0   \\
NGC  4486 & E0 & 4    & NGC 4477 & SB0$_{1/2}$/SBa   & 2  \\
 NGC 4564 & E6 &  0  & NGC 4552 & S0$_1$ & 2  \\
NGC 4589 & E2 &  2 &NGC 4649 & S0$_1$  & 1  \\
NGC 4621 & E5 &  0   &  NGC 5128 &  S0+S pec & 4  \\
NGC 4660 & E5 &   0  &NGC 5273 & S0/a & 4   \\
NGC 4696 & E3 &  2      &NGC 5631 & S0$_3$/Sa &2  \\
NGC 4697 & E6 & 3  &NGC 5846 & S0$_1$  & 1  \\
NGC 5011 & E2 & 1  &NGC 5898 & S0$_{2/3}$ & 2  \\
NGC 5018 & E4 &  3    & NGC 7192 & S0$_2$  & 1  \\
NGC 5044 &  E0 & 2  & NGC 7332 & S0$_{2/3}$ & 1  \\
NGC 5077 & E3 &  2 & IC 5063 & S0$_3$/Sa & 4 \\
\hline\hline
\end{tabular}
\end{table}

The P$_{1.4 GHz}$ vs. MIR class suggests
the following considerations. Although with a large dispersion, the class-4 
includes the more powerful radio sources  (Figure~\ref{fig8}, bottom panel) .  
Star forming, class-3, ETGs have intermediate/low radio power. 
A star formation rate of $\sim$1 M$_\odot$~yr$^{-1}$ is expected to produce 
10$^{21}$ W~Hz$^{-1}$ of radio emission in ETGs \citep{Wrobel88,Bell03}. 
The Milky Way (indicated in the plot) has a radio power   
P$_{1.4 GHz}$ $\simeq$ 4$\times$10$^{21}$ W~Hz$^{-1}$.
In class-2, the stronger radio emitters  are ETGs with X-ray cavities, 
likely connected to the AGN feedback \citep{Cavagnolo10}. X-ray cavities
are also detected in class-1 and in the E of class-0 NGC~3608 
(M$_K$=-24.80), detected at 1.4 GHz.

ETGs in class-0 set the lower boundary of the P$_{1.4 GHz}$ radio power.
Remarkable exceptions are NGC~1407 and NGC~1399. These are among 
the brightest  (M$_K \leq -25.3$) Es in the sample, 
 have radio jets  \citep{Velzen12} and X-rays detected AGN
 in their nuclei  \citep{Pellegrini05}. A misalignment of the SL and LL slits with respect to
the jet position angle may explain the class-0 spectrum of these galaxies.  
In the case of NGC~4486, the jet is sampled only by the LL slits, 
notwithstanding \citet{Buson2009} detected the synchrotron emission.

\section{MIR classes and accretion signatures}
\label{phenomenology}
 
 \citet{Panuzzo11} suggested that an accretion event may cause a 
 passively evolving ETG (class-0) to ignite star formation (class-3) 
 and/or  feeding the central black hole, revitalizing the AGN activity (class-4). 
 We review here possible accretion signatures, such as peculiarities in
the morphology and kinematics of the galaxy and of the dust-lane 
structure.

\subsection{Morphological and kinematical peculiarities}

In Table~B2  and Table~B3  we collect  the kinematical (column 3)
and morphological (column 4) peculiarities described in the literature
for Es and S0s, respectively. 
The kinematical notes  consider the detection of counter rotation (CR),
gas vs. gas, stars vs. gas and stars vs. stars,
and of multiple kinematical components (MC), widely believed to be associated 
with accretion events \citep[see e.g.][]{Corsini98,Krajnovic08,Bois12}. 
Other kinematical peculiarities are reported, like rotation along the minor axis, 
a phenomenon often associated with galaxy triaxiality  \citep[][]{Bertola88}. 
The morphological peculiarities include the detection of shells, tidal tails, 
peculiar isophote shapes that are believed to be associated with interaction 
and/or minor/major merger events 
\citep{Thomson90,Thomson91,Dupraz86,Hernquist87a,Hernquist87b}.    

41$^{+8}_{-7}$\% of ETGs show some form of kinematical peculiarity. The
percentage is likely a lower limit since some galaxies do not have
detailed kinematical studies. Many galaxies also show morphological
peculiarities. The above rate is consistent with the findings (53$^{+10}_{-4}$\%) 
of  \citet{vanDokkum05} obtained from the analysis  of morphological  
peculiarities in a comparable sample of nearby ETGs. 
He concludes that the majority of today's most luminous Es in the field were 
assembled through mergers of gas poor, bulge dominated systems.  However,
the high fraction  of ETGs nuclei with line emission we found (64$^{12}_6$\%)
argues that the above ``dry merger'' scenario is not so relevant.  

Although the column 4 in the Tables~B2 and B3 is largely incomplete, 
due to a still limited number of 2D kinematical studies,  it is evident that 
kinematical (and morphological) signatures of accretions are  found
also in passively evolving Es and S0s 
(e.g. NGC~3377, NGC~3608, NGC~4570,  NGC~4621, NGC~4660, NGC~5831).  
The fact that ETGs in class-0 show  kinematical/morphological 
peculiarities suggest these are long lasting structures 
with respect to spectroscopic features: galaxies were active in the recent past, 
as traced by the peculiar kinematics/structures, but their spectra have already
 evolved into a passive phase \citep[see e.g.][for shell galaxies]{Longhetti00}. 
Another possibility is  a``dry'' \citep{vanDokkum05} and {\it sterile} accretion, 
i.e. it did not induce a star formation/AGN  episode. Despite the presence of 
kinematical/morphological distortions, the spectrum remains passive, as in 
the case of an unperturbed/quiescent galaxy. 

\begin{table}
\centering
\caption{Demography of MIR spectral classes}
\begin{tabular}{lll}
\hline\hline
         & ~~E galaxies                                  & ~~S0 galaxies                                 \\
MIR classes &  $\overbrace{0 ~~~1 ~~~2 ~~~3  ~~4} $   & $\overbrace{0 ~~~1 ~~~2 ~~~3  ~~4}$         \\
\hline
     &                                           &                                   \\
Total Nr. of gal.   &  26 ~~7 ~~16 ~~4 ~~3      &  7~~~~8 ~~12 ~~4 ~~~4       \\
          &                                           &                                            \\
Nr. of gal. in cluster &      10 ~~2 ~~~1 ~~~0 ~~2&  6 ~~~3 ~~~4 ~~~3 ~~~0   \\
    &                                           &                                              \\
Nr. of gal. in LDE   &   16 ~~~5 ~~15 ~~4 ~~1     &  ~1 ~~~5 ~~~8 ~~1 ~~~4   \\
\hline
\end{tabular}
\label{tab6}

Notes: see text and Tables 1 and 2 for galaxy members of cluster or located in LDE.
\end{table}

\subsection{Dust-lane peculiarities}

In column 5 of Tables~B2 and B3 a description 
of the observed dust-lanes in the nuclear region is provided. 
The dust-lane morphology is available for 48/56 Es and
for 25/35 S0s. 50$^{+12}_{-11}$\% of Es and 32$^{+16}_{-11}$\%
S0s do not show dust-lanes.  

Dust in ETGs, including the most passive objects, is normally
produced by the collective outflow of dusty gas from evolving AGB 
 stars. The silicate emission from this
dust explains the 10 $\mu$m feature seen in MIR ETGs spectra 
\citep[][]{Knapp89,Athey02}.

However, the short lifetime of dust in the ETGs environment means 
that little dust can accumulate unless somehow shielded from the
hot gas \citep[][]{Clemens10}. In addition, the distribution of this 
dust follows that of stars, so dust-lanes cannot be produced in
this way.

The presence of dust-lanes and their asymmetries therefore points to
an external origin for the dust. An external origin is also consistent with 
(1) the lack of any strong  correlation between FIR and optical 
galactic luminosities \citep[e.g.][]{Forbes91,Goudfrooij95,Trinchieri02}; 
(2)  warm, counter-rotating gas observed in some ETGs 
(see Tables B1 and B2), (3) the distorted morphology of the dust 
\citep[e.g.][]{Finkelman12}.

In 42$^{+18}_{-13}$\% of Es and 24$^{+14}_{-9}$\% of S0s
the dust-lane morphology is irregular or chaotic. Dust-lanes are often
associated with morphological and/or kinematic  peculiarities. 
Irregular/chaotic dust-lanes are found in all MIR classes. Dust-lanes,
some quite complex, are found in class-4  (NGC 1052, 
NGC 1275, NGC 5128,  NGC 5273 and IC 5063).
Some Es, like NGC~1407, NGC~4621, NGC 4660 and NGC~5831, with
MIR class-0 spectra, do not have dust-lanes detected in the optical but 
display kinematical peculiarities, such as counter rotation or rotation 
along the apparent minor axis, suggesting that either dry accretion 
events or dust evaporation may have occurred.

\section{Summary and conclusions}
\label{conclusions}

We present homogeneously reduced and calibrated low resolution, {\it Spitzer}-IRS spectra
of a sample of 91 ETGs ($T<0$) in the {\it Revised Shapely Ames
Catalog} \citep{Sandage87}. We aim to provide a MIR atlas of well studied, nearby ETGs
 in the range $\sim 5-40~ \mu$m.


The atlas provides a measure of the atomic  and molecular emission lines and PAH 
intensities. We classify each spectrum into one of the 5 MIR spectral classes 
devised by \citet{Panuzzo11} for characterizing passive (class-0)
intermediate (class-1 and class-2), star forming (class-3) and AGN (class-4)
nuclei.  The main results are the following.

\begin{itemize}

\item{The nucleus of $\approx{1}/{3}$ (36$^{+7}_{-6}$\%) of nearby ETGs
in the atlas has a MIR class-0 spectrum, i.e. passively evolving 
according to \citet{Bressan06}. Class-0 spectra are more frequently detected among 
Es than S0s (46$^{+11}_{-10}$\% vs. 20$^{+11}_{-7}$\%). 
ETGs in class-0  set the lower boundary of the P$_{1.4 GHz}$ radio power, 
i.e. they are the most quiescent nuclei, on average.
Relevant exceptions are the massive (M$_K<$-25.2), 
class-0 Es in Fornax  NGC 1399 and NGC 1407, 
both having a high P$_{1.4GHz}$ and a radio-jet. 

About 78$^{+22}_{-18}$\% of Es and all S0s in class-0 
show no dust--lanes in the optical, although some  galaxies show
kinematical and morphological distortion, pointing to the 
occurrence of a past accretion episode. Thus, class-0 ETGs include
 either genuinely unperturbed, passively evolving galaxies 
or systems in which the effect of accretion has already quenched.

\item{Emission lines are detected in 64$^{+12}_{-6}$\% of nuclei. 
The detection of emission lines in many galaxies that show kinematical
evidence of merger events argues against the so-called ``dry mergers'' 
being an important formation mechanism for ETGs.  
H$_2$ molecular lines are also  frequently detected: the H$_2$S(1) emission line is
detected in 34$^{+10}_{-8}$\% of Es and 51$^{+15}_{-12}$\% of S0s. 
PAH features (classes 2, 3 and 4), detected in  47$^{+8}_{-7}$\% of ETGs,
tend to be less frequently found in cluster Es than in their LDE counterparts,
suggesting a triggering of the star formation activity in LDEs.
ETGs detected  in the CO ($J=1-0$) and/or  ($J=2-1$) lines 
show PAH complexes, either with normal or anomalous ratios,  
in their MIR nuclear spectra.}

\item{A  small fraction  (9$^{+4}_{-3}$\%) of ETGs shows class-3 spectra 
with normal PAH ratios,  typical of star forming galaxies.  
The ratio between ETGs with normal (class-3) vs. anomalous (class-2) 
PAHs is consistent with being produced by a star formation episode 
with typical lifetime of 200~Myr.  Class-3 ETGs have radio powers typical of 
star forming galaxies \citep{Brown11}.}

\item{The 11.3/7.7 $\mu$m PAH ratio weakly correlates with the 
galaxy X-ray nuclear luminosity, L$_{X,nuc}$, suggesting a dependence 
on the radiation field. Class-3 nuclei tend to be
less luminous in X-rays with respect to both class-4  and class-2, 
where an AGN contribution may be present.}
}
\end{itemize}

\citet{Panuzzo11} suggest that each of the five MIR 
classes is a snapshot of the evolution of ETG nuclei during an accretion episode, 
starting from, and ending with, a class-0 spectrum. 
The MIR emission line diagnostic diagram \citep{Dale06} is 
 unable to distinguish between the different powering mechanisms  ionizing the
ISM. MIR classes are successful in identifying those LINER nuclei powered by
star formation (class-3). Most ETGs in class-4  indicate the presence of an AGN.
Several mechanisms ionizing the ISM seem to be at work, especially in transition 
class 2 where (low accretion rate) AGN, shock fronts (X-ray cavities, H$_2$) 
and/or past star  formation (PAHs)  signatures combine. ETGs in class-1 may 
still show signatures of past activity like strong radio emission 
and X-ray cavities in some galaxies.
The MIR class-1 continuum is indistinguishable from that of class-0, 
suggesting that this class marks the end of the active phase in the nucleus.
 Kinematical and/or morphological scars 
of recent accretion episodes are found in a large fraction (41$^{+8}_{-7}$\%)
of our ETGs, even in passively evolving galaxies, further supporting  the 
view of an evolutionary link between the MIR classes and accretion/feedback 
phenomena.

\section*{Acknowledgments}

We thank the anonymous referee for the valuable suggestions.  
RR thanks the CEA, Laboratoire AIM, Irfu/SAP, Saclay (France) for the kind hospitality 
during the {\it Spitzer}-IRS reduction phase.  
RR acknowledges partial financial support by  contracts ASI-INAF I/016/07/0 and 
ASI-INAF I/009/10/0.  RR thanks Dr. P. Mazzei for  very useful discussion. 
OV acknowledges support from the Conacyt grant C0005-2010-01 146221.
AB acknowledges financial support by PRIN MIUR 2009. 
This research has made use of the NASA/ IPAC Infrared Science Archive, which is operated by the Jet Propulsion Laboratory, California Institute of Technology, under contract with the National Aeronautics and Space Administration.

\appendix

\section{Intensities of PAH complexes, nebular and molecular emissions}

Table~A1 lists the intensity of PAH complexes of
ETGs in the MIR spectral classes 2, 3 and 4.
Table~A2 lists the intensity of nebular and molecular line emission
of ETGs in the MIR spectral classes 1, 2, 3 and 4.
See Section~\ref{analysis} for details. 

\begin{table*}
\caption{PAH complex intensities.}
\tiny{
\begin{tabular}{lcccccc}
\hline
Ident.    &  6.22$\mu$m  &   7.7$\mu$m$^{(a)}$ &     8.6$\mu$m &      11.3$\mu$m$^{(b)}$ &
12.7$\mu$m$^{(c)}$  &   17$\mu$m$^{(d)}$ \\
&{($10^{-18}$ W m$^{-2}$)}&{($10^{-18}$ W m$^{-2}$)}&{($10^{-18}$ W m$^{-2}$)}&{($10^{-18}$ W m$^{-2}$)}&
{($10^{-18}$ W m$^{-2}$)}&{($10^{-18}$ W m$^{-2}$)}\\
\hline
\large{\bf E} & & & & & & \\
NGC 1275&\dots &1047.2 $\pm$ 226.2 &\dots &439.2 $\pm$ 141.9 &794.2 $\pm$ 250.7 &\dots\\
NGC 1297&13.3 $\pm$ 4.4&62.9 $\pm$ 3.3&13.8 $\pm$ 1.0&44.9 $\pm$ 1.4&21.8 $\pm$ 1.3&23.6 $\pm$ 2.5\\
NGC 1395&\dots&\dots&\dots	&39.6 $\pm$ 3.4&27.6 $\pm$ 3.4&\dots\\
NGC 1453&\dots&77.4 $\pm$ 10.7&10.6 $\pm$ 2.4&31.0 $\pm$ 2.8&20.8 $\pm$ 2.1&13.2 $\pm$ 1.7\\
NGC 1549&\dots&\dots&\dots&9.6 $\pm$ 3.0&\dots&4.1 $\pm$ 1.1\\
NGC 2974&88.3 $\pm$ 9.4&409.7 $\pm$ 21.1&44.0 $\pm$ 3.9&256.9 $\pm$ 7.2&113.5 $\pm$ 5.7&144.9 $\pm$ 5.3\\
NGC 3258&113.8 $\pm$ 4.9&537.6 $\pm$ 15.3&110.4 $\pm$ 4.6&119.4 $\pm$ 3.1&72.0 $\pm$ 2.9&44.1 $\pm$ 1.4\\
NGC 3268&67.5 $\pm$ 6.4&341.1 $\pm$ 9.6&48.3 $\pm$ 2.9&150.5 $\pm$ 4.3&83.6 $\pm$ 4.1&73.3 $\pm$ 2.5\\
NGC 3557&32.5 $\pm$ 10.7&134.9 $\pm$ 11.5&19.6 $\pm$ 5.8&91.5 $\pm$ 4.9&48.4 $\pm$ 5.8&32.5 $\pm$ 3.9\\
NGC 3962&31.9 $\pm$ 8.8&133.1 $\pm$ 8.8&24.6 $\pm$ 3.8&102.8 $\pm$ 4.6&37.3 $\pm$ 2.9&35.3 $\pm$ 4.8\\
NGC 4261&40.4 $\pm$ 8.6&117.3 $\pm$ 13.8&11.6 $\pm$ 2.7&58.1 $\pm$ 4.9&53.6 $\pm$ 4.3&7.4 $\pm$ 1.3\\
NGC 4374&65.7 $\pm$ 20.6&258.3 $\pm$ 28.7&51.3 $\pm$ 6.6&178.9 $\pm$ 9.4&95.1 $\pm$ 7.0&65.2 $\pm$ 4.7\\
NGC 4486   &\dots               &89.6$\pm$26.1 &\dots & \dots&30.2$\pm$9.7&\dots\\
NGC 4589&13.9 $\pm$ 4.2&52.5 $\pm$ 3.4&21.1 $\pm$ 1.1&49.8 $\pm$ 1.4&21.6 $\pm$ 2.1&27.7 $\pm$ 1.1\\
NGC 4696&\dots&\dots&\dots&11.3 $\pm$ 1.7&\dots&7.5 $\pm$ 0.5\\
NGC 4697&211.0 $\pm$ 21.2&816.8 $\pm$ 44.5&199.0 $\pm$ 16.2&263.0 $\pm$ 14.2&179.3 $\pm$ 5.0&78.4 $\pm$ 7.9\\
NGC 5018&232.1 $\pm$ 13.8&938.9 $\pm$ 39.4&233.7 $\pm$ 13.1&329.5 $\pm$ 11.9&152.1 $\pm$ 6.1&133.8 $\pm$ 7.1\\
NGC 5044   &29.4$\pm$7.0 &112.5$\pm$5.8 &32.1 $\pm$ 3.1&58.1 $\pm$ 2.1&27.1 $\pm$ 1.8&28.8 $\pm$ 0.9\\
NGC 5077&\dots&66.6 $\pm$ 6.5&22.3 $\pm$ 4.7&30.8 $\pm$ 2.5&14.0 $\pm$ 1.8&41.2 $\pm$ 2.6\\
NGC 5090&13.5 $\pm$ 4.1&80.6 $\pm$ 13.4&10.5 $\pm$ 1.6 &34.5 $\pm$ 2.4 &21.8 $\pm$ 2.4 &11.7 $\pm$ 3.1\\
IC 1459&\dots&108.4 $\pm$ 15.1&6.1 $\pm$ 1.8&84.9 $\pm$ 6.0&41.0 $\pm$ 5.4&47.1 $\pm$ 4.7\\
IC 3370    &\dots             &62.9 $\pm$ 19.6&13.1 $\pm$ 3.5 &22.2 $\pm$ 3.6 &22.5 $\pm$ 5.7 &21.5 $\pm$ 2.5\\
IC 4296&\dots&	\dots&	\dots&26.7 $\pm$ 5.8&26.0 $\pm$ 5.6&19.8 $\pm$ 5.6\\
  \large{\bf E/S0} & & & & &   & \\
NGC 1052&120.0 $\pm$ 33.0&239.0 $\pm$ 41.4&\dots&195.8 $\pm$ 40.4&\dots&254.1 $\pm$ 66.3\\
NGC 4550&59.9 $\pm$ 5.3&323.2 $\pm$ 12.0&79.7 $\pm$ 5.1&102.9 $\pm$ 3.1&33.1 $\pm$ 2.1&\dots\\
NGC 4636&\dots&\dots&\dots	&20.7 $\pm$ 5.0&	\dots&8.8 $\pm$ 0.8\\
NGC 5353&\dots&52.5 $\pm$ 11.9&5.12 $\pm$ 1.1&27.4 $\pm$ 1.5&15.7 $\pm$ 1.2&13.9 $\pm$ 1.1\\
NGC 6868&50.5 $\pm$ 12.3&241.7 $\pm$ 24.3&27.5 $\pm$ 2.9&129.6 $\pm$ 6.9&86.8 $\pm$ 4.9&45.6 $\pm$ 3.1\\
\large{\bf S0}  &                     &        &           & &  & \\
NGC 1533&\dots&\dots&\dots&31.3 $\pm$ 3.8&12.6 $\pm$ 2.1&27.4 $\pm$ 1.9\\
NGC 1553&105.6 $\pm$ 10.3&406.8 $\pm$ 35.3&48.7 $\pm$ 7.4&275.8 $\pm$ 10.5&84.4 $\pm$ 7.0&78.3 $\pm$ 5.5\\
NGC 2685&\dots&128.9 $\pm$ 10.7&12.6 $\pm$ 1.4&85.6 $\pm$ 3.2&44.1 $\pm$ 3.1&41.6 $\pm$ 2.1\\
NGC 3245&1242.0 $\pm$ 47.1&4490.2 $\pm$ 85.2&567.7 $\pm$ 19.7&863.9 $\pm$ 23.3&439.4 $\pm$ 24.1&\dots\\
NGC 4036&106.4 $\pm$ 15.6&194.1 $\pm$ 11.6&42.9 $\pm$ 5.3&130.2 $\pm$ 4.1&75.9 $\pm$ 4.1&\dots\\
NGC 4371&\dots&55.2 $\pm$ 7.8&30.7 $\pm$ 7.2&21.3 $\pm$ 2.5&8.1 $\pm$ 1.5&5.1 $\pm$ 1.1\\
NGC 4383&3691.8 $\pm$ 81.7&12597.3 $\pm$ 158.3&2985.3 $\pm$ 72.8&2710.8 $\pm$ 65.4&1486.1 $\pm$ 38.1&563.2 $\pm$ 56.9\\
NGC 4435&1304.5 $\pm$ 29.1&5906.7 $\pm$ 86.7&805.1 $\pm$ 21.5&1035.7 $\pm$ 23.7&802.2 $\pm$ 31.8&357.6 $\pm$ 15.3\\
NGC 4477&104.4 $\pm$ 14&278.6 $\pm$ 12.6&25.6 $\pm$ 1.9&133.7 $\pm$ 4.1&59.3 $\pm$ 2.8&64.7 $\pm$ 2.1\\
NGC 4552&\dots&\dots&\dots&38.3 $\pm$ 7.9&\dots&7.9 $\pm$ 1.8\\
NGC 5128 & 3337.4 $\pm$ 295.6 &18734.1 $\pm$ 909.5 & $>$1569.1 $\pm$ 330.6& $>$5603.1 $\pm$ 511.2&4251.8 $\pm$ 715.7&5230.6 $\pm$ 461.3\\
NGC 5273&\dots&346.2 $\pm$ 12.9&279.1 $\pm$ 15.1&329.5 $\pm$ 10.1&97.4 $\pm$ 5.5&320.6 $\pm$ 16.1\\
NGC 5631&\dots&87.3 $\pm$ 3.4&31 $\pm$ 1.1&45.5 $\pm$ 2.1&21.8 $\pm$ 2.1&28.9 $\pm$ 2.1\\
NGC 5898&\dots&53 $\pm$ 16.3&	\dots&46.5 $\pm$ 4.3&15.1 $\pm$ 3.3&21.7 $\pm$ 2.4\\
IC 5063&\dots&1190.4 $\pm$ 140.9&208.8 $\pm$ 68.9&462.9 $\pm$ 125.4&\dots	&\dots\\
\hline\hline
\end{tabular}
\label{tabA1}
}

{\bf Notes}. Uncertainties are 1$\sigma$.
$^{(a)}$ Includes the 7.42 $\mu$m, 7.60 $\mu$m and the 7.85 $\mu$m features.
$^{(b)}$ Includes the 11.23 $\mu$m, and the 11.33 $\mu$m  features. 
$^{(c)}$ Includes the 12.62 $\mu$m and the 12.69 $\mu$m features. 
$^{(d)}$ Includes the 16.45 $\mu$m, 17.04 $\mu$m, 17.375 $\mu$m  and the 17.87 $\mu$m features. 
We consider the 8.6 $\mu$m
and 11.3$\mu$m PAH emissions in NGC~5128 as lower limits because of the huge silicate absorption.

\end{table*}

\begin{table*}
\caption{Nebular and molecular emission line intensities}
{\tiny
\begin{tabular}{lcccccccccc}
\hline
Galaxy &	H$_2$ 0--0 S(7)&	H$_2$ 0--0 S(6)&	H$_2$ 0--0 S(5)&	$[${Ar}{II}]  &	H Pf$\alpha$&	H$_2$ 0--0 S(4)&	$[${Ar}{III}] &	H$_2$ 0--0 S(3) &$[${S}{IV}] &	H$_2$ 0--0 S(2)\\
name &	5.51 &	6.11 &	6.91  & 	6.99 & 	7.46 &	8.03 &	8.99 & 	9.66  & 	10.51&	12.28 \\
     &	$\mu$m& $\mu$m&	$\mu$m& 	$\mu$m& 	$\mu$m&	$\mu$m&	$\mu$m& 	$\mu$m& 	$\mu$m&	$\mu$m\\
\hline
\large{\bf E} & & & & & & & & &\\
NGC 1209&	\dots&	\dots&	\dots&	4.8 $\pm$ 1.0 &	\dots&	\dots&	3.5 $\pm$ 0.9 &	\dots&	\dots&	1.4 $\pm$ 0.2\\
NGC 1275&	120.4 $\pm$ 31.9&	\dots&	194.4 $\pm$ 25.0&	128.7 $\pm$ 18.4&	\dots&	\dots&	\dots&	184.9 $\pm$ 35.5&	\dots&	\dots\\
NGC 1297&	\dots&	\dots&	2.8 $\pm$ 0.3 &	3.3 $\pm$ 0.3 &	\dots&	0.5 $\pm$ 0.1 &	1.2 $\pm$ 0.3 &	5.1 $\pm$ 0.5 &	0.6 $\pm$ 0.2 &	2.1 $\pm$ 0.1\\
NGC 1395&	\dots&	\dots&	\dots&	\dots&	\dots&	\dots&	\dots&	\dots&	\dots&	1.8 $\pm$ 0.4\\
NGC 1404&	\dots&	\dots&	\dots&	\dots&	\dots&	\dots&	\dots&\dots&\dots&\dots		\\	
NGC 1453&\dots&	8.5 $\pm$ 1.7 &	\dots&	9.0 $\pm$ 1.1 &	3.9 $\pm$ 0.6 &	\dots&	\dots&	7.1 $\pm$ 1.7 &	2.2 $\pm$ 1.0 &	1.1 $\pm$ 0.2\\
NGC 1549&	\dots&	\dots&	\dots&	\dots&	\dots&	\dots&	\dots&	\dots&	\dots&	\dots\\
NGC  1700&	\dots&	\dots&	\dots&	\dots&	\dots&	\dots&	\dots&	\dots&	\dots&	\dots\\
NGC 2974&	9.0 $\pm$ 2.5 &	3.8 $\pm$ 0.5 &	16.1 $\pm$ 0.8 &	18.7 $\pm$ 1.4 &	\dots&	4.2 $\pm$ 0.3 &	\dots&	30.9 $\pm$ 1.7 &	4.1 $\pm$ 0.6 &	13.3 $\pm$ 0.8\\
NGC 3258&	\dots&	\dots&	4.1 $\pm$ 0.3 &	8.5 $\pm$ 0.5 &	\dots&	2.3 $\pm$ 0.1 &	0.8 $\pm$ 0.1 &	5.1 $\pm$ 0.8 &	0.7 $\pm$ 0.2 &	4.2 $\pm$ 0.2\\
NGC 3268&	\dots&	\dots&	6.1 $\pm$ 0.6 &	7.7 $\pm$ 0.5 &	\dots&	1.6 $\pm$ 0.1 &	0.5 $\pm$ 0.1 &	4.5 $\pm$ 0.7 &	1.6 $\pm$ 0.3 &	2.4 $\pm$ 0.2\\
NGC 3557&	\dots&	\dots&	\dots&	5.4 $\pm$ 0.6 &	\dots&	\dots&	\dots&	\dots&	\dots&	\dots\\
NGC 3962&	\dots&	\dots&	3.1 $\pm$ 0.7 &	8.6 $\pm$ 1.0 &	\dots&	4.5 $\pm$ 0.4 &	2.5 $\pm$ 0.7 &	16.6 $\pm$ 1.5 &	2.2 $\pm$ 0.6 &	4.8 $\pm$ 0.3\\	
NGC 4261 &	\dots&	\dots&	6.6 $\pm$ 1.1&	12.4  $\pm$ 1.3&	\dots&	\dots&	3.3 $\pm$ 1.1&	5.2 $\pm$ 1.7&	4.2 $\pm$ 1.0&	5.3 $\pm$ 0.4\\
NGC 4374&	\dots&	\dots&	7.4 $\pm$ 1.0 &	24.4 $\pm$ 2.2 &	\dots&	1.3 $\pm$ 0.1 &	6.9 $\pm$ 1.5 &	8.3 $\pm$ 2.0 &	5.2 $\pm$ 1.2 &	3.6 $\pm$ 0.5\\
NGC 4486&	\dots&	\dots&	\dots&	36.0 $\pm$ 3.0&	\dots&	\dots&	8.2 $\pm$ 2.6&	\dots&	\dots&	\dots\\
NGC 4589&	1.4 $\pm$ 0.4&	\dots&	\dots&	1.1 $\pm$ 0.3&	\dots&	\dots&	1.0 $\pm$ 0.3&	\dots&	\dots&	\dots\\
NGC 4696&	\dots&	\dots&	3.5 $\pm$ 0.8 &	8.4 $\pm$ 0.9 &	\dots&	\dots&	\dots&	6.4 $\pm$ 0.9 &	\dots&	1.5 $\pm$ 0.1\\
NGC 4697&	14.5 $\pm$ 3.2 &	8.2 $\pm$ 1.1 &	7.4 $\pm$ 1.2 &	12.9 $\pm$ 1.1 &	0.3 $\pm$ 0.1 &	6.4 $\pm$ 0.5 &	8.2 $\pm$ 1.2 &	8.2 $\pm$ 2.0 &	\dots&	1.6 $\pm$ 0.1\\
NGC 5011&	\dots&	\dots&	\dots&	7.2 $\pm$ 1.0 &	\dots&	\dots&	2.1 $\pm$ 0.6 &	3.6 $\pm$ 0.6 &	1.8 $\pm$ 0.5 &	\dots\\
NGC 5018&	\dots&	6.3 $\pm$ 1.0&	7.5 $\pm$ 1.0&	9.6 $\pm$ 1.0&	\dots&	4.3 $\pm$ 0.4&	3.4 $\pm$ 0.5&	4.9 $\pm$ 1.6&	\dots&	4.7 $\pm$ 0.5\\
NGC 5044&	6.4 $\pm$ 1.3 &	1.8 $\pm$ 0.4 &	14.9 $\pm$ 0.9 &	14.6 $\pm$ 0.5 &	\dots&	2.9 $\pm$ 0.2 &	1.8 $\pm$ 0.3 &	23.4 $\pm$ 1.1 &	1.2 $\pm$ 0.3 &	10.6 $\pm$ 0.4\\
NGC 5077&	\dots&	\dots&	6.2 $\pm$ 1.1 &	9.1 $\pm$ 0.8 &	3.5 $\pm$ 0.6 &	1.7 $\pm$ 0.2 &	2.6 $\pm$ 0.6 &	9.4 $\pm$ 0.9 &	1.2 $\pm$ 0.5 &	2.8 $\pm$ 0.4\\
NGC 5090&	\dots&	3.5 $\pm$ 0.9 &	7.0 $\pm$ 0.6 &	9.2 $\pm$ 0.8 &	\dots&	1.8 $\pm$ 0.2 &	3.0 $\pm$ 0.8 &	\dots&	3.1 $\pm$ 0.5 &	\dots\\
NGC 5813&	\dots&	7.5 $\pm$ 1.5 &	6.4 $\pm$ 0.6 &	15.3 $\pm$ 1.3 &	\dots&	2.9 $\pm$ 0.4 &	3.6 $\pm$ 1.0 &	6.3 $\pm$ 0.9 &	\dots&	\dots\\
NGC 7619&	\dots&	\dots&	\dots&	\dots&	\dots&	\dots&	\dots&	\dots&	\dots&	\dots\\
IC 1459&	\dots&	\dots&	5.0 $\pm$ 1.6 &	21.2 $\pm$ 1.5 &	\dots&	\dots&	\dots&	10.3 $\pm$ 2.4 &	\dots&	2.6 $\pm$ 0.5\\
IC 2006&	\dots&	\dots&	\dots&	\dots&	\dots&	\dots&	\dots&	\dots&	\dots&	\dots\\
 IC 3370&	\dots&	\dots&	4.2 $\pm$ 0.3 &	4.8 $\pm$ 0.4 &	\dots&	\dots&	2.3 $\pm$ 0.3 &	3.7 $\pm$ 0.4 &	\dots&	2.2 $\pm$ 0.2\\
 IC 4296&	\dots&	\dots&	\dots&	\dots&	\dots&	\dots&	\dots&	\dots&	\dots&	2.2 $\pm$ 0.5\\
  \large{\bf E/S0} & & & & &   & & & &\\
NGC 1052&	17.8 $\pm$ 3.1 &	8.7 $\pm$ 2.7 &	31.5 $\pm$ 7.8 &	81.0 $\pm$ 2.7 &	\dots&	6.7 $\pm$ 2.0 &	17.0 $\pm$ 2.9 &	63.3 $\pm$ 10.7 &	\dots&	11.2 $\pm$ 3.9\\
NGC 4472&	\dots&	\dots&	3.1 $\pm$ 1.0&	11.3 $\pm$ 2.5&	\dots&	\dots&	\dots&	\dots&	\dots&	\dots\\
NGC 4550&	\dots&	2.0 $\pm$ 0.4&	3.4 $\pm$ 0.4&	2.7 $\pm$ 0.3&	\dots&	\dots&	1.3 $\pm$ 0.2&	6.4 $\pm$ 0.6&	1.9 $\pm$ 0.4&	2.4 $\pm$ 0.3\\
NGC 4636&	\dots&	\dots&	\dots&	\dots&	\dots&	\dots&	\dots&	\dots&	\dots&	\dots\\
NGC 5353&	\dots&	\dots&	\dots&	\dots&	\dots&	\dots&	\dots&	2.0 $\pm$ 0.6&	\dots&	1.1 $\pm$ 0.1\\
NGC 6868&	7.0 $\pm$ 2.0 &	3.5 $\pm$ 1.0 &	20.2 $\pm$ 1.4 &	16.3 $\pm$ 1.0 &	\dots&	6.0 $\pm$ 0.5 &	4.0 $\pm$ 0.6 &	23.9 $\pm$ 1.5 &	1.4 $\pm$ 0.4 &	8.9 $\pm$ 0.3\\
\large{\bf S0}  &                     &        &           & &  & & & &\\
 NGC 584&	\dots&	\dots&	\dots&	\dots&	\dots&	\dots&	\dots&	\dots&	\dots&	\dots\\
NGC 1366&	\dots&	\dots&	\dots&	\dots&	\dots&	\dots&	\dots&	\dots&	\dots&	\dots\\
NGC 1533&	\dots&	\dots&	\dots&	2.2 $\pm$ 0.5 &	3.8 $\pm$ 1.2 &	\dots&	\dots&	3.6 $\pm$ 0.7 &	\dots&	1.8 $\pm$ 0.4\\
NGC 1553&	\dots&	1.5 $\pm$ 0.3 &	5.9 $\pm$ 0.8 &	10.6 $\pm$ 1.2 &	\dots&	3.7 $\pm$ 0.4 &	\dots&	10.8 $\pm$ 2.7 &	3.7 $\pm$ 0.8 &	6.2 $\pm$ 0.6\\
NGC 2685&	\dots&	\dots&	\dots&	7.6 $\pm$ 1.0&	2.3 $\pm$ 0.6&	2.4 $\pm$ 0.3&	4.2 $\pm$ 0.9&	7.9 $\pm$ 1.0&	3.0 $\pm$ 0.6&	2.7 $\pm$ 0.3\\
NGC 3245&	\dots&	\dots&	25.0 $\pm$ 1.5&	36.8 $\pm$ 1.8&	18.1 $\pm$ 8.0& 	\dots&	\dots&	\dots&	\dots&	\dots\\
NGC 4036&	\dots&	24.6 $\pm$ 3.0&	20.5 $\pm$ 1.2&	7.5 $\pm$ 1.1&	\dots&	10.6 $\pm$ 1.0&	8.8 $\pm$ 1.2&	35.7 $\pm$ 2.1&	1.5 $\pm$ 0.4&	13.8 $\pm$ 1.0\\
NGC 4371&	\dots&	\dots&	\dots&	\dots&	\dots&	\dots&	\dots&	\dots&	\dots&	1.2 $\pm$0.3\\
NGC 4382&	\dots&	\dots&	\dots&	\dots&	\dots&	\dots&	\dots&	\dots&	\dots&	\dots\\
NGC 4383&	\dots&	\dots&	\dots&	91.3 $\pm$ 5.0&	\dots&	\dots&	83.1 $\pm$ 3.1 &	\dots&	81.3 $\pm$ 17.2&	\dots\\
NGC 4435&	\dots&	\dots&	25.0 $\pm$ 1.5&	40.0 $\pm$ 2.0&	\dots&	22.2 $\pm$ 2.0&	\dots&	\dots&	\dots&	\dots\\
NGC 4477&	\dots&	4.0 $\pm$ 1.0&	9.0 $\pm$ 1.0&	18.9 $\pm$ 1.3&	\dots&	3.2 $\pm$ 0.4&	4.9 $\pm$ 0.7&	12.8 $\pm$ 1.7&	2.9 $\pm$ 0.7&	7.0 $\pm$ 0.5\\
NGC 4552&	\dots&	\dots&	\dots&	\dots&	\dots&	\dots&	\dots&	10.4 $\pm$ 3.4 &	\dots&	2.8 $\pm$ 0.9\\
NGC 4649&	\dots&	\dots&	\dots&	\dots&	\dots&	\dots&	\dots&	\dots&	\dots&	\dots\\
NGC 5128	&\dots	&83.4$\pm$ 22.2	&246.6 $\pm$ 40.2	&804.3 $\pm$ 92.9	&\dots	&54.5 $\pm$ 8.9	&82.9 $\pm$ 26.2	&231.7 $\pm$ 71.3	&\dots	&213.7 $\pm$ 69.3\\		
NGC 5273&	\dots&	\dots&	\dots&	\dots&	\dots&	22.9 $\pm$ 2.0&	\dots&	\dots&	30.4 $\pm$ 8.0&	9.5 $\pm$ 0.8\\
NGC 5631&	\dots&	\dots&	6.3 $\pm$ 1.3&	3.8 $\pm$ 1.0&	5.9 $\pm$ 1.2&	\dots&	3.4 $\pm$ 1.1&	11.0 $\pm$ 1.5&	\dots&	\dots\\
NGC 5846&	\dots&	\dots&	\dots&	\dots&	\dots&	\dots&	\dots&	0.5 $\pm$ 0.1 &	\dots&	0.6 $\pm$ 0.1\\
NGC 5898&	\dots&	\dots&	\dots&	\dots&	\dots&	\dots&	\dots&	4.1 $\pm$ 1.4 &	\dots&	0.9 $\pm$ 0.2\\
NGC 7192&	\dots&	\dots&	\dots&	2.6 $\pm$ 0.8 &	\dots&	\dots&	\dots&	\dots&	\dots&	\dots\\
NGC 7332&	\dots&	\dots&	\dots&	\dots&	\dots&	\dots&	\dots&	3.9 $\pm$ 0.2 &	\dots&	\dots\\
IC 5063&	\dots&	\dots&	133.8 $\pm$ 27.7 &	106.5 $\pm$ 29.6 &	\dots&	\dots&	176.3 $\pm$ 44.8 &	244.9 $\pm$ 49.9 &	788.5 $\pm$ 123.9 &	\dots\\
\hline\hline
\end{tabular}

{\bf Notes}. Uncertainties are 1$\sigma$. Values are in unit of 10$^{-18}$ [W~m$^{-2}$]. Neither atomic nor molecular emission lines are detected in Passive ETGs. $^d$ The two lines are blended in LL1 spectra: the values reported are the result of a 
line de-blending.
\label{tabA2}
}
\end{table*}

\begin{table*}
\addtocounter{table}{-1}
\caption{Nebular and molecular emission line intensities ({\it cont.})}
{\tiny
\begin{tabular}{lcccccccccc}
\hline
Galaxy &	 $[${Ne}{II}] &	 $[${Ne}{V}] &$[${Ne}{III}] &H$_2$ 0--0 S(1)&$[${Fe}{II}]&$[${S}{III}] &$[${Ar}{III}]&$[${Ne}{V}]&$[${O}{IV}]$^d$&$[${Fe}{II}]$^d$ \\
name &	12.81 $\mu$m &	14.32 $\mu$m& 	15.55   &	17.03  &	17.94   &	18.71   &	21.83   &	24.32   &	25.89  &	25.99   \\
&	$\mu$m &	 $\mu$m& 	  $\mu$m &	  $\mu$m &	  $\mu$m &	  $\mu$m &	  $\mu$m &	  $\mu$m &	  $\mu$m &	  $\mu$m \\
\hline
\large{\bf E} & & & & & & & & &\\
NGC 1209&	3.1 $\pm$  0.3 	&\dots&	3.0 $\pm$  0.2 	&\dots&	0.9 $\pm$  0.1 	&1.5 $\pm$  0.1 	&\dots&	\dots&	0.4 $\pm$  0.1 	&2.1 $\pm$  0.2\\
NGC 1275&	507.1 $\pm$ 50.1&	\dots&	398.6 $\pm$ 98.8&	\dots&	\dots&	\dots&	\dots&	\dots&	\dots&	\dots\\
NGC 1297&	1.9 $\pm$  0.1 	&\dots&	1.2 $\pm$  0.1 	&7.5 $\pm$  0.2 	&\dots&	0.4 $\pm$  0.1 	&\dots&	\dots&	0.8 $\pm$  0.1 	&1.0 $\pm$  0.1\\
NGC 1395&	2.8 $\pm$0.2&	\dots&	\dots&	\dots&	\dots&	\dots&	\dots&	\dots&	\dots&	\dots\\
NGC 1404&	5.9 $\pm$ 2.0&	\dots&	4.3 $\pm$ 1.5&	\dots&	\dots&	\dots&	\dots&	\dots&	\dots&	\dots\\
NGC 1453&	3.6 $\pm$  0.4 	&\dots&	7.5 $\pm$  0.5 	&2.2 $\pm$  0.2 	&\dots&	2.3 $\pm$  0.3 	&\dots&	\dots&	\dots&	\dots\\
NGC 1549&	2.5 $\pm$ 0.4&	\dots&	4.4 $\pm$ 0.5&	0.8 $\pm$ 0.2&	\dots&	1.6 $\pm$ 0.4&	\dots&	\dots&	\dots&	\dots\\
NGC  1700&	\dots&	\dots&	2.8 $\pm$0.9&	\dots&	\dots&	\dots&	\dots&	\dots&	\dots&	\dots\\
NGC 2974&	35.0 $\pm$  1.0 	&\dots&	27.5 $\pm$  1.2 	&21.8 $\pm$  1.1 	&\dots&	13.3 $\pm$  0.9 	&\dots&	7.3 $\pm$  0.9 	&2.1 $\pm$  0.2 	&10.3 $\pm$  0.6\\
NGC 3258&	15.2 $\pm$  0.5 	&\dots&	4.6 $\pm$  0.3 	&6.6 $\pm$  0.3 	&1.2 $\pm$  0.2 	&3.7 $\pm$  0.3 	&2.2 $\pm$  0.5 
	&\dots&	1.1 $\pm$  0.2 	&1.8 $\pm$  0.4\\
NGC 3268&	10.5 $\pm$  0.4 	&\dots&	5.5 $\pm$  0.2 	&3.0 $\pm$  0.3 	&1.0 $\pm$  0.3 	&3.4 $\pm$  0.2 	&1.3 $\pm$  0.2 	&\dots&	1.0 $\pm$  0.1 	&1.9 $\pm$  0.2\\
NGC 3557&	9.6 $\pm$  0.6 	&\dots&	9.0 $\pm$  0.6 	&\dots&	1.4 $\pm$  0.2 	&2.7 $\pm$  0.3 	&\dots&	1.1 $\pm$  0.3 	&0.8 $\pm$  0.1 	&2.8 $\pm$  0.3\\
NGC 3962&	14.2 $\pm$  0.4 	&1.0 $\pm$  0.1 	&8.2 $\pm$  0.5 	&4.7 $\pm$  0.2 	&2.8 $\pm$  0.2 	&5.3 $\pm$  0.2 	&\dots&	\dots&	3.1 $\pm$  0.2 	&3.0 $\pm$  0.2\\
NGC 4261 &	31.7 $\pm$ 1.6&	1.7 $\pm$ 0.4&	11.6 $\pm$ 0.8&	\dots&	\dots&	1.3 $\pm$ 0.3&	\dots&	\dots&	\dots&	\dots\\
NGC 4374&	32.9 $\pm$  1.3 	&1.6 $\pm$  0.3 	&26.5 $\pm$  1.1 	&8.3 $\pm$  0.6 	&1.7 $\pm$  0.3 	&14.9 $\pm$  0.5 	&2.5 $\pm$  0.6 	&\dots&	3.5 $\pm$  0.3 	&5.7 $\pm$  0.6\\
NGC 4486&	58.3 $\pm$ 2.8&	\dots&	24.1 $\pm$ 2.7&	\dots&	\dots&	19.1 $\pm$ 3.1&	\dots&	\dots&	\dots&	\dots\\
NGC 4589&	0.4 $\pm$ 0.1&	\dots&	0.3 $\pm$ 0.1&	0.3 $\pm$ 0.1&	\dots &	0.3 $\pm$ 0.1 &	\dots&	\dots&	\dots&	\dots\\
NGC 4696&	13.3 $\pm$  0.6 	&\dots&	7.6 $\pm$  0.3 	&6.4 $\pm$  0.3 	&\dots&	1.2 $\pm$  0.1 	&\dots&	\dots&	1.5 $\pm$  0.1 	&2.8 $\pm$  0.2\\
NGC 4697&	8.3 $\pm$  0.6 	&\dots&	9.1 $\pm$  1.1 	&4.0 $\pm$  0.5 	&\dots&	5.8 $\pm$  0.8 	&\dots&	\dots&	\dots&	\dots\\
NGC 5011&	2.8 $\pm$  0.3 	&0.6 $\pm$  0.2 	&3.9 $\pm$  0.3 	&\dots&	\dots&	1.8 $\pm$  0.1 	&\dots&	\dots&	0.9 $\pm$  0.1 	&0.5 $\pm$  0.1\\
NGC 5018&	19.7 $\pm$ 1.0&	\dots&	15.9 $\pm$ 1.0&	7.3 $\pm$ 0.4&	\dots&	6.0 $\pm$ 0.7&	\dots&	\dots&	2.5 $\pm$ 0.4&	3.1 $\pm$ 0.7\\
NGC 5044&	24.3 $\pm$  0.9 	&\dots&	12.2 $\pm$  0.4 	&13.4 $\pm$  0.5 	&0.6 $\pm$  0.1 	&3.1 $\pm$  0.2 	&1.0 $\pm$  0.2 	&1.2 $\pm$  0.2 	&1.5 $\pm$  0.1 	&2.1 $\pm$  0.2\\
NGC 5077&	20.6 $\pm$  1.0 	&0.6 $\pm$  0.1 	&17.9 $\pm$  0.8 	&12.7 $\pm$  0.7 	&1.4 $\pm$  0.2 	&6.3 $\pm$  0.4 	&2.9 $\pm$  0.6 	&\dots&	1.7 $\pm$  0.1 	&8.4 $\pm$  0.5\\
NGC 5090&	20.7 $\pm$  0.6 	&0.9 $\pm$  0.1 	&9.4 $\pm$  0.6 	&1.6 $\pm$  0.3 	&3.8 $\pm$  0.4 	&4.0 $\pm$  0.3 	&4.1 $\pm$  0.4 	&\dots&	2.7 $\pm$  0.3 	&2.2 $\pm$  0.2\\
NGC 5813&	6.3 $\pm$  0.3 	&\dots&	7.0 $\pm$  0.3 	&2.3 $\pm$  0.4 	&\dots&	3.0 $\pm$  0.2 	&\dots&	\dots&	\dots&	\dots\\

NGC 7619&	\dots&	\dots&	1.4 $\pm$ 0.6&	\dots&	\dots&	\dots&	\dots&	\dots&	\dots&	\dots\\
IC 1459&	51.4 $\pm$  2.1 	&\dots&	38.7 $\pm$  1.7 	&1.4 $\pm$  0.3 	&3.6 $\pm$  0.6 	&9.9 $\pm$  0.7 	&\dots&	4.7 $\pm$  1.1 	&2.8 $\pm$  0.3 	&12.3 $\pm$  1.0\\
IC 2006&	\dots&	\dots&	3.4 $\pm$1.6&	\dots&	\dots&	\dots&	\dots&	\dots&	\dots&	\dots\\
 IC 3370&	2.1 $\pm$  0.1 	&\dots&	1.4 $\pm$  0.1 	&4.4 $\pm$  0.2 	&\dots&	0.6 $\pm$  0.1 	&\dots&	0.3 $\pm$  0.1 	&0.7 $\pm$  0.1 	&0.9 $\pm$  0.1\\
 IC 4296&	24.1 $\pm$  1.4 	&0.5 $\pm$  0.1 	&10.9 $\pm$  0.7 	&1.7 $\pm$  0.5 	&1.4 $\pm$  0.3 	&4.2 $\pm$  0.6 	&\dots&	\dots&	3.9 $\pm$  0.4 	&2.9 $\pm$  0.2\\
 \large{\bf E/S0} & & & & &   & & & &\\
NGC 1052&	250.3 $\pm$  9.7 	&\dots&	180.4 $\pm$  9.3 	&47.5 $\pm$  9.7 	&43.6 $\pm$  12.6 	&71.4 $\pm$  9.4 	&\dots&	\dots&	27.2 $\pm$  3.9 	&50.1 $\pm$  9.3\\
NGC 4472&	2.7 $\pm$ 0.7&	\dots &	\dots &	\dots &	\dots &	\dots &	\dots &	\dots &	\dots &	\dots\\
NGC 4550&	4.3 $\pm$ 0.3&	\dots &	\dots &	\dots&	\dots&	\dots&	\dots&	\dots &	\dots &	\dots\\
NGC 4636&	12.9 $\pm$  0.8 	&\dots &	15.1 $\pm$  0.8 	&2.4 $\pm$  0.4 	&\dots&	6.1 $\pm$  0.5 	&\dots&	\dots&	\dots&	\dots\\
NGC 5353&	2.5 $\pm$ 0.2&	\dots&	3.3 $\pm$ 0.3&	2.1 $\pm$ 0.1&	\dots &	1.7 $\pm$  0.2 &	\dots &	\dots&	0.7 $\pm$ 0.1&	1.8 $\pm$ 0.2\\
NGC 6868&	30.0 $\pm$  1.0 	&1.1 $\pm$  0.1 	&23.8 $\pm$  0.8 	&9.6 $\pm$  0.6 	&1.5 $\pm$  0.2 	&7.8 $\pm$  0.4 	&1.3 $\pm$  0.4 	&2.0 $\pm$  0.4 	&4.8 $\pm$  0.3 	 &6.6 $\pm$  0.4\\
\large{\bf S0}  &                     &        &           & &  & & & &\\
NGC 584&	3.1 $\pm$ 1.0&	\dots&	6.5 $\pm$ 0.7&	\dots&	\dots&	\dots&	\dots&	\dots&	\dots&	\dots\\
NGC 1366&	1.9 $\pm$  0.4 	&\dots&	3.4 $\pm$  0.4 	&\dots&	\dots&	1.7 $\pm$  0.2 	&\dots&	\dots&	0.5 $\pm$  0.1 	&0.5 $\pm$  0.1\\
NGC 1533&	5.3 $\pm$  0.3 	&\dots&	6.6 $\pm$  0.4 	&4.5 $\pm$  0.2 	&\dots&	5.7 $\pm$  0.4 	&1.6 $\pm$  0.2 	&\dots&	0.6 $\pm$  0.1 	&3.0 $\pm$  0.3\\
NGC 1553&	25.1 $\pm$  0.9 	&0.4 $\pm$  0.1 	&22.3 $\pm$  1.6 	&5.7 $\pm$  0.5 	&\dots&	8.5 $\pm$  0.9 	&\dots&	\dots&	1.3 $\pm$  0.2 	&6.1 $\pm$  0.9\\
NGC 2685&	8.6 $\pm$ 0.5&	\dots&	6.9 $\pm$ 0.4&	5.2 $\pm$ 0.3	& \dots&	0.5 $\pm$ 0.1&	\dots&	\dots&	1.4 $\pm$ 0.2&	2.4 $\pm$ 0.2\\
NGC 3245&	104.6 $\pm$ 13.6&	\dots&	\dots&	\dots&	\dots&	\dots&	\dots&	\dots&	\dots&	\dots\\
NGC 4036&	34.2 $\pm$ 1.4&	\dots&	\dots&	\dots&	\dots&	\dots&	\dots&	\dots&	\dots&	\dots\\
NGC 4371&	1.8 $\pm$ 0.2&	\dots&	0.7 $\pm$ 0.2&	\dots&	\dots&	\dots&	\dots&	\dots&	\dots&	\dots\\
NGC 4382&	\dots&	\dots&	3.4 $\pm$  0.9&	\dots&	\dots&	\dots&	\dots&	\dots&	\dots&	\dots\\
NGC 4383&	359 $\pm$ 11.1&	\dots&	373.0 $\pm$ 16.3&	24.6 $\pm$ 8.0&	\dots&	305.5 $\pm$ 19.3&	\dots&	\dots&	26.9 $\pm$ 7.6&	25.6 $\pm$ 8.0\\
NGC 4435&	85.9 $\pm$ 2.3&	\dots&	22.2 $\pm$ 2.6&	17.6 $\pm$ 2.2&	\dots&	22.4 $\pm$ 1.7&	\dots&	\dots&	\dots&	\dots\\
NGC 4477&	20.2 $\pm$ 1.0&	1.1 $\pm$ 0.3&	19.7 $\pm$ 1.0&	12.8 $\pm$ 0.6&	1.5 $\pm$ 0.2&	9.3 $\pm$ 0.4&	\dots&	\dots&	4.5 $\pm$ 0.4&	5.6 $\pm$ 0.4\\
NGC 4552&	13.5 $\pm$  1.1 	&0.3 $\pm$  0.1 	&10.4 $\pm$  1.1 	&\dots&	\dots&	10.0 $\pm$  1.0 	&\dots&	\dots&	\dots&	\dots\\
NGC 4649&	2.2 $\pm$ 0.8&	\dots&	3.7 $\pm$ 0.8&	\dots&	\dots&	\dots&	\dots&	\dots&	\dots&	\dots\\
NGC 5128&	2057.8 $\pm$ 118.1	&201.4 $\pm$ 64.2	&1297.1 $\pm$ 148.6	&634.7 $\pm$ 60.2	&\dots&	710.9 $\pm$ 62.2	&\dots&	\dots&	541.3 $\pm$ 75.3	&551.3 $\pm$ 109.2\\
NGC 5273&	27.0 $\pm$1.3&	\dots&	16.4 $\pm$ 1.6&	7.5 $\pm$ 1.2&	\dots&	9.7 $\pm$ 1.5&	\dots&	\dots&	\dots&	\dots\\
NGC 5631&	3.4 $\pm$ 0.1&	\dots&	3.2 $\pm$ 0.1&	4.6 $\pm$ 0.2&	\dots&	1.0 $\pm$ 0.1&	\dots&	2.4 $\pm$ 0.2&	1.4 $\pm$0.1&	1.7 $\pm$0.1\\
NGC 5846&	12.5 $\pm$  0.9 	&\dots&	9.1 $\pm$  0.6 	&2.0 $\pm$  0.2 	&\dots&	3.8 $\pm$  0.3 	&\dots&	\dots&	\dots&	\dots\\
NGC 5898&	4.2 $\pm$  0.3 	&\dots&	8.3 $\pm$  0.5 	&3.1 $\pm$  0.3 	&\dots&	3.8 $\pm$  0.4 	&\dots&	1.2 $\pm$  0.4 	&0.5 $\pm$  0.1 	&2.1 $\pm$  0.2\\
NGC 7192&	2.4 $\pm$  0.4 	&\dots&	1.0 $\pm$  0.2 	&0.3 $\pm$  0.1 	&0.3 $\pm$  0.1 	&0.3 $\pm$  0.1 	&1.0 $\pm$  0.2 	&\dots&	\dots&	0.6 $\pm$  0.1\\
NGC 7332&	\dots	&\dots&	9.0 $\pm$  0.9 	&1.7 $\pm$  0.3 	&\dots&	3.4 $\pm$  0.5 	&\dots&	1.2 $\pm$  0.4 	&3.4 $\pm$  0.2 	&4.7 $\pm$  0.3\\
IC 5063&	238.7 $\pm$  48.2 	&335.2 $\pm$  68.7 	&1103.6 $\pm$  88.9 	&168.1 $\pm$  36.9 	&\dots&	321.7 $\pm$  71.2 	&\dots&	402.8 $\pm$  107.1 	&477.1 $\pm$  44.8 	&673.0 $\pm$  51.4\\
\hline\hline
\end{tabular}
}
\end{table*}

\begin{table*}
\addtocounter{table}{-1}
\caption{Nebular and molecular emission line intensities ({\it cont.})}
{\tiny
\begin{tabular}{lccccc}
\hline
Galaxy &	H$_2$ 0--0 S(0)&	$[${S}{III}] &	$[${Si}{II}] &	$[${Fe}{II}]&	$[${Ne}{III}]\\
name &	28.22 &	33.48 &	34.82 &	35.35 &	36.01\\
&	 $\mu$m&	 $\mu$m&	 $\mu$m&	$\mu$m&	 $\mu$m\\
\hline
\large{\bf E} & & & & & \\
NGC 1209&	\dots&	3.5 $\pm$  0.2 &	11.3 $\pm$  0.3 &	\dots&	1.1 $\pm$  0.1\\
NGC 1275&	\dots&	456.0 $\pm$ 122.0&	847.4 $\pm$ 127.8&	\dots&	\dots\\
NGC 1297&	3.0 $\pm$  0.1 &	1.5 $\pm$  0.1 &	5.5 $\pm$  0.2 &	0.3 $\pm$  0.1 &	\dots\\
NGC 1395&	\dots&	\dots&	\dots&	\dots&	\dots\\
NGC 1404&	\dots&	\dots&	\dots&	\dots&	\dots\\
NGC 1453&	\dots&	\dots&	\dots&	\dots&	\dots\\
NGC 1549&	\dots&	\dots&	\dots&	\dots&	\dots\\
NGC  1700&	\dots&	\dots&	\dots&	\dots&	\dots\\
NGC 2974&	2.6 $\pm$  0.6 &	13.5 $\pm$  1.2 &	44.9 $\pm$  1.2 &	\dots&	\dots\\
NGC 3258&	\dots&	5.6 $\pm$  0.5 &	\dots&	\dots&	\dots\\
NGC 3268&	\dots&	4.8 $\pm$  0.3 &	7.6 $\pm$  0.5 &	\dots&	\dots\\
NGC 3557&	\dots&	8.4 $\pm$  0.4 &	7.1 $\pm$  0.4 &	2.7 $\pm$  0.3 &	4.3 $\pm$  0.3\\
NGC 3962&	\dots&	6.2 $\pm$  0.5 &	31.7 $\pm$  1.0 &	1.4 $\pm$  0.4 &	2.1 $\pm$  0.3\\				
NGC 4261 &	2.7 $\pm$ 0.9&	9.3 $\pm$ 1.1&	\dots&	\dots&	\dots\\
NGC 4374&	2.2 $\pm$  0.6 &	24.3 $\pm$  1.0 &	40.8 $\pm$  1.3 &	3.6 $\pm$  0.7 &	\dots\\
NGC 4486&	\dots&	\dots&	\dots&	\dots&	\dots\\
NGC 4589&	\dots&	0.3 $\pm$ 0.1&	0.4 $\pm$ 0.1&	\dots&	\dots\\
NGC 4696&	0.9 $\pm$  0.1 &	4.9 $\pm$  0.2 &	14.5 $\pm$  0.5 &	4.0 $\pm$  0.1 &	4.5 $\pm$  0.2\\
NGC 4697&	\dots&	\dots&	\dots&	\dots&	\dots\\
NGC 5011&	\dots&	3.1 $\pm$  0.2 &	5.0 $\pm$  0.2 &	\dots&	\dots\\
NGC 5018&	\dots&	12.1 $\pm$ 2.7&	18.3 $\pm$ 3.8&	\dots&	\dots\\
NGC 5044&	1.5 $\pm$  0.2 &	9.0 $\pm$  0.3& 	44.4 $\pm$  1.1 &	0.8 $\pm$  0.1 &	1.4 $\pm$  0.1\\
NGC 5077&	2.6 $\pm$  0.5 &	11.7 $\pm$  0.6 &	45.6 $\pm$  1.4 &	2.1 $\pm$  0.5 &	1.6 $\pm$  0.5\\
NGC 5090&	0.8 $\pm$  0.2 &	5.3 $\pm$  0.4 &	12.9 $\pm$  0.5 &	\dots&	\dots\\
NGC 5813&	\dots&	\dots&	\dots&	\dots&	\dots\\
NGC 7619&	\dots&	\dots&	\dots&	\dots&	\dots\\
IC 1459&	3.1 $\pm$  1.0 &	23.5 $\pm$  1.7 &	50.0 $\pm$  2.2 &	\dots&	\dots\\
IC 2006&	\dots&	\dots&	\dots&	\dots&	\dots\\
 IC 3370&	0.7 $\pm$  0.1 &	1.6 $\pm$  0.2 &	0.9 $\pm$  0.2 &	1.5 $\pm$  0.2 &	\dots\\
 IC 4296&	\dots&	10.1 $\pm$  1.1 &	15.0 $\pm$  0.8 &	2.0 $\pm$  0.5 &	\dots\\
  \large{\bf E/S0} & & & & &  \\
NGC 1052&	\dots&	84.6 $\pm$  13.3 &	207.3 $\pm$  18.4 &	\dots&	\dots\\
NGC 4472&	\dots&	\dots&	\dots &	\dots&	\dots\\
NGC 4550&	\dots&	\dots&	\dots&	\dots&	\dots\\
NGC 4636&	\dots&	\dots&	\dots&	\dots&	\dots\\
NGC 5353&	0.8 $\pm$  0.2 &	1.0 $\pm$ 0.3&	3.4 $\pm$ 0.3&	2.8 $\pm$  0.4 &	2.9 $\pm$ 0.4\\
NGC 6868&	\dots&	22.3 $\pm$  0.8 &	28.8 $\pm$  1.0 &	2.8 $\pm$  0.5 &	3.8 $\pm$  0.5\\
\large{\bf S0}  &                     &        &           & &  \\
NGC 584&	\dots &	\dots &	\dots &	\dots &	\dots\\
NGC 1366&	\dots&	1.5 $\pm$  0.1 &	2.7 $\pm$  0.2 &	1.5 $\pm$  0.2 &	\dots\\
NGC 1533&	1.8 $\pm$  0.3 &	6.3 $\pm$  0.4 &	12.0 $\pm$  0.4 &	0.5 $\pm$  0.1 &	\dots\\
NGC 1553&	\dots&	15.4 $\pm$  1.0 &	28.9 $\pm$  1.8 &	\dots&	7.5 $\pm$  0.7\\
NGC 2685&	1.7 $\pm$ 0.2&	3.8 $\pm$ 0.3&	12.9 $\pm$  0.5 &	\dots&	\dots\\
NGC 3245&	\dots&	\dots&	\dots&	\dots&	\dots\\
NGC 4036&	\dots&	\dots&	\dots&	\dots&	\dots\\
NGC 4371&	\dots&	\dots&	\dots&	\dots&	\dots\\
NGC 4382&	\dots&	\dots&	\dots&	\dots&	\dots\\
NGC 4383&	\dots&	459.9 $\pm$ 38.7&	390.9 $\pm$ 26.8&	\dots&	\dots\\
NGC 4435&	\dots&	\dots&	\dots&	\dots&	\dots\\
NGC 4477&	\dots&	16.8 $\pm$ 1.1&	31.4 $\pm$ 1.5&	\dots&	\dots\\
NGC 4552&	\dots&	\dots&	\dots&	\dots&	\dots\\
NGC 4649&	\dots&	\dots&	\dots&	\dots&	\dots\\
NGC 5128&	 & 234.9 $\pm$ 77.3	&\dots &	\dots & \dots\\			
NGC 5273&	\dots&	49.5 $\pm$ 4.1&	83.2 $\pm$ 5.8&	\dots&	\dots\\
NGC 5631&	1.7 $\pm$ 0.5&	3.6 $\pm$ 0.2&	9.4 $\pm$ 0.1&	3.7 $\pm$ 0.3&	\dots\\
NGC 5846&	\dots&	\dots&	\dots&	\dots&	\dots\\
NGC 5898&	\dots&	4.5 $\pm$  0.4 &	9.3 $\pm$  0.5 &	\dots&	2.0 $\pm$  0.4\\
NGC 7192&	\dots&	2.7 $\pm$  0.2 &	4.0 $\pm$  0.2 &	\dots&	\dots\\
NGC 7332&	1.8 $\pm$  0.3 &	4.7 $\pm$  0.3 &	8.2 $\pm$  0.7 &	1.0 $\pm$  0.3 &	1.1 $\pm$  0.2\\
IC 5063&	\dots&	480.5 $\pm$  58.9 &	524.9 $\pm$  66.5 &	\dots&	\dots\\
\hline
\end{tabular}
}
\addtocounter{table}{-1}
\end{table*}

\section{Tables of nuclear properties of ETGs}

We collect in Appendix~B a set of tables summarizing the
properties of ETGs used to characterize the MIR spectral classes in the
atlas. References are  reported in the caption of each table.

In Table~B1 we report the optical ``activity'' class (columns 3 and 8), the nuclear
(2\arcsec\ radius) X-ray luminosity, L$_{X, nuc}$ (columns 4 and 9) and the radio 
power at 1.4 GHz (columns 5 and 10) which refers to the entire galaxy.
In Table~B2 and Table~B3 we report the kinematical and morphological 
peculiarities (columns 3 and 4)  for Es and S0s, respectively. 
In column 5 we give the morphology of dust-lanes from optical observations.  
The kinematical peculiarities  refer to the nuclear part 
of the galaxy, basically within one effective radius or less, so they provide 
a description of star and gas properties that contribute to the formation 
of the present {\it Spitzer}-IRS spectrum.

\vfill \eject

\begin{table*}
\label{tabB1}
\tiny{
\caption{Optical activity class, nuclear X-ray luminosity and 1.4 GHz radio power}
\begin{tabular}{lllcclllll}
\hline
 Ident. & Morpho. & Opt.  & Log L$_{X,nuc}$    & P$_{1.4~GHz}$& Ident.  & Morpho. & Opt.  & Log L$_{X,nuc}$ & P$_{1.4 ~GHz}$\\
           & RSA&  class & [erg~s$^{-1}$] & [W~Hz$^{-1}$]       &             &RSA &  class &  [erg~s$^{-1}$]   & [W~Hz$^{-1}$]   \\
\hline
NGC 636   &  E1 & \dots & \dots & <6.4~10$^{19}$ & NGC 5090 & E2 &  LIN(H)  & \dots & \dots \\
 NGC 720  & E5 & \dots  &  38.90 &  <1.2~10$^{20}$  &NGC 5638 & E1 & IN   &  \dots    &   <4.1~10$^{19}$ \\
NGC  821 & E6 & \dots & <38.37 & <6.3~10$^{19}$  & NGC 5812 &  E0 & IN& \dots & 1.3~10$^{20}$ \\
NGC 1209 & E6 & LIN(H) & \dots & 1.9~10$^{21}$  & NGC 5813 & E1 & LIN(W); L2: & 38.80& 1.8~10$^{21}$ \\
NGC 1275 & E pec  & S1.5 & \dots &  1.4~10$^{25}$ &NGC 5831 & E4 & IN & \dots & <5.3~10$^{19}$  \\
NGC 1297 & E2 & LIN(H) & \dots & <5.9~10$^{19}$   & NGC 7619 & E3 & \dots & 40.84 &6.8~10$^{21}$  \\
NGC 1339 & E4 & \dots & \dots & <4.2~10$^{19}$   & IC 1459    & E1  & LIN(H) & 40.87  &1.3~10$^{23}$ \\
NGC 1374 & E0 & \dots & \dots & <2.3~10$^{19}$   &   IC 2006   &  E1 & Comp(W) & \dots & 4.9~10$^{19}$\\
NGC  1379 & E0 & \dots & \dots & <1.4~10$^{19}$ &  IC 3370 & E2 pec & LIN(H) &\dots & 8.8~10$^{20}$ \\
NGC 1395 & E2 & \dots  & 39.06 & 5.4~10$^{19}$ & IC 4296     &  E0 & LIN(H) & 41.20 & 5.6~10$^{24}$ \\
NGC 1399 &  E1 & & <38.96 & 1.2~10$^{23}$        &  &  & &  &   \\
NGC 1404 & E2 & \dots & 40.57 & 1.9~10$^{20}$ & NGC 1052 & E3/S0 & LIN(H); L1.9 & 41.20 & 5.1~10$^{22}$ \\
NGC  1407 & E0 & IN  & <39.14 & 8.7~10$^{21}$ & NGC 1351 & S0$_1$/E6  & \dots & \dots & 4.9~10$^{19}$  \\
NGC 1426 & E4 & IN & \dots & <8.5~10$^{19}$   & NGC 4472 & E1/S0$_1$ & S2:: & <38.67& 7.4~10$^{21}$ \\
NGC 1427 & E5 & \dots & \dots &<3.2~10$^{19}$ & NGC 4550    & E/S0  & L2 & <38.37 & <2.0~10$^{19}$ \\
NGC 1453 & E2 & LIN(H) & \dots & 9.5~10$^{21}$& NGC 4570 & E7/S0$_1$  & \dots & 38.18 & <3.5~10$^{19}$\\
NGC 1549 & E2 &  \dots  &38.46 & \dots & NGC 4636 & E0/S0$_1$ & LIN(H); L1.9 & <38.24 & 2.0~10$^{21}$ \\
NGC 1700 & E3 & \dots & 38.84 &<3.3~10$^{20}$ & NGC 5353 & S0$_1$/E7 & L2/T2 & \dots & 4.2~10$^{21}$ \\
NGC 2300 & E3 &  \dots & 40.96 & 2.4~10$^{20}$    &          NGC 6868 & E3/S0$_{2/3}$  & LIN(H) & \dots & \dots \\
NGC 2974 & E4 &  LIN(H) & 40.32 & 5.7~10$^{20}$  & NGC 584 & S0$_1$  & \dots & \dots &  1.6~10$^{21}$  \\
NGC 3193 & E2 & L2: & <39.74&<1.1~10$^{20}$   & NGC 1366 & S0$_1$ & IN & \dots &\dots \\
NGC 3258 &  E1 & Comp(H) & \dots &  6.2~10$^{21}$& NGC 1389 & S0$_1$/SB0$_1$ & IN & \dots &<5.4~10$^{19}$  \\
NGC 3268 & E2 &LIN(H) & \dots & 3.7~10$^{21}$ & NGC 1533 & SB0$_2$/SBa & LIN(H) & \dots & \dots  \\
NGC 3377 & E6 & \dots & 38.24 & <1.8~10$^{19}$& NGC 1553 & S0$_{1/2}$ pec & LIN(W) & 40.22 & \dots  \\
NGC  3379 & E0 & L2/T2:: & 38.12 &3.2~10$^{19}$ & NGC 2685 & S0$_3$ pec&  S2/T2 & \dots & 3.3~10$^{19}$ \\
NGC 3557 & E3 & LIN(W) & 40.24 & 2.0~10$^{23}$ & NGC 3245 & S0$_1$ & T2: &  39.00  & 3.5~10$^{20}$ \\
NGC 3608 &  E1 & L2/S2:& 38.21& 8.2~10$^{19}$ & NGC 4036 & S0$_3$/Sa  & L1.9 & 39.05& 5.4~10$^{20}$ \\
NGC 3818 & E5 & IN(Traces) & \dots & <2.8~10$^{20}$ & NGC 4339 & S0$_{1/2}$  & \dots & \dots & <9.7~10$^{18}$ \\
NGC 3904 & E2 &  \dots & \dots & <1.3~10$^{20}$  &NGC 4371 & SB0$_{2/3}$(r) & \dots & \dots & <3.8~10$^{19}$ \\
NGC 3962 &  E1 & LIN(H) & \dots & 3.4~10$^{20}$ &NGC 4377 & S0$_1$  & \dots & \dots & 6.0~10$^{19}$\\
NGC 4261 & E3 & L1 & 41.10 & 2.6~10$^{24}$  &NGC 4382 & S0$_1$ pec & \dots & <37.93 & <6.1~10$^{19}$ \\
NGC 4365 & E3 & \dots & 38.25 & <3.2~10$^{19}$ &NGC 4383 & S0:  & MRK star-form.  & \dots & \dots \\
NGC 4374 & E1  & LIN(H); L2 & 39.50  &2.9~10$^{23}$ & NGC 4435 & SB0$_1$  & T2/H:; no AGN & 38.45 & \dots \\
NGC 4473 & E5   &  \dots         &<38.14 &<2.2~10$^{19}$ &NGC 4442 &  SB0$_1$ & \dots & \dots & <1.4~10$^{19}$ \\
NGC 4478 & E2 & \dots  &<38.49 & \dots  &NGC 4474 & S0$_1$   & \dots & \dots & <1.7~10$^{19}$ \\
NGC  4486 & E0 &  L2  &  40.80 & 6.9~10$^{24}$  & NGC 4477 & SB0$_{1/2}$/SBa   &  S2 & \dots &<7.4~10$^{19}$\\
NGC 4564 & E6 & \dots  & 38.45 & 4.9~10$^{19}$  & NGC 4552 & S0$_1$ & Comp (W); T2: & 39.20 & 3.1~10$^{21}$ \\
NGC 4589 & E2 &  L2 & 38.90 &  2.1~10$^{21}$ &NGC 4649 & S0$_1$  & \dots & 38.09 & 9.4~10$^{20}$ \\
NGC 4621 & E5 & \dots & 38.92 & <1.3~10$^{19}$ &NGC 5128 &S0+S pec &  S2 & 41.88 &  4.4~10$^{23}$ \\
NGC 4660 & E5 & \dots & 38.22 & <3.5~10$^{19}$  &NGC 5273 & S0/a & S1.5 & 40.55 & 1.1~10$^{20}$\\
NGC 4696 & E3 & LIN(H)  & 40.00 & \dots &NGC 5631 & S0$_3$/Sa & S2/L2: & \dots & 1.2~10$^{20}$\\
NGC 4697 & E6 & LIN(W) & 38.41 & 5.2~10$^{20}$   &NGC 5846 & S0$_1$  & LIN(H); T2 &  40.80 &  1.6~10$^{21}$ \\
NGC 5011 & E2 &  LIN(W) & \dots & \dots  &NGC 5898 & S0$_{2/3}$ & LIN(W) & \dots & <1.5~10$^{20}$ \\
NGC 5018 & E4 &  LLAGN   & <39.41 & 3.5~10$^{20}$& NGC 7192 & S0$_2$  & LIN(W)  & \dots & \dots\\
NGC 5044 &  E0 & LIN(H)  & 39.44& 4.0~10$^{21}$   & NGC 7332 & S0$_{2/3}$ & IN(Traces) & <39.70 & <9.5~10$^{19}$\\
NGC 5077 & E3 & LIN(H); L1.9 & \dots & 2.8~10$^{22}$  & IC 5063 & S0$_3$/Sa & S2 & \dots & \dots \\
\hline\hline
\end{tabular}}

The optical activity (columns 3 and 8) is derived from \citet{Annibali10} who use
the following notation: LlN = LINER; AGN = AGN like emission; IN = either faint
(Traces) or no emission lines; Comp = transition between HII regions and LINERs.
W and H  indicate weak emission ($EW(H_\alpha +[NII]6584)<3 \AA$) and
strong emission line galaxies, respectively. For the optical activity class, we also use
the notation of \citet{Ho97} in S is for Seyfert, L for LINERS, T for transition objects
 and H for HII region (: indicate uncertain estimates).   
 The nuclear X-ray luminosity, L$_{X,nuc}$ (columns 4 and 9) and
the radio power at 1.4 GHz, P$_{1.4~GHz}$ (columns 5 and 10), are from \cite{Pellegrini10} and \citet{Brown11},
respectively.
\end{table*}

\begin{table*}
\label{tabB2}
\tiny{
\caption{Ellipticals: kinematical and morphological characterization from optical studies}  
\begin{tabular}{lllll}
\hline
Ident & Morpho. & Gas vs. stars kinematic & Morphological & Dust-lane \\
         & RSA  & peculiarities                    & peculiarities                          &  morphology \\
\hline
  NGC 636  &  E1  &   \dots   & \dots & asymmetric dust distribution (16) \\
  NGC~720   & E5  & \dots & Boxy outer isophotes (6) & no dust (13)\\
  NGC~821   & E6 & SC (4) & \dots & no dust (9) \\
  NGC~1209 & E6 &   \dots    & X-like struct.; NW linear feature (6) & \dots \\
  NGC~1297 & E2 &      \dots   & \dots & dust-lane (a) \\
  NGC~1275 & E pec & \dots & jet & complex patches (18) \\
  NGC~1339 & E4 & \dots & \dots & no dust (10) \\
  NGC 1374 &  E0 & \dots & \dots & no dust (10,11) \\
  NGC 1379 &  E0 & \dots & \dots & no dust  (11)    \\
  NGC~1395 & E2 & \dots & NW perpend. feature (6); low contr. shells (7) & \dots \\ 
  NGC~1399 & E1 & \dots &\dots & no dust (10,11) \\ 
  NGC~1404 & E2 & \dots & \dots & no dust (11) \\
  NGC~1407 & E0 & rotat.  min. axis  (a) & \dots &  no dust  (13) \\
  NGC~1426 & E4 & \dots & \dots & no dust(10) \\
  NGC~1427 & E5  & \dots & \dots & no dust (10,11)  \\
  NGC~1453 & E2 & g-d and g-maj t  (a) & \dots & \dots \\
  NGC~1549 & E2 & stars rotat.  min.  axis (0) & shells (7) & \dots \\
  NGC~1700 & E3  & CR s-s(1)& \dots & chaotic dust patches (10) \\
  NGC~2300 & E3 & \dots & \dots & no-dust (10) \\
  NGC~2974 & E4 & SC(4) & multiple shells (6); g-d  and g-maj t$\approx$20$^\circ$ (a) & spiral dust-lanes (10) \\
  NGC~3193 & E2 & \dots & \dots & no dust  (9) \\
  NGC~3258 & E1 & CR g-s (1)  & \dots  & no dust  (13)  \\
  NGC~3268 & E2 &\dots  & \dots  & small disk of dust (a,16) \\
   NGC~3377 & E6 & modest star rotat. min. axis (3); SC(4) & \dots & chaotic dust patches and filam. (9,10) \\
  NGC 3379   & E0 & SC (4)  &\dots & nuclear dust ring (10)  \\
  NGC~3557 & E3 & \dots & SW fan; asym.  outer isophotes (6) & nuclear ring of dust (10)\\
  NGC 3608  & E1 & CR s-s (3); KDC (4) & \dots & nuclear dust ring (10) \\ 
  NGC~3818 & E5  & \dots & \dots & \dots \\
  NGC~3904 & E2 & \dots & \dots & \dots \\
  NGC~3962 & E1 &\dots &gas disk+outer  arc-like struct. (a) & dust patches (a) \\
  NGC~4261 & E3 & \dots &  NW tidal arm/ faint SE fan (6) & dust disk (9) \\
  NGC~4365 & E3 & \dots & Faint SW fan (6)  & no dust (10) \\
  NGC~4374 & E1 &Rotat. Vel.  $\approx$0 (3); SC (4)  & \dots & dust-lane (a,9) \\
  NGC~4473 & E5  & complex moprh. of $\sigma$ and Vel. Field (3); MC (4) &  \dots & no dust (10) \\
  NGC~4478  & E2 & \dots & \dots &no dust (10) \\
  NGC 4486   & E0 & V$\approx$0 (3); SC(4)& jet& no dust (9) \\
  NGC~4589  & E2 & \dots & \dots & chaotic dust patches (10) \\
  NGC~4564 & E6  & SC (4) & \dots & no-dust (9) \\
  NGC~4621 & E5  & CR s-s inner2\arcsec\ (3); KDC (4) & \dots & no dust (10)  \\
  NGC~4660 & E5 & two disk components (3); MC (4) &  \dots & no dust (9)  \\
  NGC~4696 & E3 & \dots &  Faint outer  shells (2) & dust arc (a,11,13) \\
  NGC~4697 & E6  & \dots &   Non spherical  inner isophotes (6)& disk (23) \\
  NGC~5011 & E2  & \dots & \dots & no dust (11,13) \\
  NGC~5018 & E4 & \dots &  multiple tidal tails and shells (6,7) & dust  nuclear ring/chaotic (10)\\
  NGC~5044 & E0  & CR s-s; gas  irr.motion (a)  & gas fil. shape (a) & chaotic dust patches (15) \\
  NGC~5077 & E3 &  CR s-s (a);  g-d and  g-maj t$\approx$90$^\circ$ (a) & \dots & dust filaments (9) \\
  NGC~5090 & E2 & \dots & \dots & no dust (11,13) \\
  NGC~5638 & E1 & \dots & \dots & \dots  \\
  NGC~5812 & E0 &\dots & Tidal tail  (2) & dust nuclear disk 0.4\arcsec\ diameter (a) \\
  NGC~5813 & E1 & CR s-s; gas  irr. motion (a); KDC (4) & gas fil. shape (a) &  dust nuclear ring/chaotic (10)   \\
  NGC~5831 & E4 &  CR s-s (2,3); KDC (4) &  \dots & no dust (9)\\ 
  NGC~7619 & E3 & rotat. min. axis (17) & \dots & no dust (10)  \\
  IC~1459    & E1  & CR g-g (1) &  Multiple shells (6) & dust nuclear ring/chaotic (10) \\
  IC~2006    & E1  & CR g-s (1) & \dots & \dots\\
  IC~3370  & E2 pec & \dots &  X-like struct.;  broad N fan (6); polar ring? (5) & spiral dust-lane (10,11) \\
  IC~4296    & E0  &  CR s-s (a)  & \dots & nuclear ring (10)  \\
\hline\hline
\end{tabular}}

In column 3 we use the following notation: 
{\bf CR g-s}: counter rotation gas vs. stars;  
{\bf CR s-s}: counter rotation stars vs. stars; {\bf CR g-g}: 
counter rotation gas vs. gas; {\bf rotat. min. axis}: 
stars rotate along the galaxy minor axis; 
{\bf g-d and g-maj t}: gas disk and galaxy major  
axis are tilted by the reported angle, if provided in the literature. 
{\bf KDC}: kinematical decoupled component, not necessarily counter-rotation; 
{\bf MC}: multiple components; {\bf SC} single component 
\citep[see][]{Krajnovic08}. A description of the kinematic and morphological peculiarities  
of the galaxies and full references are reported in in \citet{Rampazzo05}  and  
\citet{Annibali06} labelled by (a).
 Further kinematical references: (0) \citet{Rampazzo88}; (1) \citet{Corsini98}; 
(2) \citet{Davies83}; (3) \citet[][]{Emsellem04}; (4) \citet[][]{Krajnovic08}; 
 (5) \cite{Sil'chenko09}
Morphological peculiarities are from: (6) \citet{Tal09}; (7) \citet{MC83}; (8) \citet{Pierfederici04}.
The dust detection and  structure in the optical bands are from 
(9) \citet{Zhang08}; (10) \citet{Lauer05};
(11) \citet{Sadler95}; (12) \citet{Coccato04}; (13) \citet{Veron88}; (14) \citet{Morganti07}; (15) \citet{Temi07};
(16) \citet{Ferrari02}; (17) \citet{Pu10}; (18) \citet{Tremblay07}; 
(19) \citet{Wiklind01}; (20) \citet{Simoes07}; (21) \citet{Pinkney03}; (22) \citet{Patil07}; (23) \citet{vanDokkum95}; (24)
\citet{Kormendy09}; \citet{Whitmore90}.
\end{table*}

\begin{table*}
\label{tabB3}
\tiny{
\caption{E/S0s and S0s: kinematical and morphological characterization from optical studies}  
\begin{tabular}{lllll}
\hline
Ident & Morpho.& Gas vs. stars kinematic & Morphological & Dust-lane \\
         & RSA  & peculiarities                    & peculiarities                 &  morphology \\
\hline
NGC~1052 & E3/S0  &  CR g-g (1,a)   & \dots & dust (9) \\
NGC~1351 & S0$_1$/E6 & \dots & \dots & no dust (11) \\
NGC~4472 &  E1/S0$_1$ & CR s-s (1) & \dots & chaotic dust patches (10) \\
NGC~4550 &  E/S0  &  CR -ss (1);  Slow rot  (3); sC (4) & \dots & asymmetric dust patches (19) \\
NGC~4570 &  E7/S0$_1$ & MC (4)& \dots & no dust (20) \\
 NGC~4636 & E0/S01(6)   & gas irr. motion (a)  & \dots & dust lanes (9) \\
NGC~5353 &  S0$_1$/E7 & \dots & \dots & \dots \\
NGC~6868 & E3/S02/3(3)  & CR g-s (a)  & \dots & dust patches (a,11) \\
NGC ~584 & S0$_1$  & \dots & Non-spherical isophotes (6) & chaotic dust patches (10) \\
 NGC~1366 & E7/S01(7)    & \dots      & \dots    & \dots \\
 NGC~1389 & S01(5)/SB01  & \dots     & \dots    &   no-dust (11) \\
 NGC~1533 & SB02(2)/SBa   & \dots     & \dots    &   no-dust (11) \\
  NGC~1553 & S01/2(5)pec    & \dots & shells (7) & no-dust (13) \\
 NGC~2685 & S0$_3$ pec & $\sigma$ double peak (7); SC (4) gas min.axis grad (12) & polar ring (25) & polar dust lanes (7) \\
NGC~3245 & S0$_1$ & \dots & \dots & \dots \\
NGC~4036 & S0$_3$/Sa & gas min.axis grad (12) & \dots & irr.dust structures (20) \\
NGC~4339 & S0$_{1/2}$ & \dots & \dots & \dots \\
NGC~4371 & SB0$_{2/3}$(r) & \dots & \dots & dust disk (20)\\
NGC~4377 & S0$_1$   & \dots & \dots & \dots \\ 
NGC~4382 & S0$_1$ pec & CR s-s (3); MC(4) & shells  (24) & no dust (10) or weak (20) \\
NGC~4383 & S0: & gas min.axis grad (12) & \dots & \dots \\
NGC~4435 & SB0$_1$ & \dots & Interact. with NGC 4438 & dust nuclear disk (12)\\
NGC~4442 & SB0$_1$  & \dots & \dots & \dots \\
NGC~4474 & S0$_1$ & \dots & \dots & \dots \\
NGC~4477 &  SB0$_{1/2}$/SBa  & CR s-s (1); SC (4) & kin. and phot. axes misaligned (3) & dust spiral (20) \\
NGC~4552 & S01(0)      &  KDC (4) & shells (7) & chaotic dust patches (10) \\
NGC~4649 & S0$_1$  &asym. rot. curve (21) & \dots & no dust (10) \\
NGC~5128 & S0+S pec & gas min.axis grad (12) & Many shells (7), warps (11) & strong dust lane (11) \\
NGC~5273 & S0/a &  & \dots & weak dust features (20) \\
NGC~5631 & S0$_3$/Sa & CR (g-s) (5) & \dots & \dots \\
NGC~5846 & S0$_1$(0)   & gas irr. motion (a); SC(4)   & Faint outer shells (6) & dust filaments extended (a) \\
NGC~5898 & S0$_{2/3}$(0) &  CR g-s (1); gas min.axis grad (12) &  Three  spiral arm-like tidal tails (6) & dust patches (22) \\
NGC~7192 & S0$_{2}$(0)  & CR s-s (a) &   Shell (6)  & no dust (13) \\
NGC~7332 & S0$_{2/3}$   & CR g-g (1); KDC (4); gas min.axis grad (12) & \dots & \dots \\
IC~5063    & S0$_{3}$(3)pec/Sa  &  gas min.axis grad (12) & \dots  & dust-lane (14) \\
\hline\hline
\end{tabular}}

\noindent{References and legenda as in Table~B2.}
\end{table*}

\bsp

\label{lastpage}

\end{document}